\documentclass[11pt]{article}
\usepackage[a4paper, margin=1in]{geometry}
\usepackage{setspace}
\usepackage{xcolor}
\usepackage{etoolbox}
\usepackage{amsmath, amssymb, mathtools, amsthm, mathrsfs, bbm, dsfont, bm}
\usepackage{booktabs, multirow, siunitx, pdflscape, float, threeparttable}
\usepackage{microtype, graphicx, subcaption, multirow, makecell, calc, comment, url}
\usepackage[normalem]{ulem}
\usepackage[authoryear,square]{natbib}

\usepackage[colorlinks=true,linkcolor=blue,citecolor=blue,urlcolor=blue]{hyperref}
\usepackage[capitalize,nameinlink,noabbrev]{cleveref}
\usepackage{aliascnt}
\usepackage{enumitem}
\usepackage{algorithm, algpseudocode}

\theoremstyle{plain}
\newtheorem{innerTheorem}{Theorem}[section]

\newaliascnt{innerLemma}{innerTheorem}
\newtheorem{innerLemma}[innerLemma]{Lemma}
\aliascntresetthe{innerLemma}

\newaliascnt{innerCorollary}{innerTheorem}
\newtheorem{innerCorollary}[innerCorollary]{Corollary}
\aliascntresetthe{innerCorollary}

\newaliascnt{innerProposition}{innerTheorem}
\newtheorem{innerProposition}[innerProposition]{Proposition}
\aliascntresetthe{innerProposition}

\theoremstyle{definition}
\newaliascnt{innerDefinition}{innerTheorem}
\newtheorem{innerDefinition}[innerDefinition]{Definition}
\aliascntresetthe{innerDefinition}

\newtheorem{innerAssumption}{Assumption}

\newaliascnt{innerExample}{innerTheorem}
\newtheorem{innerExample}[innerExample]{Example}
\aliascntresetthe{innerExample}

\theoremstyle{remark}
\newaliascnt{innerRemark}{innerTheorem}
\newtheorem{innerRemark}[innerRemark]{Remark}
\aliascntresetthe{innerRemark}

\crefname{innerTheorem}{Theorem}{Theorems}
\crefname{innerLemma}{Lemma}{Lemmas}
\crefname{innerCorollary}{Corollary}{Corollaries}
\crefname{innerProposition}{Proposition}{Propositions}
\crefname{innerDefinition}{Definition}{Definitions}
\crefname{innerAssumption}{Assumption}{Assumptions}
\crefname{innerExample}{Example}{Examples}
\crefname{innerRemark}{Remark}{Remarks}

\NewDocumentEnvironment{theorem}{m m}
  {\ifstrempty{#1}{\begin{innerTheorem}}{\begin{innerTheorem}[#1]}\label{thm:#2}}
  {\end{innerTheorem}}
\NewDocumentEnvironment{lemma}{m m}
  {\ifstrempty{#1}{\begin{innerLemma}}{\begin{innerLemma}[#1]}\label{lem:#2}}
  {\end{innerLemma}}
\NewDocumentEnvironment{corollary}{m m}
  {\ifstrempty{#1}{\begin{innerCorollary}}{\begin{innerCorollary}[#1]}\label{cor:#2}}
  {\end{innerCorollary}}
\NewDocumentEnvironment{proposition}{m m}
  {\ifstrempty{#1}{\begin{innerProposition}}{\begin{innerProposition}[#1]}\label{prop:#2}}
  {\end{innerProposition}}
\NewDocumentEnvironment{definition}{m m}
  {\ifstrempty{#1}{\begin{innerDefinition}}{\begin{innerDefinition}[#1]}\label{def:#2}}
  {\end{innerDefinition}}
\NewDocumentEnvironment{assumption}{m m}
  {\ifstrempty{#1}{\begin{innerAssumption}}{\begin{innerAssumption}[#1]}\label{assump:#2}}
  {\end{innerAssumption}}
\NewDocumentEnvironment{example}{m m}
  {\ifstrempty{#1}{\begin{innerExample}}{\begin{innerExample}[#1]}\label{ex:#2}}
  {\end{innerExample}}
\NewDocumentEnvironment{remark}{m m}
  {\ifstrempty{#1}{\begin{innerRemark}}{\begin{innerRemark}[#1]}\label{rem:#2}}
  {\end{innerRemark}}

\newcommand{\E}[1]{\mathbb{E}\left[#1\right]}
\newcommand{\Var}[1]{\mathrm{Var}\left[#1\right]}
\newcommand{\Prob}[1]{\mathbb{P}\left(#1\right)}
\newcommand{\abs}[1]{\left|#1\right|}

\newcommand{\given}{\,\middle|\,}
\newcommand\cA{\mathcal{A}}

\newcommand\cF{\mathcal{F}}

\newcommand\cH{\mathcal{H}}

\newcommand\cL{\mathcal{L}}

\newcommand\cP{\mathcal{P}}

\newcommand\cT{\mathcal{T}}
\newcommand\cU{\mathcal{U}}
\newcommand\cV{\mathcal{V}}

\newcommand\bM{\mathbb{M}}

\newcommand\bP{\mathbb{P}}

\newcommand\bR{\mathbb{R}}

\title{Reinforcement Learning for Speculative Trading under Exploratory Framework}
\author{
  Yun Zhao\thanks{ Department of Mathematics, Imperial College, London SW7 2AZ, UK. Email: \texttt{yun.zhao23@imperial.ac.uk}. Supported by Roth Scholarship.}
  \and
  Alex S.L. Tse\thanks{ Department of Mathematics, University College London, London WC1H 0AY, UK. Email: \texttt{alex.tse@ucl.ac.uk}.}
  \and
  Harry Zheng\thanks{ Department of Mathematics, Imperial College, London SW7 2AZ, UK. Email: \texttt{h.zheng@imperial.ac.uk}.}}
\date{}

\begin{document}

\maketitle

\begin{abstract}
We study a speculative trading problem within the exploratory reinforcement learning (RL) framework of \citet{wangReinforcementLearningContinuous2020}. The problem is formulated as a sequential optimal stopping problem over entry and exit times under general utility function and price process. We first consider a relaxed version of the problem in which the stopping times are modeled by the jump times of Cox processes driven by bounded, non-randomized intensity controls. Under the exploratory formulation, the agent's randomized control is characterized via the probability measure over the jump intensities, and their objective function is regularized by Shannon's differential entropy. This yields a system of the exploratory HJB equations and Gibbs distributions in closed-form as the optimal policy. Error estimates and convergence of the RL objective to the value function of the original problem are established. Finally, an RL algorithm is designed, and its implementation is showcased in a pairs-trading application.
\end{abstract}



\section{Introduction}

Machine learning in finance is a rapidly growing field that is gradually transforming how financial institutions operate and make decisions. One of its most promising sub-fields is the use of deep and reinforcement learning (RL) for quantitative finance applications, where examples include derivatives pricing, portfolio optimization and algorithmic trading. RL adopts a model-free perspective: agents choose actions in an unknown environment by learning from interactions, and they optimally balance the exploration of the environment and the exploitation of the learned knowledge. RL has shown great success in various fields, including playing games such as Go \citep{silverMasteringGameGo2016, silverMasteringGameGo2017}, training large language models \citep{ouyangTrainingLanguageModels, baiConstitutionalAIHarmlessness2022, guoDeepSeekR1IncentivizesReasoning2025}, controlling robotics \citep{levineEndtoEndTrainingDeep, haarnojaSoftActorCriticAlgorithms2019}, managing energy \citep{chenReinforcementLearningSelective2022}, and many more.

Central to RL is the balance between curiosity and greediness: if the agent explores too much, they may never exploit the learned knowledge and in turn fails to consistently perform the action that delivers the highest reward; if the agent exploits too much, they may never sufficiently explore the unknown environment and thus may fail to discover a state or an action that gives better rewards. It is therefore crucial to develop a systematic way to balance exploration and exploitation in RL algorithms. In discrete time, entropy-regularized RL formulations address this by adding the entropy of the exploration strategy to the objective, with a temperature parameter governing the importance of exploration \citep{ziebartMaximumEntropyInverse, nachumBridgingGapValue, foxTamingNoiseReinforcement, neuUnifiedViewEntropyRegularized}.
With continuous time, state and action space, \citet{wangReinforcementLearningContinuous2020} introduce an exploratory formulation of a stochastic control framework for RL in which Shannon's entropy quantifies the cost of exploration. The policy is a probability measure over the action space and the reward is penalized by the exploration cost. A distributionally dispersed policy yields higher entropy (less penalty) which encourages exploration, whereas a deterministic policy (i.e., a Dirac measure) incurs infinite negative entropy and thus is always sub-optimal.
\citet{wangReinforcementLearningContinuous2020} focus on the linear-quadratic (LQ) setting and show that the optimal policy is Gaussian, aligning with the widespread use of Gaussian exploration in RL algorithms.
The exploratory framework has since been extended to general exploratory HJB equations \citep{tangExploratoryHJBEquations2022},
policy gradient and Q-learning algorithms \citep{jiaPolicyGradientActor2021, jiaQLearningContinuousTime2022}, mean-variance portfolio selection \citep{wangContinuoustimeMeanVariance2020}, mean-field games \citep{guoEntropyRegularizationMean2021},
jump diffusions \citep{gaoReinforcementLearningJumpDiffusions2025}, and optimal stopping problems \citep{dongRandomizedOptimalStopping2024a,daiLearningOptimallyStop2025}, among others.

RL is well suited to algorithmic trading because it can learn trading strategies from data without strong assumptions on the market dynamics. 
This paper extends the framework of \citet{wangReinforcementLearningContinuous2020} to a sequential stopping problem in the context of speculative trading. In such problem, the agent optimally times the entry and exit of an investment opportunity and their goal is to maximize the expected utility associated with the round-trip profit net of any transaction costs. Our setup can encapsulate many different trading applications. One example is the model of \citet{tseSpeculativeTradingProspect2023}, in which the agent needs to decide the optimal time to first purchase and then to sell an indivisible risky asset to maximize their expected utility under Prospect Theory preference. Another example is pairs-trading (e.g., \cite{leungOPTIMALMEANREVERSION2015}) where the price spread of two strongly co-moving stocks can be described by some mean-reverting process, and the agent needs to strategically enter and exit the trade to maximize profit.

In a nutshell, this paper presents a theoretical and methodological framework featuring the following three elements simultaneously: (i) a dynamic optimization problem involving sequential (rather than one-off) stopping; (ii) a continuous-time, exploratory RL framework under general price dynamics and utility functions which bypasses modeling assumptions; and (iii) a suitable formulation of the randomized policy that captures the spirit of exploratory RL in a classical sense. More specifically, our work contributes to the growing literature of continuous-time RL approach for optimal stopping on both the theoretical and computational front, as summarized below.

\begin{itemize}
	\item \textbf{Theory.} We extend the continuous-time RL framework in \citet{wangReinforcementLearningContinuous2020}, \citet{dongRandomizedOptimalStopping2024a} and \citet{daiLearningOptimallyStop2025} to a sequential (entry and exit) optimal stopping problem under general diffusion dynamics and utility functions. Following the randomization idea of \citet{dongRandomizedOptimalStopping2024a}, we model the stopping times via the jump times of two Cox processes driven by deterministic intensity processes. Upon introducing two new auxiliary variables which respectively describe the current trading regime (before entry, after entry but before exit, and after exit) and the reference entry price of an existing open position, we reformulate the sequential stopping problem as a relaxed Markovian control problem over two intensities processes bounded above some constant $M$.
	
	While the stopping times are randomized, the corresponding controlled intensities are still deterministic. To encourage exploration, we propose a regularized version of the problem where the control is now replaced by the probability measure over the intensities and the agent's reward is penalized by Shannon's differential entropy governed by a temperature parameter $\eta$. We then derive a system of HJB equations for three value functions, one for each possible trading regime. We also obtain the optimal exploration densities in closed-form as Gibbs distributions, which parameters are driven by the economic advantage of opening or closing the position.
	
	Furthermore, we establish error bounds for the RL objective relative to the original value function in terms of the intensities bound $M$ and the temperature parameter $\eta$, and prove convergence when $M\to\infty$ and $\eta \downarrow 0$ with $M^2\abs{\eta\ln\eta}\to 0$. Another novelty of our approach concerns the definition of the entropy considered. Instead of taking entropy with respect to product Lebesgue measure as in \citet{wangReinforcementLearningContinuous2020}, we take the entropy relative to the uniform distribution, so it is nonpositive and gives monotone convergence of the value function. Our analysis relies on probabilistic techniques, in contrast to the PDE-based arguments in the extant literature such as \citet{dongRandomizedOptimalStopping2024a,daiLearningOptimallyStop2025}.

	\item \textbf{Numerics.} 
	We demonstrate the methodologies developed using a concrete example of pairs-trading. The underlying price dynamics is an  Ornstein--Uhlenbeck (OU) process that represents the stochastic spread between two risky assets. After solving the HJB system by finite differences method iteratively, we validate the impact of $M$ and $\eta$ on our error bounds. Our numerical results confirm the distinctive regularization effects of $M$ and $\eta$ for the learning problem. A low $M$ prevents the agent from entering or exiting the trade in time to capture the profit, while a high $\eta$ induces excessive exploration and weak exploitation.
	
	We further investigate the optimal entry and exit decision via examining the densities of the stopping intensities to illustrate the trading behaviors of the agent. We also quantify the effect of mean reversion speed and volatility: strong mean-reversion is economically (dis)advantageous if the spread is far from (close to) its long-run mean, while higher volatility generally increases the expected trading profit.
	
	Finally, we design an offline, model-free RL algorithm with policy iteration and deep neural network value approximation, and we demonstrate that it closely matches the benchmark solution from the HJB solver. 
\end{itemize}

We conclude the introduction by discussing some related works.
Optimal stopping has a long and rich history in both the areas of stochastic analysis and financial mathematics. See for example \citet{peskirOptimalStoppingFreeboundary2006} for an overview of the solution techniques, which typically involve characterizing the value function via a variational inequality and/or a free boundary problem.
The classical solution methods tend to suffer from the curse of dimensionality. As a result, there has now been a large literature on the use of deep learning to solve an optimal stopping problem. \citet{sirignanoDGMDeepLearning2018, zhaoNeuralNetworkConvergence2025a} solve variational inequalities for high-dimensional optimal stopping via deep neural networks. \citet{beckerDeepOptimalStopping2019, beckerSolvingHighdimensionalOptimal2021} approximate the optimal stopping rule in high dimensions using backward induction or forward schemes. \citet{wangDeepLearningFree2021a, reppenNeuralOptimalStopping2025} estimate the free boundaries with neural networks. These methods however have limitations, including for example the reliance on specific state dynamics assumptions to deduce the variational inequalities to be solved, time-discretization errors that arise from approximating the stopping time over a discrete time-grid, and the strong structural/regularity assumptions imposed on the stopping region.

In terms of model-free RL approach, \citet{dongRandomizedOptimalStopping2024a} is the first to apply the exploratory formulation of \citet{wangReinforcementLearningContinuous2020} to stopping problems. 
They randomize the stopping time using Bernoulli variables, so that the policy is an intensity process that continuously modulates the probability of stopping. 
This yields a continuous control formulation and an offline RL algorithm. 
Their convergence analysis, however, is restricted to American put pricing (equivalent to a specific piecewise linear utility function) under geometric Brownian motion (GBM). Furthermore, the policy is a non-randomized intensity process: the only source of randomness is an exogenous Bernoulli variable, so the agent is not choosing a distribution over actions and hence it does not constitute exploration in the traditional sense of RL. In contrast, we consider sequential stopping under general diffusion dynamics and utility functions, and we randomize the intensity itself using a probability measure and penalize exploration via Shannon's differential entropy. The agent determines the optimal sampling strategy over the candidate policies by directly balancing the exploration-exploitation tradeoff, which aligns better with the exploratory spirit of RL. We also cap the intensity by a constant $M$ to limit over-exploitation. With this approach, we get a numerically more stable HJB equation system because the source term is log-exponential, not exponential in \citet{dongRandomizedOptimalStopping2024a}.

Another related work is \citet{daiLearningOptimallyStop2025}. They start with a penalized PDE associated with the variational inequality and reformulate the problem as a binary-action stochastic control problem. Then an exploratory formulation with Bernoulli entropy regularization is applied to the latter where the control is now a probability measure over the binary action. There are some important differences between their approach and ours. First, we consider sequential (entry and exit) stopping rather than one-off stopping. Second, the parameterization of the stopping strategy is different. In \citet{daiLearningOptimallyStop2025}, the reformulated stochastic control problem yields a binary optimal control that partitions the state space into the stopping set and the continuation set. In turn, the randomized policy obtained in their exploratory formulation can be viewed as the classification probability of whether a given state is in the stopping set or not. Such classification probability nonetheless does not explicitly suggest how exploration shall be performed in the context of the original optimal stopping task. Meanwhile, our stopping times are parameterized by controlled intensities of Cox processes, and our exploration policy is a probability measure over the capped intensity range rather than a classification probability. Since the intensity directly corresponds to the stopping rate, our formulation naturally and directly results in an exploratory stopping rule for the original problem. Although the state and policy space are now higher-dimensional, we are still able to obtain a closed-form optimal policy and the resulting HJB system is decoupled such that the computational cost remains comparable. More generally, echoing Footnote 2 of \citep{daiLearningOptimallyStop2025}, the practical relevance of our approach via stopping intensity might become even more significant for a continuous-time optimal stopping problem under which a randomized stopping time is optimal (without entropy regularization). A prime example of such setup is optimal stopping under probability weighting, where \cite{henderson2017randomized, henderson2018probability} characterize the optimal strategy via non-degenerate stopping intensity functions.

The remainder of the paper is organized as follows. \cref{sec:stopping-to-control} sets up the speculative trading framework as a sequential optimal stopping problem, where we describe how the state space shall be suitably augmented and introduce the notion of randomized stopping times via Cox processes with controlled intensities. We also derive the error estimate in terms of the upper bound of the intensity processes. \cref{sec:entropy-regularization} introduces the exploratory (entropy-regularized) formulation of the problem that replaces non-randomized intensity policies with probability measures. We derive the HJB system for the control problem, and prove the error estimates and convergence of the RL objective to the original one. \cref{sec:OU-S-shaped-utility} presents numerical results for a pairs-trading application under the OU process, including the finite-difference solution of the HJB system and the model-free RL algorithm. \cref{sec:conclusion} concludes. Several technical proofs are deferred to the appendix.

\section{Speculative trading and intensity-relaxed problem} \label{sec:stopping-to-control}

\subsection{Original problem - three stage optimal stopping}
Let $\left(\Omega, \cF, (\cF_t)_{t\ge 0}, \bP\right)$ be a filtered probability space satisfying the usual conditions and $W=(W_t)_{t\ge 0}$ be a standard Brownian motion adapted to $(\cF_t)_{t\ge 0}$.
The underlying signal $P$ follows a general one-dimensional diffusion process:
\begin{equation} \label{eq:SDE-P}
  dP_t = \mu (P_t) dt + \sigma (P_t) dW_t, \quad t>0, \quad P_0 = p.
\end{equation}

In the speculative trading problem, the agent seeks the optimal time to enter and then exit a trading opportunity. The entry time $\tau$ and exit time $\nu$ are stopping times in $\cT := [0,\infty]$. When the trade is initiated at time $\tau$, a cost proportional to $P_{\tau}$ is  paid. When the trade is closed at time $\nu$, a payoff proportional to $P_{\nu}$ is received. The goal of the agent is to find the optimal pair $(\tau,\nu)$ to maximize their discounted expected utility over the round-trip profit net of any possible transaction costs. Mathematically, we define the value function at inception (i.e., before the entry of the trade) as 
\begin{equation}\label{eq:V-orig}
  V_{\text{orig}}(p) = \sup_{\tau,\nu \in \cT: \tau \leq \nu} \E{e^{-\rho \nu}U(\gamma P_{\nu}-\iota P_{\tau}-\Psi-R) \mathds{1}_{\{\nu<\infty\}} \given P_0 = p},
\end{equation}
where $U:\bR \to \bR$ is  a continuous increasing utility function,  $\rho > 0$ is a  subjective discount rate, and $\gamma, \iota, \Psi, R$ are constants.

\begin{remark}{}{scaling-of-the-signal}
  $\gamma, \iota$ can be viewed as some scaling parameters that describe the possible mismatches between the signal values and the actual traded prices. For example, if $P>0$ represents the price of some asset, we can set $\gamma \in (0,1]$ and $\iota \ge 1$ to represent the proportional transaction costs on sale and purchase respectively. Meanwhile, $\Psi\ge 0$ can be viewed as the fixed transaction costs in a round-trip trade, and $R\in\mathbb{R}$ can be viewed as some reference point against which the profit is evaluated (c.f. a Prospect Theory preference model). One could also set $R=-w_0$ where $w_0$ represents the agent's initial wealth. Then $\gamma P_{\nu}-\iota P_{\tau}-\Psi-R=\gamma P_{\nu}-\iota P_{\tau}-\Psi+w_0$ represents the agent's total wealth when the trade is closed.
\end{remark}

\begin{remark}{}{financial-interpretation-for-negative-value-of-signals}
  The signal process $P$ may take negative values.
  For example, under a pairs-trading strategy, the signal can be $P_t = S^A_t - S^B_t$, where $S^A, S^B$ are the price processes of two stocks.
  Then, opening position at time $\tau$ refers to buying $S^A$ and selling $S^B$, and closing position at time $\nu$ refers to liquidating the trade via selling $S^A$ and buying $S^B$.
  If proportional transaction costs $\gamma$ and $\iota$ are present, then
  the payoff shall be $\gamma(S^B_\tau + S^A_\nu) - \iota(S^A_\tau + S^B_\nu)$.
  This cannot be correctly captured by our one-dimensional model because $\gamma P_{\nu}-\iota P_{\tau} = \gamma (S^A_\nu - S^B_\nu) - \iota (S^A_\tau - S^B_\tau)$.
  The problem of pairs-trading with proportional transactions cost would require us to consider a two-dimensional model to describe the joint dynamics of $(S^A,S^B)$. When we study a pairs-trading application in \cref{sec:OU-S-shaped-utility}, we focus on the special case of $\gamma=\iota=1$ such that it becomes sufficient to model the one-dimensional process  $P_t = S^A_t - S^B_t$.
\end{remark}

\begin{remark}{}{discounting}
We  require $\rho>0$ to be sufficiently high to ensure the problem yields a finite value function. See \cref{assump:continuity-growth} and \cref{lem:wellposedness} below. Our theoretical analysis  depends heavily on the assumption of large positive $\rho$. More generally, consideration of zero discount rate will lead to two fundamental subtleties. First, it is no longer priori clear whether the value function is finite without further assumptions on $(P_t)_{t\geq 0}$ and $U(\cdot)$. Second, from an economic modeling viewpoint one needs to carefully distinguish the payoff to the agent in the cases of $\tau=\nu=\infty$ and $\tau<\infty=\nu$, which can arise as an optimal solution. It is because trading losses can no longer be infinitely deferred and discounted away such that an agent might not be willing to enter the trade at all if the investment opportunity is bad. In the model of \citet{tseSpeculativeTradingProspect2023} with zero discount rate,
	the objective is defined as
	\begin{equation*}
		\sup_{\tau,\nu \in \cT: \tau \leq \nu} \E{\cU(\tau,\nu) \given P_0 = p},
	\end{equation*}
	where
	\begin{equation*}
		\cU(\tau,\nu)=
		\begin{cases}
			U(\gamma P_{\nu}-\iota P_{\tau}-\varPsi-R), & \tau<\infty,\;\nu<\infty,\\
			U(-R), & \tau=\nu=\infty,\\
			U(-\iota P_{\tau}-\varPsi-R), & \tau<\infty,\;\nu=\infty.
		\end{cases}
	\end{equation*}
	In particular, the fixed transaction cost $\Psi$ only materializes if the agent decides to enter a trade (i.e., $\tau<\infty$) while the reference point $R$ is always present as a part of the agent's preference specification. When discounting is introduced, the payoff in the second and the third case shall be replaced by zero to capture the behaviors that utility value realized in the distant future should have zero net present value. This inspires our definition of the objective as per \labelcref{eq:V-orig}. 
\end{remark}

We introduce the following standing assumptions concerning the diffusion process and the utility function.
\begin{assumption}{}{continuity-growth}
	\mbox{}
	\begin{enumerate}
		\item The drift and diffusion coefficients $\mu(\cdot),\sigma(\cdot)$ are Lipschitz continuous, i.e., there exists a constant $C>0$ such that for any $x,y \in \bR$,
		\begin{equation*}
			\abs{\mu(x)-\mu(y)} + \abs{\sigma(x)-\sigma(y)} \leq C\abs{x-y}.
		\end{equation*}
		
		\item The utility function $U$ is H\"older continuous, i.e., there exist constants $C>0$ and $0<\kappa \leq 1$ such that for any $x,y \in \bR$,
		\begin{equation} \label{eq:utility-holder}
			\abs{U(x)-U(y)} \leq C\abs{x-y}^{\kappa}.
		\end{equation}
		
		\item The subjective discount rate $\rho$ is sufficiently large such that $\rho>\hat{A}$, where $\hat{A}>0$ is some constant to be defined in \labelcref{eq:A_hat} in \cref{appendix:proof-wellposedness}.
	\end{enumerate}
\end{assumption}

\begin{lemma}{}{wellposedness}
		Under \cref{assump:continuity-growth}, $V_{\text{orig}}(p)\in[0,\infty)$ for all $p$.
		\begin{proof}
			See  \cref{appendix:proof-wellposedness}.
		\end{proof}
\end{lemma}

\subsection{Intensity-relaxed problem} \label{sec:intensity-relaxed-problem}
To facilitate the connection to a standard exploratory RL framework, we first convert the entry and exit decisions into continuous controls. Given $M>0$, define $\bM:=[0,M]$.
Choose two processes $\bm{\alpha}:=(\alpha_t)_{t\ge0}$ and $\bm{\beta}:=(\beta_t)_{t\ge0}$ such that $\alpha_t, \beta_t \in \bM$.
Denote $u_t:=(\alpha_t,\beta_t)$ and $\bm{u}:=(u_t)_{t \geq 0}$, which will be our control processes.

Starting from the signal process, we augment it to the controlled process $X^{\bm{u}} = (X_t^{\bm{u}})_{t\ge0}$ where $X_t^{\bm{u}} = (P_t, J_t^{\bm{u}}, B_t^{\bm{u}}) \in \bR\times\{0,1,2\}\times\bR$. Economically, $J_t^{\bm{u}} \in \{0,1,2\}$ is an indicator representing the current stage of the speculative trading problem: $J_t^{\bm{u}}=0$ indicates that the agent has not initiated the  position yet; $J_t^{\bm{u}}=1$ indicates that the agent is holding the position and has not closed the trade yet; and $J_t^{\bm{u}}=2$ indicates that the agent has already completed the round-trip transaction (and there is no further decision to be made thereafter). $B_t^{\bm{u}}$ records the signal value when the position is initiated (we manually set $B_t^{\bm{u}}=0$ if never opening). The initial state of the problem is $X_0^{\bm{u}}=x:=(p,0,0)$.

$X^{\bm{u}}$ is controlled in the following way. When $J_t^{\bm{u}}=0$, the agent can choose to enter the market by setting an intensity $\alpha_t \in \bM$, and the probability of entering the market in the time interval $[t, t+dt)$ is $\alpha_t dt$. When $J_t^{\bm{u}}=1$, the agent can choose to exit the market by setting an intensity $\beta_t \in \bM$, and the probability of exiting the market in the time interval $[t, t+dt)$ is $\beta_t dt$. When $J_t^{\bm{u}}=2$, the round-trip trade has already been completed where the agent cannot re-enter the market again and there is no further decision to be made.
To formally model the events of entry and exit,
we enlarge the probability space to support two independent exponential random variables $E^a, E^b \sim \text{Exp}(1)$, with $W, E^a, E^b$ mutually independent.
These serve as the canonical randomness sources for entry and exit timing.
Define the unit-rate Poisson seeds as
\begin{equation} \label{eq:poisson-seeds}
  \Pi^a(t) := \mathds{1}_{\{E^a \leq t\}}, \quad \Pi^b(t) := \mathds{1}_{\{E^b \leq t\}}.
\end{equation}
The triple $(W,E^a,E^b)$ constitutes the primitive randomness of the model. The control $\bm{u}$ is an $(\cF_t)$-predictable process that determines the intensity but does not introduce additional randomness. It enters the state dynamics only through the cumulative effective intensities used to time-change the fixed seeds.
Define the effective intensities $\lambda^{\bm{\alpha}}_t$ and $\lambda^{\bm{\beta}}_t$ as
\begin{equation} \label{eq:effective-intensities}
  \lambda^{\bm{\alpha}}_t := \alpha_t \mathds{1}_{\{J_{t-}^{\bm{u}}=0\}}, \quad \lambda^{\bm{\beta}}_t := \beta_t \mathds{1}_{\{J_{t-}^{\bm{u}}=1\}},
\end{equation}
i.e., the entry intensity $\lambda^{\bm{\alpha}}_t$ is effective (strictly positive) only if the agent has not yet opened the position ($J_{t-}^{\bm{u}}=0$), and the exit intensity $\lambda^{\bm{\beta}}_t$ is effective (strictly positive) only if the agent is currently holding an open position ($J_{t-}^{\bm{u}}=1$).
The cumulative effective intensities $\Lambda^{\bm{\alpha}}_t$ and $\Lambda^{\bm{\beta}}_t$ are defined as
\begin{equation} \label{eq:cumulative-effective-intensities}
  \Lambda^{\bm{\alpha}}_t := \int_0^t \lambda^{\bm{\alpha}}_s ds, \quad \Lambda^{\bm{\beta}}_t := \int_0^t \lambda^{\bm{\beta}}_s ds,
\end{equation}
which are non-decreasing and absolutely continuous processes that accumulate the effective intensities up to time $t$, acting as the random clocks for time-change.
The Cox counting processes $N^{\bm{\alpha}}_t$ and $N^{\bm{\beta}}_t$ are constructed by time-change of the unit-rate Poisson seeds:
\begin{equation} \label{eq:cox-counting-processes}
  N^{\bm{\alpha}}_t := \Pi^a(\Lambda^{\bm{\alpha}}_t), \quad N^{\bm{\beta}}_t := \Pi^b(\Lambda^{\bm{\beta}}_t).
\end{equation}
There are two compensated martingales defined as
\begin{equation*}
  M^{\bm{\alpha}}_t := N^{\bm{\alpha}}_t - \int_0^t \lambda^{\bm{\alpha}}_s ds, \quad M^{\bm{\beta}}_t := N^{\bm{\beta}}_t - \int_0^t \lambda^{\bm{\beta}}_s ds.
\end{equation*}

\textbf{State dynamics.} The state dynamics are given by
\begin{equation} \label{eq:state-dynamics}
\begin{aligned}
  dP_t &= \mu (P_t) dt + \sigma (P_t) dW_t, &P_0 &= p, \\
  dJ_t^{\bm{u}} &= \mathds{1}_{\{J_{t-}^{\bm{u}}=0\}} dN^{\bm{\alpha}}_t + \mathds{1}_{\{J_{t-}^{\bm{u}}=1\}} dN^{\bm{\beta}}_t, &J_0^{\bm{u}} &= 0, \\
  dB_t^{\bm{u}} &= P_t \mathds{1}_{\{J_{t-}^{\bm{u}}=0\}} dN^{\bm{\alpha}}_t, &B_0^{\bm{u}} &= 0.
\end{aligned}
\end{equation}
In the above, the indicator $J_t^{\bm{u}}$ jumps from 0 to 1 when the agent opens the position, and jumps from 1 to 2 when the agent closes the position. The record $B_t^{\bm{u}}$ is updated to the current signal $P_t$ at the moment that the position is opened, and remains unchanged otherwise.
\begin{remark}{}{pathwise-construction}
  For any admissible control $\bm{u}$, the state process $X^{\bm{u}}$ is constructed from the same primitive randomness $(W,E^a,E^b)$. Different controls yield different state trajectories by time-changing the same Poisson seeds with different cumulative effective intensities. In particular, for two controls $\bm{u}_1, \bm{u}_2$, the corresponding state process $X^{\bm{u}_1},X^{\bm{u}_2}$ live on the same probability space and can be compared pathwise. This is the standard setup in stochastic control and analogous to different controls sharing the same Brownian motion.
  This common-seed construction is necessary to compare different admissible controls pathwise on the same probability space. It does not restrict the marginal law of any single controlled system. For a fixed admissible control, its law is the same as in the standard Cox construction with an independent exponential seed.
\end{remark}

\textbf{Reward defined with instantaneous utility.} The reward function is given by the instantaneous utility derived when the position is closed. Specifically, when the agent closes the position at time $t$, the realized utility is $U(\gamma P_t - \iota B_t^{\bm{u}} - \Psi - R)$. This reward materializes only when $J_{t-}^{\bm{u}}=1$ and $J_t^{\bm{u}}=2$ (i.e., the moment that $N^{\bm{\beta}}_t$ jumps). Since $J^{\bm{u}}$ jumps from $1$ to $2$ over the time interval $[t,t+\Delta t)$ with probability $\beta_t \mathds{1}_{\{J_{t-}^{\bm{u}}=1\}}\Delta t$, the expected instantaneous reward rate at time $t$ is $\beta_t U(\gamma P_t - \iota B_t^{\bm{u}} - \Psi - R) \mathds{1}_{\{J_{t-}^{\bm{u}}=1\}}$.
Following this idea, define
\begin{equation*}
  G(p,b):=U(\gamma p - \iota b - \Psi - R), \quad p,b \in \bR,
\end{equation*}
and
\begin{equation} \label{eq:instantaneous-reward-function}
  r(x,u) := r((p,j,b),(\alpha,\beta)) := \beta G(p,b) \mathds{1}_{\{j=1\}}, \quad x=(p,j,b) \in \bR \times \{0,1,2\} \times \bR, \quad u=(\alpha,\beta) \in \bM^2.
\end{equation}
The instantaneous reward rate is given by
\begin{equation*}
  r_t^{\bm u} := r(X_{t-}^{\bm{u}},u_t) = r((P_{t-},J_{t-}^{\bm{u}},B_{t-}^{\bm{u}}),(\alpha_t,\beta_t)) = \beta_t G(P_{t-},B_{t-}^{\bm{u}}) \mathds{1}_{\{J_{t-}^{\bm{u}}=1\}}, \quad t \geq 0,
\end{equation*}
where $X_{0-}^{\bm{u}} = X_0^{\bm{u}}$ by convention and thus $r_0^{\bm{u}}=0$.
Since $P$ and $U(\cdot)$ are continuous, and $B_{t-}^{\bm{u}}=B_t^{\bm{u}}$ if $J_{t-}^{\bm{u}}=1$, we have
\begin{equation*}
  r_t^{\bm u} = \beta_t G(P_t,B_t^{\bm{u}}) \mathds{1}_{\{J_{t-}^{\bm{u}}=1\}}.
\end{equation*}
The first form of value function with intensity control is defined as
\begin{align*}
  V_{\text{inte,1}}(p) &:= \sup_{\bm{u} \in \bM^2} \E{ \int_0^{\infty} e^{-\rho t} r_t^{\bm u} dt \given P_0=p}.
\end{align*}

\textbf{Reward defined with two stopping times.} Alternatively, we can define the reward function by two stopping times $\tau^{\bm{\alpha}}$ and $\nu^{\bm{\alpha},\bm{\beta}}$, which are defined as
\begin{equation*}
  \tau^{\bm{\alpha}} := \inf\{t \geq 0: N^{\bm{\alpha}}_t = 1\}, \quad \nu^{\bm{\alpha},\bm{\beta}} := \inf\{t \geq 0: N^{\bm{\beta}}_t = 1\},
\end{equation*}
where $\inf \emptyset := \infty$.
From the definition of $N^{\bm{\alpha}}_t$ and $N^{\bm{\beta}}_t$, we know that $\nu^{\bm{\alpha},\bm{\beta}} \geq \tau^{\bm{\alpha}}$. The second form of value function with intensity control is defined as
\begin{equation*}
  V_{\text{inte,2}}(p) := \sup_{\bm{\alpha}, \bm{\beta} \in \bM} \E{e^{-\rho \nu^{\bm{\alpha},\bm{\beta}}} G(P_{\nu^{\bm{\alpha},\bm{\beta}}}, P_{\tau^{\bm{\alpha}}}) \mathds{1}_{\{\nu^{\bm{\alpha},\bm{\beta}}<\infty\}} \given P_0 = p}.
\end{equation*}
It can be shown that $V_{\text{inte,1}}(p) = V_{\text{inte,2}}(p)$, see \cref{thm:equivalence-two-formulations} below.

\textbf{Admissible control set.} Define the admissible set of intensity controls with initial signal $p$ as
  \begin{equation*}
  \cU(p) := \Bigl\{ \bm u=(\alpha_t,\beta_t)_{t\ge0} : 
  \bm u \text{ is } (\cF_t)\text{-predictable and } \bM^2\text{-valued, and \labelcref{item:U-well-posedness,item:U-integrability-instantaneous-utility,item:U-integrability-two-stopping-times} below hold} \Bigr\},
  \end{equation*}
  where
  \begin{enumerate}[label=(U\arabic*), ref=(U\arabic*), start=1]
    \item \label{item:U-well-posedness}(Well-posedness) 
    The state dynamic \labelcref{eq:state-dynamics} of $X^{\bm{u}}$ admits a unique strong solution;

    \item \label{item:U-integrability-instantaneous-utility}(Integrability with instantaneous utility) 
    $\E{\int_0^\infty e^{-\rho t}\abs{r_t^{\bm u}} dt \given P_0=p}<\infty$.

    \item \label{item:U-integrability-two-stopping-times}(Integrability with stopping times) 
    $\E{e^{-\rho \nu^{\bm{\alpha},\bm{\beta}}} \abs{G(P_{\nu^{\bm{\alpha},\bm{\beta}}}, P_{\tau^{\bm{\alpha}}})} \mathds{1}_{\{\nu^{\bm{\alpha},\bm{\beta}}<\infty\}} \given P_0=p} < \infty$.
  \end{enumerate}

\begin{theorem}{}{equivalence-two-formulations}
  The two formulations of the value function with intensity control are equivalent: 
  \begin{equation*}
    V_{\text{inte,1}}(p) = V_{\text{inte,2}}(p), \quad \forall p \in \bR.
  \end{equation*}

  \begin{proof}
    We need to show that for any admissible control $\bm{u} = (\bm{\alpha},\bm{\beta})$, the expected cumulative reward defined with instantaneous utility is equal to the expected reward defined with two stopping times, i.e.,
   \begin{equation*}
      \E{ \int_0^{\infty} e^{-\rho t} \beta_t G(P_t,B_t^{\bm{u}}) \mathds{1}_{\{J_{t-}^{\bm{u}}=1\}} dt \given P_0=p} = \E{e^{-\rho \nu^{\bm{\alpha},\bm{\beta}}} G(P_{\nu^{\bm{\alpha},\bm{\beta}}}, P_{\tau^{\bm{\alpha}}}) \mathds{1}_{\{\nu^{\bm{\alpha},\bm{\beta}}<\infty\}} \given P_0 = p}.
    \end{equation*}
    Denote $H_t^{\bm{u}} := e^{-\rho t} G(P_t,B_t^{\bm{u}}) \mathds{1}_{\{J_{t-}^{\bm{u}}=1\}}$, which is integrable and predictable ($P_t$ is continuous, $\mathds{1}_{\{J_{t-}^{\bm{u}}=1\}}$ is predictable, $B_t^{\bm{u}}$ is constant if $J_{t-}^{\bm{u}}=1$). Then, we have
    \begin{align*}
      \E{ \int_0^{\infty} e^{-\rho t} \beta_t G(P_t,B_t^{\bm{u}}) \mathds{1}_{\{J_{t-}^{\bm{u}}=1\}} dt \given P_0=p} &= \E{ \int_0^{\infty} H_t^{\bm{u}} \lambda^{\bm{\beta}}_t dt \given P_0=p} = \E{ \int_0^{\infty} H_t^{\bm{u}} dN^{\bm{\beta}}_t \given P_0=p},
    \end{align*}
   where the last equality follows from that $M^{\bm{\beta}}_t=N^{\bm{\beta}}_t-\int_0^t \lambda^{\bm{\beta}}_s ds$ is the compensated martingale with zero expectation, see Proposition 14.2.I in \citet{daleyIntroductionTheoryPoint2008}.
    Since $N^{\bm{\beta}}_t$ jumps at most once at $\nu^{\bm{\alpha},\bm{\beta}}$, and if it jumps, $J_{\nu^{\bm{\alpha},\bm{\beta}}-}^{\bm{u}}=1$ by our formulation, we have
   \begin{equation*}
      \int_0^\infty H_t^{\bm{u}} dN^{\bm{\beta}}_t = H_{\nu^{\bm{\alpha},\bm{\beta}}}^{\bm{u}} \Delta N^{\bm{\beta}}_{\nu^{\bm{\alpha},\bm{\beta}}} \mathds{1}_{\{\nu^{\bm{\alpha},\bm{\beta}}<\infty\}} = e^{-\rho \nu^{\bm{\alpha},\bm{\beta}}} G(P_{\nu^{\bm{\alpha},\bm{\beta}}},B_{\nu^{\bm{\alpha},\bm{\beta}}}^{\bm{u}}) \mathds{1}_{\{\nu^{\bm{\alpha},\bm{\beta}}<\infty\}}
    \end{equation*}
    and $B_{\nu^{\bm{\alpha},\bm{\beta}}}^{\bm{u}}=P_{\tau^{\bm{\alpha}}}$, 
    therefore,
   \begin{align*}
      \E{ \int_0^{\infty} e^{-\rho t} \beta_t G(P_t,B_t^{\bm{u}}) \mathds{1}_{\{J_{t-}^{\bm{u}}=1\}} dt \given P_0=p} 
      = & \E{e^{-\rho \nu^{\bm{\alpha},\bm{\beta}}} G(P_{\nu^{\bm{\alpha},\bm{\beta}}},P_{\tau^{\bm{\alpha}}}) \mathds{1}_{\{\nu^{\bm{\alpha},\bm{\beta}}<\infty\}} \given P_0 = p}.
    \end{align*}
    Taking supremum over all admissible controls $(\bm{\alpha},\bm{\beta})$, we obtain $V_{\text{inte,1}}(p) = V_{\text{inte,2}}(p)$.
  \end{proof}
\end{theorem}

From the equivalence, we denote 
\begin{equation} \label{eq:V-inte}
V_{\text{inte}}(p) := V_{\text{inte,1}}(p) = V_{\text{inte,2}}(p).
\end{equation}
By choosing $\bm{\beta} \equiv 0$, we get $V_{\text{inte}} \geq 0$.

\subsection{Error estimate of the intensity-relaxed problem}
Next, we give an error estimate between the original problem \labelcref{eq:V-orig} and the intensity-relaxed problem \labelcref{eq:V-inte} in terms of the intensity cap $M$. We first present \cref{lem:moment-estimate-sde,lem:bound-signal-change-exp-distributed-time} which respectively give the estimates of the conditional moment of the signal and the mean of the absolute change in the signal.

\begin{lemma}{}{moment-estimate-sde}
  Under \cref{assump:continuity-growth}, for any stopping time $S$ and constant $r \geq 0$, there exist constants $C_r, A_r > 0$ such that
  \begin{equation*}
    \E{ \abs{P_{S+t}}^r \given \cF_S} \leq C_r(1+\abs{P_S}^r)e^{A_r t}, \quad \forall t \geq 0,
  \end{equation*}
  where $C_r, A_r$ only depend on $\mu(\cdot),\sigma(\cdot)$ and $r$, but not on $P_S$ nor $t$.
  
  \begin{proof}
    Since \labelcref{eq:SDE-P} is time-homogeneous with Lipschitz continuous coefficients, from Theorems 2.3.1 and 2.9.5 in \citet{maoStochasticDifferentialEquations2008}, there exists a unique solution $P$ to the SDE \labelcref{eq:SDE-P}, and it is a strong Markov process.
    By Theorem 2.4.1 and Corollary 2.4.5 in \citet{maoStochasticDifferentialEquations2008}, for any deterministic initial value $P_0=p$, $\E{\abs{P_{t}}^r \given \cF_0} \leq C_r(1+\abs{p}^r)e^{A_r t}$ for some constants $C_r, A_r > 0$ that only depend on $\mu(\cdot),\sigma(\cdot)$ and $r$, not on $p$ nor $t$. The result follows using the strong Markov property of $P$.
  \end{proof}
\end{lemma}

\begin{lemma}{}{bound-signal-change-exp-distributed-time}
  Suppose \cref{assump:continuity-growth} holds. Let $S$ and $T$ be any two stopping times such that $ \left. T-S \right| \cF_S \sim \text{Exp}(M)$ where the constant $M$ satisfies $M \geq \max\{1,2A_2,2A_4\}$ ($A_2$ and $A_4$ are constants defined via $A_r$ introduced in \cref{lem:moment-estimate-sde}).
  Then, there exists a constant $C > 0$ such that
  \begin{equation*}
    \E{\abs{P_T - P_S}} \leq C \frac{1+\E{\abs{P_S}}}{\sqrt{M}},
  \end{equation*}
  where $C$ only depends on $\mu(\cdot),\sigma(\cdot)$, but not on $M$.
  
  \begin{proof}
    See \cref{appendix:proof-bound-signal-change-exp-distributed-time}.
  \end{proof}
\end{lemma}

The main result of this section is the following error estimate between the value function of the original sequential stopping problem and the relaxed control problem over the bounded intensities processes.

\begin{theorem}{}{error-estimate-original-continuous-control}
  Let \cref{assump:continuity-growth} holds.
  Suppose for some $\xi>1 $, $\E{\abs{P_{\tau^*}}^\xi}+\E{\abs{P_{\nu^*}}^\xi }< +\infty$, where $(\tau^*, \nu^*)$ are the optimal stopping times of the original problem \eqref{eq:V-orig}. Then, we have
  \begin{equation*}
    0 \leq V_{\text{orig}}(p) - V_{\text{inte}}(p) \leq C M^{-\kappa/2},
  \end{equation*}
  for some constant $C > 0$ and any $M \geq \max\{1,2A_\xi,2A_2,2A_4\}$ (recall the definition of $A_r$ in \cref{lem:moment-estimate-sde}), where $\kappa \in (0,1]$ is the H\"older exponent of the utility function $U$. $C$ depends on $\mu(\cdot), \sigma(\cdot), U(\cdot), \xi, \E{\abs{P_{\tau^*}}^\xi}$ and $\E{\abs{P_{\nu^*}}^\xi}$, but not on $M$. $C$ also implicitly depends on $p$ through $\E{\abs{P_{\tau^*}}^\xi}, \E{\abs{P_{\nu^*}}^\xi}$.
  \begin{proof}
    From \cref{thm:equivalence-two-formulations}, we know that $V_{\text{inte}}(p) = V_{\text{inte,2}}(p)$. Therefore, we only need to estimate the error between $V_{\text{orig}}(p)$ and $V_{\text{inte,2}}(p)$. Since the stopping times in $V_{\text{inte,2}}(p)$ are all in $\cT$, we know $V_{\text{orig}}(p) \geq V_{\text{inte,2}}(p)$. Thus, we only need to find a pair of intensities such that the error is within $CM^{-\kappa/2}$.

    If $\nu^*=\infty$, then $V_{\text{orig}}(p) = 0$. Since $V_{\text{inte}}(p) \geq 0$, we get $V_{\text{inte}}(p) = 0$.
    Therefore, we only need to consider $\tau^*, \nu^* < \infty$.
    Define a pair of intensities as
    \begin{equation*}
      \alpha_t^{M} := M \mathds{1}_{[\tau^*,+\infty)}(t), \quad \beta_t^{M} := M \mathds{1}_{[\nu^*,+\infty)}(t).
    \end{equation*}
    With the intensities defined above, we can replicate the constructions in \labelcref{eq:effective-intensities,eq:cumulative-effective-intensities,eq:cox-counting-processes},
    to define a new version of Cox processes $N^{\bm{\alpha},M}$ and $N^{\bm{\beta},M}$ driven by $(\alpha_t^{M},\beta_t^{M})$.
    Define the stopping times as
    \begin{equation*}
      \tau^{M} := \inf\{t \geq 0: N^{\bm{\alpha},M}_t = 1\}, \quad \nu^{M} := \inf\{t \geq 0: N^{\bm{\beta},M}_t = 1\}.
    \end{equation*}
    It could happen that $\tau^M>\nu^*$. Then, the exit intensity only becomes effective after $\tau^M$.
    $\tau^{M}-\tau^*$ and $\nu^{M} - (\tau^{M}\vee \nu^*)$ are independent and follow $\text{Exp}(M)$ distribution.

    From \cref{lem:bound-signal-change-exp-distributed-time}, taking $T=\tau^M$ and $S=\tau^*$, or $T=\nu^M$ and $S=\tau^M\vee\nu^*$, we have
    \begin{equation} \label{eq:bound-apply-lemma}
      \E{\abs{P_{\tau^{M}} - P_{\tau^*}}} \leq C\frac{1+\E{\abs{P_{\tau^*}}}}{\sqrt{M}}, \quad \E{\abs{P_{\nu^{M}} - P_{\tau^M\vee\nu^*}}} \leq C\frac{1+\E{\abs{P_{\tau^M\vee\nu^*}}}}{\sqrt{M}},
    \end{equation}
    for some constant $C>0$.
    Since $\E{\abs{P_{\tau^M\vee\nu^*}}} \leq \E{\abs{P_{\tau^M}}} + \E{\abs{P_{\nu^*}}}$ and $\E{\abs{P_{\tau^M}}} \leq \E{\abs{P_{\tau^{M}} - P_{\tau^*}}} + \E{\abs{P_{\tau^*}}} \leq C(1+\E{\abs{P_{\tau^*}}})$, we have $\E{\abs{P_{\tau^M\vee\nu^*}}} \leq C(1+\E{\abs{P_{\tau^*}}}+\E{\abs{P_{\nu^*}}})$. Then,
    \begin{equation*}
      \E{\abs{P_{\nu^{M}} - P_{\tau^M\vee\nu^*}}} \leq C\frac{1+\E{\abs{P_{\tau^*}}}+\E{\abs{P_{\nu^*}}}}{\sqrt{M}}.
    \end{equation*}
    On the other hand,
    \begin{align*}
      \E{\abs{P_{\tau^M\vee\nu^*} - P_{\nu^*}}} = & \E{\abs{P_{\tau^M} - P_{\nu^*}} \mathds{1}_{\{\tau^M > \nu^*\}}} \\
      = & \E{\E{\abs{P_{\tau^M} - P_{\nu^*}} \mathds{1}_{\{\tau^M > \nu^*\}} \given \cF_{\nu^*}}} \\
      = & \E{\Prob{\tau^M > \nu^* \given \cF_{\nu^*}} \E{\abs{P_{\tau^M} - P_{\nu^*}} \given \cF_{\nu^*}, \tau^M>\nu^*}} \\
      \leq & \E{\E{\abs{P_{\tau^M} - P_{\nu^*}} \given \cF_{\nu^*}, \tau^M>\nu^*}} \leq C\frac{1+\E{\abs{P_{\nu*}}}}{\sqrt{M}},
    \end{align*}
    where the last inequality follows the memoryless property of exponential distribution and \cref{lem:bound-signal-change-exp-distributed-time}. Combining the above two inequalities, we get
    \begin{equation} \label{eq:nu-M-nu-star}
      \E{\abs{P_{\nu^M}-P_{\nu^*}}} \leq C\frac{1+\E{\abs{P_{\tau^*}}}+\E{\abs{P_{\nu^*}}}}{\sqrt{M}}.
    \end{equation}

    Since $U$ is H\"older continuous, $G(p,b)=U(\gamma p-\iota b - \Psi - R)$ is also H\"older continuous in $p$ and $b$.
    Therefore,
    \begin{align*}
      \E{\abs{G(P_{\nu^{M}}, P_{\tau^{M}}) - G(P_{\nu^*}, P_{\tau^*})}} & \leq C \left(\E{\abs{P_{\nu^{M}} - P_{\nu^*}}^\kappa} + \E{\abs{P_{\tau^{M}} - P_{\tau^*}}^\kappa}\right)\\
      & \leq C \left(\E{\abs{P_{\nu^{M}} - P_{\nu^*}}}\right)^\kappa + C \left(\E{\abs{P_{\tau^{M}} - P_{\tau^*}}}\right)^\kappa \\
      & \leq C \frac{(1+\E{\abs{P_{\tau^*}}}+\E{\abs{P_{\nu^*}}})^\kappa+(1+\E{\abs{P_{\tau^*}}})^\kappa}{(\sqrt{M})^\kappa} \\
      & \leq C M^{-\kappa/2} \left(1+(\E{\abs{P_{\tau^*}}})^\kappa+(\E{\abs{P_{\nu^*}}})^\kappa\right),
    \end{align*}
    for some constant $C>0$, where we used Jensen's inequality, \labelcref{eq:bound-apply-lemma,eq:nu-M-nu-star}.
    It follows that,
    \begin{align*}
      & \quad \abs{\E{e^{-\rho \nu^{M}}G(P_{\nu^{M}}, P_{\tau^{M}})} - \E{e^{-\rho \nu^*}G(P_{\nu^*}, P_{\tau^*})}} \\
      & \leq \E{\abs{e^{-\rho \nu^{M}} - e^{-\rho \nu^*}} \abs{G(P_{\nu^{M}}, P_{\tau^{M}})}} + \E{e^{-\rho \nu^*} \abs{G(P_{\nu^{M}}, P_{\tau^{M}}) - G(P_{\nu^*}, P_{\tau^*})}} \\
      & \leq \rho \E{\abs{ \nu^{M} - \nu^*} \abs{G(P_{\nu^{M}}, P_{\tau^{M}})}} + \E{\abs{G(P_{\nu^{M}}, P_{\tau^{M}}) - G(P_{\nu^*}, P_{\tau^*})}} \\
      & \leq \rho \left(\E{\abs{ \nu^{M} - \nu^*}^{\xi/(\xi-1)}}\right)^{1-1/\xi} \left(\E{\abs{G(P_{\nu^{M}}, P_{\tau^{M}})}^{\xi}}\right)^{1/\xi} + C M^{-\kappa/2} \left(1+(\E{\abs{P_{\tau^*}}})^\kappa+(\E{\abs{P_{\nu^*}}})^\kappa\right),
    \end{align*}
    where we used the fact that $\abs{e^{-x} - e^{-y}} \leq \abs{x - y}$ for $x,y \geq 0$ and H\"older's inequality.

    $\tau^M-\tau^*$ and $\nu^M-(\nu^*\vee\tau^M)$ can be expressed as $E^{(1)}/M, E^{(2)}/M$, where $E^{(1)}$ and $E^{(2)}$ are two independent standard exponential variables. Then, $\nu^M-\nu^* \leq (E^{(1)}+E^{(2)})/M$.
    Since $E^{(1)}+E^{(2)} \sim \text{Gamma}(2,1)$,
    and the moment of Gamma distribution is finite, we have
    \begin{align*}
      \E{\abs{ \nu^{M} - \nu^*}^{\xi/(\xi-1)}} \leq \E{\abs{ \frac{(E^{(1)}+E^{(2)})}{M}}^{\xi/(\xi-1)}} \leq CM^{-\xi/(\xi-1)}.
    \end{align*}
    Therefore,
    \begin{equation*}
      \left(\E{\abs{ \nu^{M} - \nu^*}^{\xi/(\xi-1)}}\right)^{1-1/\xi} \leq C M^{-1}.
    \end{equation*}
    From the linear growth of $U$,
    \begin{equation*}
      \abs{G(P_{\nu^{M}}, P_{\tau^{M}})}^{\xi} = \abs{U(\gamma P_{\nu^{M}} - \iota P_{\tau^{M}} - \Psi - R)}^{\xi} \leq C(1+\abs{P_{\nu^{M}}}^\xi + \abs{P_{\tau^{M}}}^\xi),
    \end{equation*}
    for some constant $C>0$.

    From the construction of the processes and the stopping times, given $\cF_{\tau^*}$, $\tau^M - \tau^*$ is independent of $\left(P_{\tau^*+t}\right)_{t\geq0}$. Therefore,
    \begin{align*}
      \E{\abs{P_{\tau^M}}^\xi} &= \E{\E{\abs{P_{\tau^M}}^\xi \given \cF_{\tau^*}}} \\
      &= \E{\int_0^\infty \E{\abs{P_{\tau^*+t}}^\xi \given \cF_{\tau^*}} M e^{-Mt} dt} \\
      &\leq \E{C_\xi \left( 1+\abs{P_{\tau^*}}^\xi\right) \int_0^\infty M e^{(A_\xi-M)t} dt} \\
      &= C_\xi \left(1+\E{\abs{P_{\tau^*}}^\xi}\right) \frac{M}{M-A_\xi} \leq C\left(1+\E{\abs{P_{\tau^*}}^\xi}\right),
    \end{align*}
    where $C=2C_\xi$, and we used \cref{lem:moment-estimate-sde} and $M\geq 2A_\xi$. Similarly, $\E{\abs{P_{\nu^{M}}}^\xi} \leq C\left(1+\E{\abs{P_{\nu^*\vee\tau^M}}^\xi}\right)$.

    Since $\E{\abs{P_{\nu^*\vee\tau^M}}^\xi} \leq \E{\abs{P_{\nu^*}}^\xi} + \E{\abs{P_{\tau^M}}^\xi} \leq C\left(1+\E{\abs{P_{\tau^*}}^\xi}+\E{\abs{P_{\nu^*}}^\xi}\right)$, we get
    \begin{equation*}
      \E{\abs{P_{\nu^{M}}}^\xi} \leq C\left(1+\E{\abs{P_{\tau^*}}^\xi}+\E{\abs{P_{\nu^*}}^\xi}\right).
    \end{equation*}
    Therefore,
    \begin{equation*}
      \left(\E{\abs{G(P_{\nu^{M}}, P_{\tau^{M}})}^{\xi}}\right)^{1/\xi} \leq C \left(1+\E{\abs{P_{\nu^{M}}}^\xi} + \E{\abs{P_{\tau^{M}}}^\xi}\right)^{1/\xi} \leq C \left(1+\E{\abs{P_{\nu^*}}^\xi} + \E{\abs{P_{\tau^*}}^\xi}\right)^{1/\xi},
    \end{equation*}
    for some constant $C>0$.
    Combining the above estimates, we obtain
    \begin{align*}
      & \quad \abs{\E{e^{-\rho \nu^{M}}G(P_{\nu^{M}}, P_{\tau^{M}})} - \E{e^{-\rho \nu^*}G(P_{\nu^*}, P_{\tau^*})}} \\
      & \leq C \left(M^{-1} \left(1+\E{\abs{P_{\nu^*}}^\xi} + \E{\abs{P_{\tau^*}}^\xi}\right)^{1/\xi} + M^{-\kappa/2} \left(1+(\E{\abs{P_{\tau^*}}})^\kappa+(\E{\abs{P_{\nu^*}}})^\kappa\right)\right) \\
      & \leq C M^{-\kappa/2},
    \end{align*}
    for some constant $C>0$.
    Note that $\E{e^{-\rho \nu^*}G(P_{\nu^*}, P_{\tau^*})} \geq \E{e^{-\rho \nu^{M}}G(P_{\nu^{M}}, P_{\tau^{M}})}$.
    Thus, we have proved there exists a pair of intensities $(\alpha^M, \beta^M)$ such that the error is within $CM^{-\kappa/2}$.
  \end{proof}
\end{theorem}

\begin{remark}{}{if-sublinear-growth}
  If $U$ has sublinear growth, i.e., there exist constants $C>0$ and $q \in (0,1)$ such that for any $x \in \bR$,
  \begin{equation*}
    \abs{U(x)} \leq C(1+\abs{x}^q),
  \end{equation*}
  then, \cref{thm:error-estimate-original-continuous-control} still holds when $\xi=1$.
\end{remark}

\begin{remark}{}{no-common-seeds-used}
  In the proof of \cref{thm:error-estimate-original-continuous-control}, we only use the property that the auxiliary entry and exit seeds are exogenous and independent of Brownian motion. The global common-seed setup across different admissible controls is not needed for this theorem.
\end{remark}

When the intensity cap $M$ tends to infinity, the value function of the intensity-control problem converges to that of the original sequential stopping problem. This observation is not surprising. By standard theories of optimal stopping of Markovian processes, the state space can be partitioned into a continuation set and a stopping set, and the optimal strategy is to stop when the process first enters the stopping set. A non-degenerated randomized stopping rule generally results in worse value because any linear combination of the stopped and continuation values is weakly dominated by the maximum of these two values. Under the intensity formulation, the agent would therefore want to exert infinite stopping rate whenever the current state is in the stopping set in order to stop with certainty. The introduction of an upper cap $M$ precludes infinite stopping rates or equivalently stopping with certainty. This restriction lowers the value function, but its performance gap against the original value function will vanish as the bound $M$ is relaxed to infinity. In the context of RL, one could view a finite value of $M$ as a mechanism to enforce exploration such that the agent cannot simply use a greedy strategy of stopping immediately whenever they are in the stopping set.

\section{Entropy-regularized problem with randomized intensities} \label{sec:entropy-regularization}

In this section, we apply the exploratory framework in \citet{wangReinforcementLearningContinuous2020} to formulate the intensity-relaxed speculative trading problem as an RL task. We let the agent choose intensity according to a probability measure, as to reflect the essence of RL that the agent will randomly explore the unknown environment. The cost of exploration is measured by Shannon's differential entropy.
At time $t\geq 0$, for the two intensities $\alpha_t$ and $\beta_t$, we introduce the randomized intensities with probability measures $(\pi_t^{\bm{\alpha}}, \pi_t^{\bm{\beta}})$ defined on the domain $\bM = [0,M]$. Denote $\pi_t :=(\pi_t^{\bm{\alpha}}, \pi_t^{\bm{\beta}})$,
$\bm{\pi} := (\pi_t)_{t \geq 0}$,
$\cP(\bM)$ as the set of all Borel probability measures defined on $\bM$, and $\cP_{\text{ac}}(\bM):=\{\mu\in\cP(\bM):\mu\ll d\lambda\}$ as the set of absolutely continuous probability measures on $\bM$ with respect to the Lebesgue measure $\lambda$.
We require $\pi_t \in \cP_{\text{ac}}(\bM) \times \cP_{\text{ac}}(\bM)$ for all $t \geq 0$. We would call $\bm{\pi}$ as the randomized intensity process.
Under $\bm{\pi}$, denote the exploratory state process as $X^{\bm{\pi}}=(X_t^{\bm{\pi}})_{t \geq 0}$ with $X_t^{\bm{\pi}} := (P_t, J_t^{\bm{\pi}}, B_t^{\bm{\pi}})$, which is defined as follows.

\subsection{Exploratory state dynamics}

Consider the intensities as the expected values of the randomized controls, i.e., at time $t$, define
\begin{equation*}
  \bar{\alpha}_t := \int_\bM \lambda \pi_t^{\bm{\alpha}}(\lambda) d\lambda, \quad \bar{\beta}_t := \int_\bM \lambda \pi_t^{\bm{\beta}}(\lambda) d\lambda,
\end{equation*}
and the corresponding effective intensities
\begin{equation*}
  \bar{\lambda}_t^{\bm{\alpha}} := \bar{\alpha}_t \mathds{1}_{\{J_{t-}^{\bm{\pi}} = 0\}}, \quad \bar{\lambda}_t^{\bm{\beta}} := \bar{\beta}_t \mathds{1}_{\{J_{t-}^{\bm{\pi}} = 1\}}.
\end{equation*}
The effective intensities $\bar{\lambda}_t^{\bm{\alpha}}$ and $\bar{\lambda}_t^{\bm{\beta}}$ capture the contribution of the randomized controls to the jump processes.
The cumulative effective intensities are
\begin{equation} \label{eq:cumulative-hazard-randomized}
  \bar{\Lambda}_t^{\bm{\alpha}} := \int_0^t \bar{\lambda}_s^{\bm{\alpha}} ds, \quad \bar{\Lambda}_t^{\bm{\beta}} := \int_0^t \bar{\lambda}_s^{\bm{\beta}} ds,
\end{equation}
and the corresponding jump processes are
\begin{equation*}
  N_t^{\bm{\alpha},\bm{\pi}} := \Pi^a(\bar{\Lambda}_t^{\bm{\alpha}}), \quad N_t^{\bm{\beta},\bm{\pi}} := \Pi^b(\bar{\Lambda}_t^{\bm{\beta}}).
\end{equation*}
With the above formulations, the exploratory state process $X^{\bm{\pi}} = (X_t^{\bm{\pi}})_{t \geq 0} = (P_t, J_t^{\bm{\pi}}, B_t^{\bm{\pi}})_{t \geq 0}$ has the following dynamics:
\begin{equation} \label{eq:state-dynamics-entropy-regularized}
  \begin{aligned}
    dP_t &= \mu(P_t) dt + \sigma(P_t) dW_t, &P_0&=p, \\
    dJ_t^{\bm{\pi}} &= \mathds{1}_{\{J_{t-}^{\bm{\pi}} = 0\}} dN_t^{\bm{\alpha},\bm{\pi}} + \mathds{1}_{\{J_{t-}^{\bm{\pi}} = 1\}} dN_t^{\bm{\beta},\bm{\pi}}, &J_0^{\bm{\pi}}&=0, \\
    dB_t^{\bm{\pi}} &= P_t \mathds{1}_{\{J_{t-}^{\bm{\pi}} = 0\}} dN_t^{\bm{\alpha},\bm{\pi}}, &B_0^{\bm{\pi}}&=0.
  \end{aligned}
\end{equation}

\begin{remark}{}{same-poisson-seeds-in-exploratory}
  The same unit-rate Poisson seeds $\Pi^a, \Pi^b$ from \labelcref{eq:poisson-seeds} are used for all randomized controls. This is consistent with the construction in \cref{sec:intensity-relaxed-problem}. The seeds are part of the probability space, and the randomized control $\bm{\pi}$ affects the state dynamics only through the average intensities $\bar{\alpha}_t, \bar{\beta}_t$ that determine the time-change.
\end{remark}

\subsection{Exploratory value function}
We define previously in \labelcref{eq:instantaneous-reward-function}  the reward function as $r(x,u) = r((p,j,b),(\alpha,\beta)) = \mathds{1}_{\{j=1\}} G(p,b) \beta$, for $x=(p,j,b) \in \bR \times \{0,1,2\} \times \bR$ and $u=(\alpha,\beta) \in \bM^2$.
Under a pair of probability measures $\pi=(\pi_1,\pi_2) \in \cP_{\text{ac}}(\bM) \times \cP_{\text{ac}}(\bM)$, the expected reward is
\begin{equation*}
  \iint_{\bM^2} r((p,j,b),(\lambda_1,\lambda_2)) \pi_1(\lambda_1) \pi_2(\lambda_2) d\lambda_1 d\lambda_2 = \mathds{1}_{\{j=1\}} G(p,b) \int_\bM \lambda \pi_2(\lambda) d\lambda.
\end{equation*}
Therefore, we define
\begin{align*}
  &\tilde r(x,\pi) = \tilde r((p,j,b),(\pi_1,\pi_2)) := \mathds{1}_{\{j=1\}} G(p,b) \int_\bM \lambda \pi_2(\lambda) d\lambda, \\
  &x=(p,j,b) \in \bR \times \{0,1,2\} \times \bR, \quad \pi=(\pi_1,\pi_2) \in \cP_{\text{ac}}(\bM) \times \cP_{\text{ac}}(\bM).
\end{align*}
The instantaneous reward is now given by
\begin{align*}
  \tilde r_t^{\bm{\pi}} &:= \tilde r(X_{t-}^{\bm{\pi}},\pi_t) = \tilde r((P_{t-}, J_{t-}^{\bm{\pi}}, B_{t-}^\pi), (\pi_t^{\bm{\alpha}},\pi_t^{\bm{\beta}})) = \mathds{1}_{\{J_{t-}^{\bm{\pi}}=1\}} G(P_{t-}, B_{t-}^\pi) \int_\bM \lambda \pi_t^{\bm{\beta}}(\lambda) d\lambda.
\end{align*}
Since $P$ is continuous, and $B_{t-}^\pi=B_t^{\bm{\pi}}$ if $J_{t-}^{\bm{\pi}}=1$, we have
\begin{align*}
  &\tilde r_t^{\bm{\pi}} = \mathds{1}_{\{J_{t-}^{\bm{\pi}}=1\}} G(P_t, B^{\bm{\pi}}_t) \int_\bM \lambda \pi_t^{\bm{\beta}}(\lambda) d\lambda.
\end{align*}
Define the entropy of a pair of probability measures $\pi=(\pi_1,\pi_2) \in \cP_{\text{ac}}(\bM) \times \cP_{\text{ac}}(\bM)$ as
\begin{equation*}
  \cH(\pi) = \cH(\pi_1,\pi_2) := -\iint_{\bM^2} \pi_1(\lambda_1) \pi_2(\lambda_2) \ln\left(M^2\pi_1(\lambda_1) \pi_2(\lambda_2)\right) d\lambda_1 d\lambda_2 \leq 0.
\end{equation*}
With simple computations, the entropy can be decomposed as
\begin{equation*}
  \cH(\pi) = -\int_\bM \pi_1(\lambda_1) \ln(M\pi_1(\lambda_1)) d\lambda_1 - \int_\bM \pi_2(\lambda_2) \ln(M\pi_2(\lambda_2)) d\lambda_2.
\end{equation*}
The entropy-regularized value function is defined as
\begin{equation*}
  V_{\text{ent}}^\eta(p) := \sup_{\bm{\pi} \in \cA(p)} \E{ \int_0^{\infty} e^{-\rho t} \left(\tilde r_t^{\bm{\pi}} + \eta \cH(\pi_t)\right) dt \given P_0=p},
\end{equation*}
with the admissible control set defined as
\begin{equation} \label{eq:admissible-control-set-randomized}
\cA(p):=\left\{\bm{\pi}=(\pi^{\bm{\alpha}}_t,\pi^{\bm{\beta}}_t)_{t\ge0}: \pi_t^{\bm{\alpha}},\pi_t^{\bm{\beta}}\in \cP_{\text{ac}}(\bM), \text{and \labelcref{item:A-predictability,item:A-well-posedness,item:A-reward-integrability,item:A-entropy-finiteness-integrability} below hold}\right\},
\end{equation}
where

\begin{enumerate}[label=(A\arabic*), ref=(A\arabic*), start=0]
\item \label{item:A-predictability}
(Predictability) For each bounded Borel measurable function $\varphi:\bM\to\bR$, the processes $t\mapsto \int_{\bM}\varphi(\lambda)\,\pi_t^{\bm{\alpha}}(d\lambda)$ and $t\mapsto \int_{\bM}\varphi(\lambda)\,\pi_t^{\bm{\beta}}(d\lambda)$ are $(\cF_t)$-predictable.

\item \label{item:A-well-posedness}
(Well-posedness) The state dynamic \eqref{eq:state-dynamics-entropy-regularized} of $X^{\bm{\pi}}$ admits a unique strong solution.

\item \label{item:A-reward-integrability}
(Reward integrability) 
$\E{\int_0^\infty e^{-\rho t}\abs{\tilde r_t^{\bm{\pi}}} dt \given P_0=p}<\infty$.

\item \label{item:A-entropy-finiteness-integrability}
(Entropy finiteness and integrability) $\cH(\pi_t)\in\bR$ a.e. $(t,\omega)$, 
$\E{\int_0^\infty e^{-\rho t} \abs{\cH(\pi_t)} dt \given P_0=p}<\infty$.
\end{enumerate}

\begin{remark}{}{nonpositive-entropy}
Different from \citet{wangReinforcementLearningContinuous2020}, we evaluate the entropy relative to the uniform distribution on $\bM^2$ instead of the unnormalized product Lebesgue measure. Since these two definitions differ only by an additive constant $-2\ln M$, they lead to the same optimizer. The advantage of our normalization is that it yields $\cH(\pi)\leq 0$ with equality attained only when $\pi_1$ and $\pi_2$ are both uniform distributions on $\bM$, according to Gibbs' inequality in continuous case.
Consequently, this allows us to show that the entropy-regularized value function converges monotonically upward as $\eta \downarrow 0$, which is convenient for tuning the exploration parameter in practice.
\end{remark}

\subsection{System of HJB equations for the entropy-regularized value function} \label{sec:HJB-general}
The HJB equation approach requires us to identify the value function not only at the initial time but also at all possible states. Therefore, we augment the initial state $p$ to
\begin{equation*}
x=(p,j,b) \in \bR \times \{0,1,2\} \times \bR.
\end{equation*}
On the augmented state space, we use $\cV_{\text{ent}}^\eta(x)$ instead of $V_{\text{ent}}^\eta(p)$ to denote the value function, with the following definition:
\begin{align*}
  \cV_{\text{ent}}^\eta(x) := \sup_{\bm{\pi} \in \cA(x)} \E{ \int_0^{\infty} e^{-\rho t} \left(\tilde r_t^{\bm{\pi}} + \eta \cH(\pi_t)\right) dt \given X_0^{\bm{\pi}} = x}, \quad x=(p,j,b) \in \bR \times \{0,1,2\} \times \bR.
\end{align*}
The admissible control set $\cA(x)$ has the same conditions as $\cA(p)$ but with the initial state $p$ replaced by $x$.
For convenience, we use the short-hands
\begin{equation*}
\cV_0(p):=\cV_{\text{ent}}^\eta(p,0,0),\qquad \cV_1(p,b):=\cV_{\text{ent}}^\eta(p,1,b),\qquad \cV_2(p,b):=\cV_{\text{ent}}^\eta(p,2,b) = 0.
\end{equation*}
Note that $\cV_2\equiv 0$ since the state of $j=2$ corresponds to the case that the agent has already completed the round-trip transaction and the problem has terminated.

From the dynamic programming principle and generalized It\^o's lemma, $\cV_{\text{ent}}^\eta(x)$ satisfies the HJB equation
\begin{align*}
  0 &= \sup_{\pi^{\bm{\alpha}},\pi^{\bm{\beta}} \in \cA(x)} \left\{\eta \cH(\pi) - \rho \cV_{\text{ent}}^\eta(x) + \cL_P \cV_{\text{ent}}^\eta(x) \right. \\
  &\quad \left. + \mathds{1}_{\{j=0\}} \left(\int_\bM \lambda \pi^{\bm{\alpha}}(\lambda) d\lambda\right) \left(\cV_1(p,p) - \cV_0(p)\right) + \mathds{1}_{\{j=1\}} \left(\int_\bM \lambda \pi^{\bm{\beta}}(\lambda) d\lambda\right) \left(G(p,b) - \cV_1(p,b)\right) \right\},
\end{align*}
for all $x=(p,j,b) \in \bR \times \{0,1,2\} \times \bR$, where $\cL_P$ is the infinitesimal generator of the process $P$,
\begin{equation*}
  \cL_P v := \mu(p) \frac{\partial v}{\partial p} + \frac{1}{2} \sigma^2(p) \frac{\partial^2 v}{\partial p^2}.
\end{equation*}
When $j=2$, the equation holds trivially since $\cV_2 \equiv 0$.
When $j=1$,
\begin{align} \label{eq:hjb_1}
  \rho \cV_1(p,b) - \cL_P \cV_1(p,b) &= \sup_{\pi^{\bm{\beta}} \in \cA(x)} \left\{-\eta \int_\bM \pi^{\bm{\beta}}(\lambda) \ln(M\pi^{\bm{\beta}}(\lambda)) d\lambda + \left(\int_\bM \lambda \pi^{\bm{\beta}}(\lambda) d\lambda\right)(G(p,b)-\cV_1(p,b))\right\} \nonumber\\
  &= \eta \ln\left(\frac{e^y-1}{y}\right), \quad y := \frac{G(p,b)-\cV_1(p,b)}{\eta} M,
\end{align}
with the optimal density being a Gibbs distribution, given by
\begin{equation*}
  \pi^{\bm{\beta},*}(\lambda;p,b) = \frac{e^{\frac{G(p,b)-\cV_1(p,b)}{\eta} \lambda}}{\int_0^M e^{\frac{G(p,b)-\cV_1(p,b)}{\eta} \lambda} d\lambda} = \frac{\frac{G(p,b)-\cV_1(p,b)}{\eta} e^{\frac{G(p,b)-\cV_1(p,b)}{\eta} \lambda}}{e^{\frac{G(p,b)-\cV_1(p,b)}{\eta} M}-1}, \quad \lambda \in \bM,
\end{equation*}
and the optimal average intensity is
\begin{equation*}
  \int_0^M \lambda \pi^{\bm{\beta},*}(\lambda;p,b) d\lambda = \frac{M}{1-e^{-y}} - \frac{\eta}{G(p,b)-\cV_1(p,b)}.
\end{equation*}
When $j=0$,
\begin{align} \label{eq:hjb_2}
  \rho \cV_0(p) - \cL_P \cV_0(p) &= \sup_{\pi^{\bm{\alpha}} \in \cA(x)} \left\{-\eta \int_\bM \pi^{\bm{\alpha}}(\lambda) \ln(M\pi^{\bm{\alpha}}(\lambda)) d\lambda + \left(\int_\bM \lambda \pi^{\bm{\alpha}}(\lambda) d\lambda\right) (\cV_1(p,p)-\cV_0(p)) \right\} \nonumber\\
  &= \eta \ln\left(\frac{e^z-1}{z}\right), \quad z := \frac{\cV_1(p,p)-\cV_0(p)}{\eta} M.
\end{align}
The optimal density $\pi^{\bm{\alpha},*}(\lambda;p)$ is similar to $\pi^{\bm{\beta},*}(\lambda;p,b)$ but with $G(p,b)-\cV_1(p,b)$ replaced by $\cV_1(p,p)-\cV_0(p)$.

\subsection{Error estimate of the entropy-regularized problem}
  It is important to know whether the RL objective would yield a solution that is close to that of the original sequential stopping problem.
  We first analyze the error of the entropy-regularized problem compared to the intensity-relaxed one in terms of $\eta$ and $M$. 
  Then, together with \cref{thm:error-estimate-original-continuous-control}, we show that as $\eta \downarrow 0$ and $M\to\infty$ along a specific path, it converges to the original value function.

  We first show the intensity-relaxed value function is equal to the value function with randomized intensity but without entropy regularization.
  Under the augmented state space, for the intensity-relaxed problem,
  define the performance functional and value function as
  \begin{equation*}
  J(x,\bm u):=\E{\int_0^\infty e^{-\rho t}\,r_t^{\bm u} dt \given X_0=x},\qquad
  \cV_{\text{inte}}(x):=\sup_{\bm u\in\cU(x)} J(x,\bm u),
  \end{equation*}
  where $\cU(x)$ has the same conditions as $\cU(p)$ but with the initial state $p$ replaced by $x$.
  Next, consider the problem with randomized intensities, but without entropy regularization. Define
  \begin{align*}
  \widetilde J(x,\bm{\pi}):=\E{\int_0^\infty e^{-\rho t} \tilde r_t^{\bm{\pi}} dt \given X_0=x},\qquad
  \widetilde{\cV}(x):=\sup_{\bm{\pi}\in\widetilde{\cA}(x)} \widetilde J(x,\bm{\pi}),
  \end{align*}
  where $\widetilde{\cA}(x)$ consists of measures in $\cP(\bM)$ such that \labelcref{item:A-predictability,item:A-well-posedness,item:A-reward-integrability} in \labelcref{eq:admissible-control-set-randomized} hold. Note that without the requirement \labelcref{item:A-entropy-finiteness-integrability},  singular measures are contained in the set $\widetilde{\cA}(x)$.

  \begin{lemma}{}{equality-value-functions}
      $\cV_{\text{inte}}(p,j,b) = \widetilde{\cV}(p,j,b)$ for all $(p,j,b) \in \bR \times \{0,1,2\} \times \bR$.
      \begin{proof}
        For any $\cV_{\text{inte}}$-admissible control $\bm{u} = (\alpha_t,\beta_t)_{t \geq 0}$, let $\bm{\pi}^{\bm{u}} = (\delta_{\alpha_t}, \delta_{\beta_t})_{t \geq 0}$, where $\delta_x$ is the Dirac delta function at $x$.
    Then, $\bm{\pi}^{\bm{u}}$ is an admissible randomized control for $\widetilde{\cV}$, and the corresponding exploratory state process $X^{\bm{\pi}^{\bm{u}}}$ has the same distribution as $X^{\bm{u}}$.
    Therefore, $J(\bm{u}) = \widetilde J(\bm{\pi}^{\bm{u}})$, and we have
    \begin{equation*}
      \cV_{\text{inte}} \leq \widetilde{\cV}.
    \end{equation*}
    Conversely, for any $\widetilde{\cV}$-admissible control $\bm{\pi} = (\pi_t^{\bm{\alpha}}, \pi_t^{\bm{\beta}})_{t \geq 0}$, let $\bm{u}^{\bm{\pi}} = (\alpha_t^{\bm{\pi}}, \beta_t^{\bm{\pi}})_{t \geq 0}$, where $\alpha_t^{\bm{\pi}} := \int_\bM \lambda \pi_t^{\bm{\alpha}}(d\lambda)$, $\beta_t^{\bm{\pi}} := \int_\bM \lambda \pi_t^{\bm{\beta}}(d\lambda)$.
    Then, $\bm{u}^{\bm{\pi}}$ is an admissible control for $\cV_{\text{inte}}$, and the corresponding state process $X^{\bm{u}^{\bm{\pi}}}$ has the same distribution as $X^{\bm{\pi}}$.
    Therefore, we have
    \begin{equation*}
      \cV_{\text{inte}} \geq \widetilde{\cV}.
    \end{equation*}
    Combining the above two inequalities, we obtain $\cV_{\text{inte}} = \widetilde{\cV}$.
      \end{proof}
  \end{lemma}
  For the entropy-regularized problem, define, with $\eta>0$,
  \begin{align*}
  J^\eta(x,\bm{\pi}):=\E{\int_0^\infty e^{-\rho t} \left(\tilde r_t^{\bm{\pi}}+\eta \cH(\pi_t)\right) dt \given X_0=x},
  \end{align*}
  so $\cV_{\text{ent}}^\eta(x)=\sup_{\bm{\pi}\in\cA(x)} J^\eta(x,\bm{\pi})$.
  By construction, $\cA(x)\subsetneq \widetilde{\cA}(x)$.
  The inclusion is strict since $\widetilde{\cA}(x)$ allows singular measures:
  for example the Dirac control $\pi^{\bm{\alpha}}_t=\delta_a$, $\pi^{\bm{\beta}}_t=\delta_b$ ($a,b\in \bM$)
  lies in $\widetilde{\cA}(x)$ but not in $\cA(x)$ since it is not absolutely continuous, and $\cH(\delta_a,\delta_b)=-\infty$.
  We will show the error estimate between $\widetilde{\cV}$ and $\cV_{\text{ent}}^\eta$ because they have similar forms. The technical difficulty mainly comes from the difference between the admissible sets.
  \begin{assumption}{}{assumption-convergence-entropy-regularized-value-function}
    Assume that
    \begin{align} \label{eq:assumption-convergence-entropy-regularized-value-function}
      \exists \zeta>0, r>2: K_r:=\int_0^\infty e^{-(\rho-\zeta)t} \E{\abs{P_t}^r \given P_0=p} dt < \infty.
    \end{align}
  \end{assumption}

  \begin{remark}{}{remark1-assumption-2}
    Since $K_r$ is increasing as $\zeta$ increases, we can always choose $\zeta$ such that $0<\zeta<\rho$. For $0\leq p<q$, we can prove $K_p \leq \int_0^\infty e^{-(\rho-\zeta)t} dt + \int_0^\infty e^{-(\rho-\zeta)t} \E{\abs{P_t}^q \given P_0=p} dt$. Combining the two properties, if $K_q<\infty$, then $K_p\leq C+K_q<\infty$. It means for all $r \in [0,2]$, $K_r<\infty$ is implied by \labelcref{eq:assumption-convergence-entropy-regularized-value-function}.
  \end{remark}

    \begin{remark}{}{sufficient-condition-for-integrability}
    Consider the set
    \begin{align*}
    \Omega := \{A>0 : & \exists C>0 \text{ independent of $p$ and $t$ s.t. } \\
    & \E{\abs{P_t}^r \given P_0=p} \leq C (1+p^2) e^{At} \text{ for some $r>2$ and any $t\geq0$}\},
    \end{align*}
    which is non-empty by \cref{lem:moment-estimate-sde}. Then, a sufficient condition for \cref{assump:assumption-convergence-entropy-regularized-value-function} to hold is $\rho > \inf\Omega$.
  \end{remark}

  \begin{theorem}{}{convergence-entropy-regularized-value-function}
    Let $M \geq 1$, $0<\eta<1$ and $\frac{2r}{(r-2)\kappa} \geq -\frac{\ln M}{\ln\eta}$, where $r>2$ is a constant associated with Condition \labelcref{eq:assumption-convergence-entropy-regularized-value-function} and $\kappa\in(0,1]$ is the H\"older exponent of $U(\cdot)$ as per \labelcref{eq:utility-holder}.
    Under \cref{assump:continuity-growth,assump:assumption-convergence-entropy-regularized-value-function},
    it holds that
    \begin{align*}
      0 \leq \cV_{\text{inte}}(p,0,0) - \cV_{\text{ent}}^\eta(p,0,0) \leq C M^2 \abs{\eta\ln\eta}, \quad \forall p \in \bR,
    \end{align*}
    for some constant $C>0$ independent of $\eta$ and $M$ but depends on $p$. In particular, the following monotone convergence holds:
    \begin{align*}
    \cV_{\text{ent}}^\eta(p,0,0) \uparrow \cV_{\text{inte}}(p,0,0) \text{ as }\eta\downarrow0.
    \end{align*}
  \begin{proof}
    
    From \cref{lem:equality-value-functions}, we have $\cV_{\text{inte}} = \widetilde{\cV}$.
    Thus, it suffices to prove the convergence of $\cV_{\text{ent}}^\eta$ to $\widetilde{\cV}$ as $\eta \downarrow 0$.
    The rough idea of the proof is to start with an $\varepsilon$-optimal control for $\widetilde{\cV}$ and modify it to be $\cV_{\text{ent}}^\eta$-admissible. The modification precludes singular measures but also introduces additional errors. The estimation of the error is the main technical challenge and summarized in \cref{lem:approximation-errors-modifying-controls}.

    In the following, we do not explicitly write the initial state $x=(p,0,0)$ in conditional expectations or $\widetilde{J}$ for simplicity.
    
    For any $\cV_{\text{ent}}^\eta$-admissible randomized control $\bm{\pi} = (\pi_t)_{t \geq 0}$,
    \begin{align*}
      \int_0^\infty e^{-\rho t} \left(\tilde r_t^{\bm{\pi}} + \eta \cH(\pi_t)\right) dt \leq \int_0^\infty e^{-\rho t} \tilde r_t^{\bm{\pi}} dt,
    \end{align*}
    since $\cH(\pi_t) \leq 0$ for all $t \geq 0$.
    Taking expectation and supremum over $\cA(x)$, we obtain
    \begin{equation*}
      \cV_{\text{ent}}^\eta \leq \widetilde{\cV}.
    \end{equation*}
    Moreover, when $\eta_1 \geq \eta_2$, we have $\cV_{\text{ent}}^{\eta_1} \leq \cV_{\text{ent}}^{\eta_2}$.
    Therefore, $\cV_{\text{ent}}^\eta$ is increasing as $\eta \downarrow 0$ and is bounded above by $\widetilde{\cV} = \cV_{\text{inte}}$.

    For any $\varepsilon > 0$, there exists a $\widetilde{\cV}$-admissible randomized control $\widetilde{\bm{\pi}}^\varepsilon$ such that
    \begin{equation*}
      \widetilde{\cV} - \varepsilon \leq \widetilde J(\widetilde{\bm{\pi}}^\varepsilon) \leq \widetilde{\cV}.
    \end{equation*}
    As we mentioned before, the admissible control set for $\widetilde{\cV}$ is broader than that of $\cV_{\text{ent}}^\eta$ because the former includes atomic measures. To ensure $\widetilde{\bm{\pi}}^\varepsilon$ is also admissible for $\cV_{\text{ent}}^\eta$, we can modify it according to the following procedure.
    Consider a class of probability density functions on $\bM$ denoted as $\Phi_{m,\delta}$ for some $m \in [0,M]$ and $\delta \in (0,1/2)$. $m$ is the approximate mean of the density function, and $\delta$ controls the spread of the density function.
    Define
    \begin{equation*}
      \Phi_{m,\delta}(\lambda) := \begin{cases}
        \frac{M-m}{\delta M^2}, & \lambda \in [0, \delta M], \\
        \frac{m}{\delta M^2}, & \lambda \in [M-\delta M, M], \\
        0, & \text{otherwise}.
      \end{cases}
    \end{equation*}
    Note that $\Phi_{m,\delta}$ is a probability density function on $\bM$ with mean
    $m+\frac{\delta(M-2m)}{2}$, 
    which gives
    \begin{equation} \label{eq:mean-error-phi}
      \abs{\int_0^M \lambda \Phi_{m,\delta}(\lambda) d\lambda - m} \leq \frac{\delta M}{2}.
    \end{equation}
    The entropy of $\Phi_{m,\delta}$ is
    \begin{equation*}
      -\int_0^M \Phi_{m,\delta}(\lambda) \ln(M\Phi_{m,\delta}(\lambda)) d\lambda = \ln \delta - \frac{M-m}{M} \ln(\frac{M-m}{M}) - \frac{m}{M} \ln(\frac{m}{M}).
    \end{equation*}
    Since $-w_1 \ln w_1 - w_2 \ln w_2 \in [0,\ln 2]$ for any $w_1,w_2 \geq 0$ with $w_1+w_2=1$, we have
    \begin{equation} \label{eq:entropy-bound-phi}
      -\int_0^M \Phi_{m,\delta}(\lambda) \ln(M\Phi_{m,\delta}(\lambda)) d\lambda \in [\ln \delta, \ln(2\delta)].
    \end{equation}
    For $\widetilde{\bm{\pi}}^\varepsilon = (\widetilde{\pi}^{\varepsilon, \bm{\alpha}}_t, \widetilde{\pi}^{\varepsilon, \bm{\beta}}_t)_{t \geq 0}$, denote their means as $\widetilde{m}_t^{\varepsilon, \bm{\alpha}}$ and $\widetilde{m}_t^{\varepsilon, \bm{\beta}}$ respectively.
    Then, define the modified probability density functions
    \begin{equation*}
      \pi_t^{\varepsilon,\eta,\bm{\alpha}} := \Phi_{\widetilde{m}_t^{\varepsilon, \bm{\alpha}}, \delta(\eta)}, \quad \pi_t^{\varepsilon,\eta,\bm{\beta}} := \Phi_{\widetilde{m}_t^{\varepsilon, \bm{\beta}}, \delta(\eta)},
    \end{equation*}
    where $\delta(\eta)$ will be chosen later such that
    $\delta(\eta) \downarrow 0$ as $\eta \downarrow 0$.
    Let $\bm{\pi}^{\varepsilon,\eta} = (\pi_t^{\varepsilon,\eta,\bm{\alpha}}, \pi_t^{\varepsilon,\eta,\bm{\beta}})_{t \geq 0}$ with average intensities denoted as $m_t^{\varepsilon,\eta,\bm{\alpha}}$ and $m_t^{\varepsilon,\eta,\bm{\beta}}$.
    Then, $\bm{\pi}^{\varepsilon,\eta}$ is a $\cV_{\text{ent}}^\eta$-admissible randomized control.
    We have the following result about the approximation error due to modified controls:
    \begin{lemma}{}{approximation-errors-modifying-controls}
      Let $M \geq 1$ and $M\delta(\eta)\leq1$. Then, we have 
      \begin{align*}
      \abs{\widetilde J(\bm{\pi}^{\varepsilon,\eta})-\widetilde J(\widetilde{\bm{\pi}}^\varepsilon)} \leq C M^2 \delta(\eta)^{\kappa-\kappa/r} + CM^{1+\kappa/2-\kappa/r} \delta(\eta)^{\kappa/2-\kappa/r},
      \end{align*} 
      for some finite constant $C>0$ depending only on $(\rho,\mu,\sigma,G)$ and the initial condition.
  
      \begin{proof}
        See \cref{appendix:proof-approximation-errors-modifying-controls}.
      \end{proof}
      
    \end{lemma}
    By the entropy bound of $\Phi_{m,\delta}$ in \labelcref{eq:entropy-bound-phi}, for all $t \geq 0$, $\cH(\pi_t^{\varepsilon,\eta})\geq 2\ln\delta(\eta)$.
    Therefore,
    \begin{align*}
      \eta \E{\int_0^\infty e^{-\rho t}\cH(\pi_t^{\varepsilon,\eta}) dt}
      \geq 2\eta \ln\delta(\eta)\int_0^\infty e^{-\rho t} dt
      = \frac{2}{\rho} \eta \ln\delta(\eta).
    \end{align*}
    Putting the above together with \cref{lem:approximation-errors-modifying-controls}, when $M\delta(\eta)\leq 1$, we have
    \begin{align*}
    \begin{aligned}
      J^\eta(\bm{\pi}^{\varepsilon,\eta})
      &= \E{\int_0^\infty e^{-\rho t}\left(\tilde r_t + \eta \cH(\pi_t^{\varepsilon,\eta})\right) dt} \\
      &= \widetilde J(\bm{\pi}^{\varepsilon,\eta}) + \eta \E{\int_0^\infty e^{-\rho t}\cH(\pi_t^{\varepsilon,\eta}) dt} \\
      &\geq \widetilde J(\widetilde{\bm{\pi}}^\varepsilon) - CM^2 \delta(\eta)^{\kappa-\kappa/r} - CM^{1+\kappa/2-\kappa/r} \delta(\eta)^{\kappa/2-\kappa/r} + \frac{2}{\rho} \eta\ln\delta(\eta).
    \end{aligned}
    \end{align*}
    By the $\varepsilon$-optimality of $\widetilde{\bm{\pi}}^\varepsilon$,
    $\widetilde J(\widetilde{\bm{\pi}}^\varepsilon)\geq \widetilde{\cV}-\varepsilon$. Hence,
    \begin{align*}
      J^\eta(\bm{\pi}^{\varepsilon,\eta}) \ge \widetilde{\cV} - \varepsilon - CM^2 \delta(\eta)^{\kappa-\kappa/r} - CM^{1+\kappa/2-\kappa/r} \delta(\eta)^{\kappa/2-\kappa/r} + \frac{2}{\rho} \eta\ln\delta(\eta).
    \end{align*}
    Taking supremum over $\cA(x)$, and since $\varepsilon>0$ is arbitrary, we get
    \begin{align*}
      \cV_{\text{ent}}^\eta \geq \widetilde{\cV} - CM^2 \delta(\eta)^{\kappa-\kappa/r} - CM^{1+\kappa/2-\kappa/r} \delta(\eta)^{\kappa/2-\kappa/r} + \frac{2}{\rho} \eta\ln\delta(\eta).
    \end{align*}
    
    The error bound above could be further optimized with respect to $\delta=\delta(\eta)$. Now we specialize to $\delta(\eta) := \eta^\beta$ for some $\beta>0$ such that $\delta(\eta) \downarrow 0$ as $\eta \downarrow 0$. Recall that $M \delta(\eta)\leq 1$ which implies the restriction $\beta\geq -\ln M/\ln \eta$.
    Consider the minimization problem
    \begin{align*}
      \min_{\beta\geq -\ln M / \ln\eta} \left\{ CM^2 \eta^{\beta(\kappa-\kappa/r)} + CM^{1+\kappa/2-\kappa/r} \eta^{\beta(\kappa/2-\kappa/r)} - \frac{2}{\rho} \beta \eta \ln\eta \right\}.
    \end{align*}
    We can now take $\beta = \frac{2r}{(r-2)\kappa} \geq -\frac{\ln M}{\ln \eta}$ so that $\beta(\kappa-\kappa/r) > \beta(\kappa/2-\kappa/r)=1$.
    Since the convergence speed of $\eta\ln\eta$ is faster than any $\eta^k$ for $k \in (0,1)$, and slower than any $\eta^k$ for $k \geq 1$, we get that for sufficiently small $\eta$ and $M\geq1$,
    \begin{align*}
      \widetilde{\cV} - \cV_{\text{ent}}^\eta &\leq \abs{\eta\ln\eta} \left(CM^2 + CM^{1+\kappa/2-\kappa/r}+C\right) \\
      &\leq \abs{\eta\ln\eta} \left(CM^{2}+C\right) \leq C M^2 \abs{\eta\ln\eta}.
    \end{align*}
    
  \end{proof}
  \end{theorem}

  \begin{remark}{}{choice-of-M-and-eta}
    From \cref{thm:convergence-entropy-regularized-value-function}, we see that the choice of $M$ affects the convergence rate of $\cV_{\text{ent}}^\eta$ to $\cV_{\text{inte}}$ as $\eta \downarrow 0$.
    A larger $M$ enlarges the action space $\mathbb{M}=[0,M]$ for exploration, which can potentially lead to better performance in the speculative trading problem.
    However, it also slows down the convergence rate, as indicated by the term $M^2 \abs{\eta\ln\eta}$ in the error bound.
    Combining \cref{thm:error-estimate-original-continuous-control,thm:convergence-entropy-regularized-value-function}, the overall error between $V_{\text{orig}}$ and $\cV_{\text{ent}}^\eta$ can be bounded by
    \begin{align*}
      0 \leq V_{\text{orig}}(p) - \cV_{\text{ent}}^\eta(p,0,0) \leq C \left(M^2\abs{\eta\ln\eta} + M^{-\kappa/2}\right),
    \end{align*}
    for some constant $C>0$ independent of $\eta$ and $M$.
    Therefore, to ensure that the original problem can be well approximated with randomized intensities and entropy regularization, we need to choose $M$ and $\eta$ such that both terms on the right-hand side are small.
    In particular, we need to jointly choose $(M,\eta)$ such that $M^2 \abs{\eta\ln\eta} \to 0$ as $M\to\infty$ and $\eta \downarrow 0$.
    
  \end{remark}

\section{A pairs-trading example}\label{sec:OU-S-shaped-utility}
\subsection{Model setup}
Following \cref{rem:financial-interpretation-for-negative-value-of-signals}, we choose  $\gamma=\iota = 1$ while set $\Psi, R \geq 0$. The signal $P$ is assumed to follow an OU process:
\begin{align} \label{eq:OU-process}
  dP_t = \theta (\bar p - P_t) dt + \sigma dW_t, \qquad P_0 = p_0 \in \bR,
\end{align}
where $\theta,\sigma>0, \bar p\in\bR$ are constants. As introduced earlier, $P$ can represent the price spread of two risky assets such that $\tau$ and $\nu$ respectively correspond to the time to initiate and to close the pairs-trading opportunity.

We can explicitly verify \cref{assump:assumption-convergence-entropy-regularized-value-function}. The OU process admits an explicit solution
\begin{equation*}
  P_t = \bar p + (p_0 - \bar p) e^{-\theta t} + \sigma \int_0^t e^{-\theta (t-s)} dW_s.
\end{equation*}
Hence, $P_t$ is Gaussian with mean  $\E{P_t} = \bar p + (p_0 - \bar p) e^{-\theta t}$ and variance 
$\Var{P_t} = \frac{\sigma^2}{2\theta} \left(1 - e^{-2\theta t}\right)$.
Since $\abs{\E{P_t}} \leq \abs{\bar p}+\abs{p_0-\bar p}$ and $\Var{P_t} \leq \frac{\sigma^2}{2\theta}$, it follows that $\sup_{t\geq0}\E{\abs{P_t}^r} < \infty$ for any $r > 0$.
Hence, \cref{assump:assumption-convergence-entropy-regularized-value-function} is satisfied for any $\zeta\in(0,\rho)$.
Finally, we consider the S-shaped power utility 
\begin{align} \label{eq:S-shaped-power-utility}
  U(x) = \begin{cases}
    x^{\varpi}, & x \geq 0, \\
    -k \abs{x}^{\varpi}, & x < 0,
  \end{cases}
\end{align}
where $\varpi \in (0,1], k > 0$ are constants. The value $\varpi$ represents the degree of risk aversion and risk seeking over the gain and loss regimes, while $k$ captures the agent's loss aversion level. With the above choices, \cref{assump:continuity-growth} is satisfied.

\subsection{The system of HJB equations}
Under \labelcref{eq:OU-process}, the infinitesimal generator of $P$ is
\begin{align*}
  \cL_p v = \theta (\bar p - p) \frac{\partial v}{\partial p} + \frac{1}{2} \sigma^2 \frac{\partial^2 v}{\partial p^2}.
\end{align*}
With \labelcref{eq:S-shaped-power-utility} and $\gamma=\iota=1$, the realized utility as a function of current signal $p$ and past signal $b$ is
\begin{align*}
  G(p,b) = U( p - b - \Psi - R) = \begin{cases}
    ( p - b - \Psi - R)^{\varpi}, &  p - b - \Psi - R \geq 0, \\
    -k \abs{ p - b - \Psi - R}^{\varpi}, &  p - b - \Psi - R < 0.
  \end{cases}
\end{align*}
Following the previous notations, we use $\cV_0, \cV_1$ and $\cV_2$ to denote the value functions in the three stages of the sequential trading problem (before entry, after entry but before exit, and after exit). Denote $\Delta_1 (p) := \cV_1(p,p) - \cV_0(p)$ and $\Delta_2 (p,b) := G(p,b) - \cV_1(p,b)$ as the advantages (i.e., the marginal values in certainty equivalent terms) for market entry and exit, respectively.
Let $\pi^{\bm{\alpha}}_t(\cdot \ ;\omega)$, $\pi^{\bm{\beta}}_t(\cdot \ ;\omega) \in \cP_{\text{ac}}(\bM)$, $t\in[0,\infty)$, $\omega\in\Omega$,
be the exploration densities for market entry and exit, respectively.
From \cref{sec:HJB-general}, the HJB system for $\cV_0(p)$ and $\cV_1(p,b)$ is given by
\begin{align*}
  \begin{cases}
    \rho \cV_0(p) - \cL_p \cV_0(p) = \eta \ln\left(\eta \frac{\exp\left(M\Delta_1(p)/\eta\right)-1}{M\Delta_1(p)}\right), \\
    \rho \cV_1(p,b) - \cL_p \cV_1(p,b) = \eta \ln\left(\eta \frac{\exp\left(M\Delta_2(p,b)/\eta\right)-1}{M\Delta_2(p,b)}\right).
  \end{cases}
\end{align*}
When $\Delta_1(p)$ or $\Delta_2(p,b)$ equals zero, the corresponding right-hand side is defined as zero by continuity.
When $\eta \downarrow 0$, the right-hand side converges to $M \Delta_1(p)^+$ or $M \Delta_2(p,b)^+$, respectively, recovering the HJB system for the exploration problem without entropy regularization.
With $P_t(\omega)=p$ and $B_t(\omega)=b$,
the optimal exploration densities are given by
\begin{align*}
  \pi_t^{\bm{\alpha},*}(\lambda; \omega) = \pi^{\bm{\alpha},*}(\lambda;p) &= \frac{\exp\left(\lambda \Delta_1(p)/\eta\right)}{\int_\bM \exp\left(\tilde\lambda \Delta_1(p)/\eta\right) d\tilde\lambda} = \frac{\Delta_1(p)}{\eta} \frac{\exp\left(\lambda \Delta_1(p)/\eta\right)}{\exp\left(M\Delta_1(p)/\eta\right)-1}, \\
  \pi_t^{\bm{\beta},*}(\lambda;\omega) = \pi^{\bm{\beta},*}(\lambda;p,b) &= \frac{\exp\left(\lambda \Delta_2(p,b)/\eta\right)}{\int_\bM \exp\left(\tilde\lambda \Delta_2(p,b)/\eta\right) d\tilde\lambda} = \frac{\Delta_2(p,b)}{\eta} \frac{\exp\left(\lambda \Delta_2(p,b)/\eta\right)}{\exp\left(M\Delta_2(p,b)/\eta\right)-1}.
\end{align*}
When $\Delta_1(p)$ or $\Delta_2(p,b)$ equals zero, the corresponding optimal density coincides with the uniform distribution on $\bM$ by continuity. When $\eta \downarrow 0$, the optimal densities converge to delta measures: $\pi^{\bm{\alpha},*}(\lambda;p)$ converges to $\delta_M(\lambda) \mathds{1}_{\{\Delta_1(p)>0\}} + \delta_0(\lambda) \mathds{1}_{\{\Delta_1(p)<0\}} + \frac{1}{M} \mathds{1}_{\{\Delta_1(p)=0\}}$, and $\pi^{\bm{\beta},*}(\lambda;p,b)$ converges to $\delta_M(\lambda) \mathds{1}_{\{\Delta_2(p,b)>0\}} + \delta_0(\lambda) \mathds{1}_{\{\Delta_2(p,b)<0\}} + \frac{1}{M} \mathds{1}_{\{\Delta_2(p,b)=0\}}$.
The optimal intensities follow from their mean values as
\begin{align*}
  \int_\bM \lambda \pi^{\bm{\alpha},*}(\lambda;p) d\lambda &= \begin{cases}
    \frac{M}{1-\exp\left(-M\Delta_1(p)/\eta\right)} - \frac{\eta}{\Delta_1(p)}, & \Delta_1(p) \neq 0, \\
    \frac{M}{2}, & \Delta_1(p) = 0,
  \end{cases} \\
  \int_\bM \lambda \pi^{\bm{\beta},*}(\lambda;p,b) d\lambda &= \begin{cases}
    \frac{M}{1-\exp\left(-M\Delta_2(p,b)/\eta\right)} - \frac{\eta}{\Delta_2(p,b)}, & \Delta_2(p,b) \neq 0, \\
    \frac{M}{2}, & \Delta_2(p,b) = 0.
  \end{cases}
\end{align*}
When $\eta \downarrow 0$, the optimal intensities converge to $M \mathds{1}_{\{\Delta_1(p)>0\}} + \frac{M}{2} \mathds{1}_{\{\Delta_1(p)=0\}}$ and $M \mathds{1}_{\{\Delta_2(p,b)>0\}} + \frac{M}{2} \mathds{1}_{\{\Delta_2(p,b)=0\}}$, respectively.

\subsection{Numerical results}

The baseline numerical parameters of the problem are now specialized to $\iota=\gamma=1$, $\Psi=0$, $R=1$, $\rho=0.05$, $\theta=0.1$, $\bar p=0$, $\sigma=0.2$, $\varpi=0.5$, and $k=2$.

\subsubsection{Solution to the HJB System}

We first compute the value functions numerically by solving the HJB system \eqref{eq:hjb_1} and \eqref{eq:hjb_2} directly using the finite difference method. Implementation of this method requires us to have perfect knowledge about the model dynamics. Nonetheless, later in \cref{sec:offline-policy-iteration-OU-S-shaped-utility} we will present an alternative, model-independent RL algorithm which only requires the simulated trajectories of the trading environment as inputs. The numerical results obtained in this subsection could therefore be considered as the benchmark solution for investigating the quality of the solutions obtained by the RL algorithm.

Denote the range of $(p,b)$ as $[p_{\min}, p_{\max}] \times [b_{\min}, b_{\max}]$. We set $p_{\min} = -4$, $p_{\max} = 4$, $b_{\min} = -4$, and $b_{\max} = 4$.
Since the diagonal values $\cV_1(p,p)$ are needed to solve $\cV_0(p)$, we choose the same range for both $p$ and $b$. 
The boundary conditions are set as follows:
\begin{itemize}
  \item For $\cV_0(p)$, at $p=p_{\min}$, the agent would enter the market immediately since the signal is very low.
  Hence, we set $\cV_0(p_{\min}) = \cV_1(p_{\min}, p_{\min})$.
  
  At $p=p_{\max}$, the agent would not enter the market since the signal is too high. They will postpone the entry decision until the signal becomes lower, so the sensitivity of the value function to signal changes is very low.
  Hence, we set $\cV_0'(p_{\max}) = 0$.
  
  \item For $\cV_1(p,b)$, at $p=p_{\min}$, the agent would not exit the market since the signal is too low. They will keep holding the asset until a higher signal is observed, so the sensitivity of the value function to signal changes is very low.
  Hence, we set $\frac{\partial \cV_1}{\partial p}(p_{\min},b) = 0$.

  At $p=p_{\max}$, the agent would exit the market immediately since the signal is very high.
  Hence, we set $\cV_1(p_{\max},b) = G(p_{\max},b)$.
\end{itemize}
We can solve for $\cV_1$ first since it is independent of $\cV_0$, and then we solve for $\cV_0$ using the estimated $\cV_1$ (more precisely, only the diagonal values $\cV_1(p,p)$ are needed).
For each $b_j$, we solve a tridiagonal system in the $p$-grid.
Following \citet{dongRandomizedOptimalStopping2024a}, we solve the HJB system using finite difference method with iteration and first-order approximation to linearize the nonlinear (log-exponential) term appearing in the HJB equation.
The grid size for both $p$ and $b$ is set as $0.05$.
The algorithm iterates until the maximum change of the value functions at all grid points is less than $10^{-6}$.

\begin{figure}[H]
  \centering
  \begin{subfigure}{0.4\textwidth}
    \includegraphics[width=\textwidth]{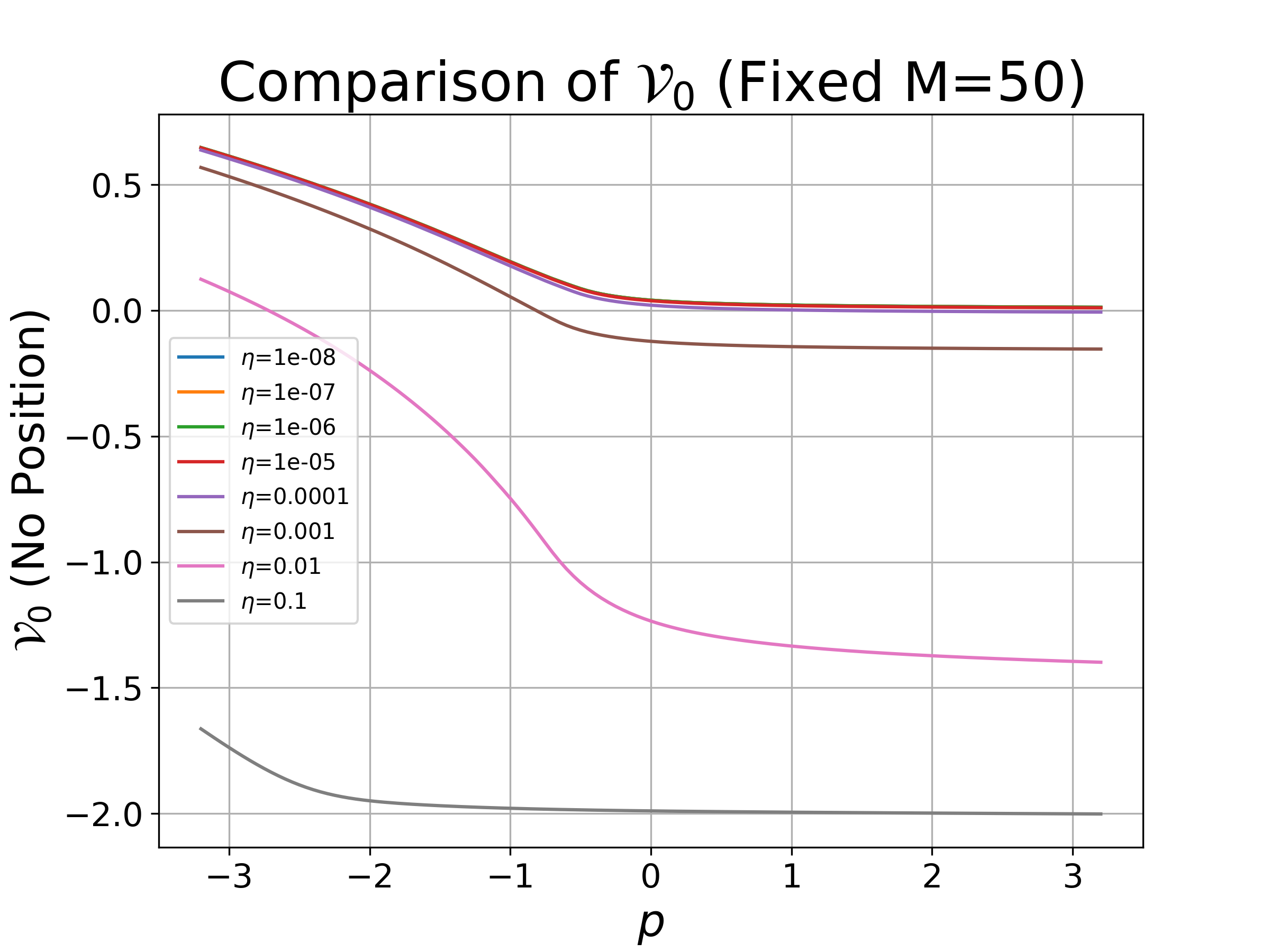}
    \caption{$M=50$}
  \end{subfigure}
  \begin{subfigure}{0.4\textwidth}
    \includegraphics[width=\textwidth]{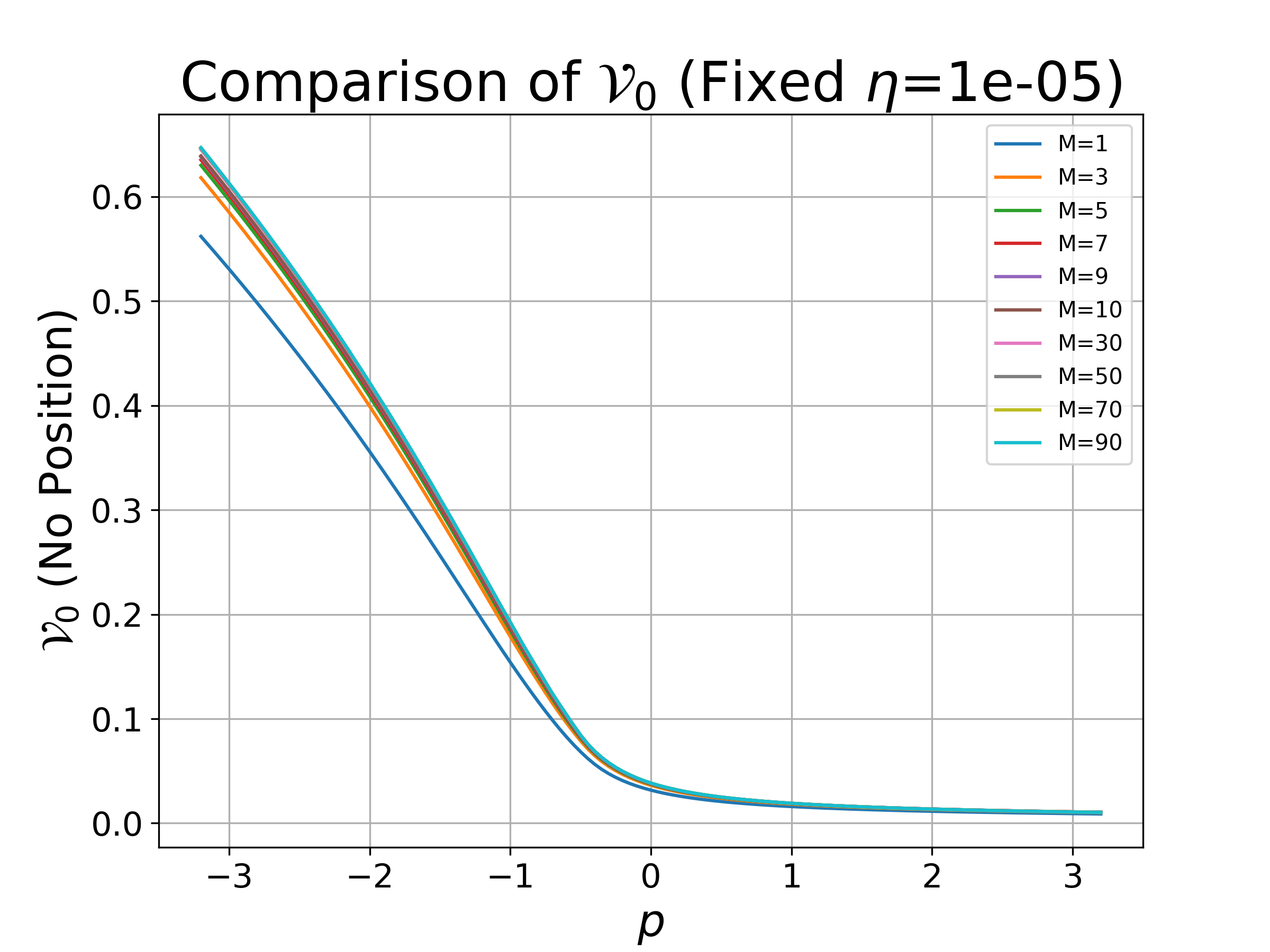}
    \caption{$\eta=10^{-5}$}
  \end{subfigure}
  
  \caption{Value function $\cV_0(p)$ under different $(M,\eta)$}
  \label{fig:HJB-V0-different-M-eta}
\end{figure}

\begin{figure}[H]
  \centering
  \begin{subfigure}{0.4\textwidth}
    \includegraphics[width=\textwidth]{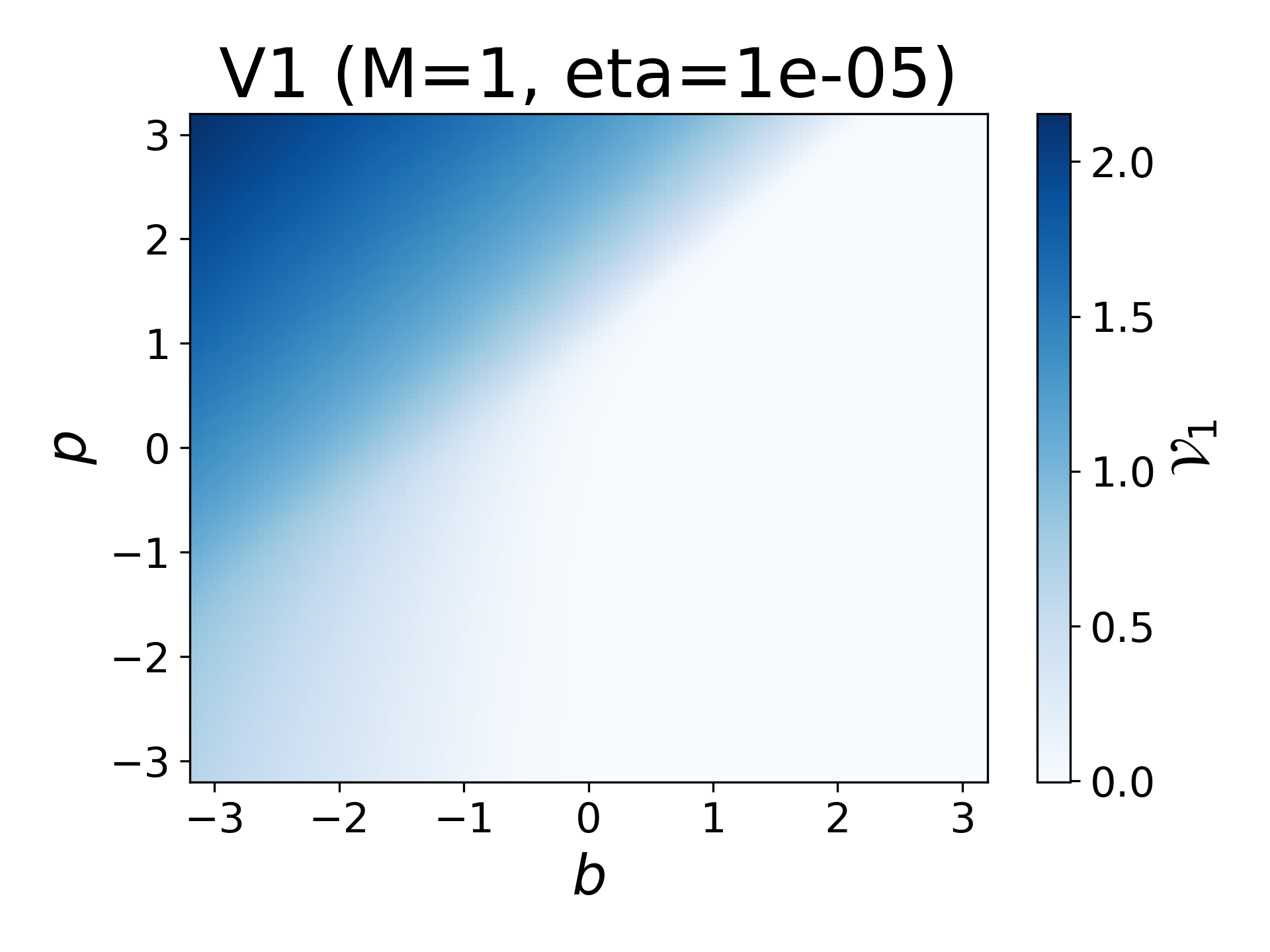}  
    \caption{$M=1$, $\eta=10^{-5}$}
  \end{subfigure}
  \begin{subfigure}{0.4\textwidth}
    \includegraphics[width=\textwidth]{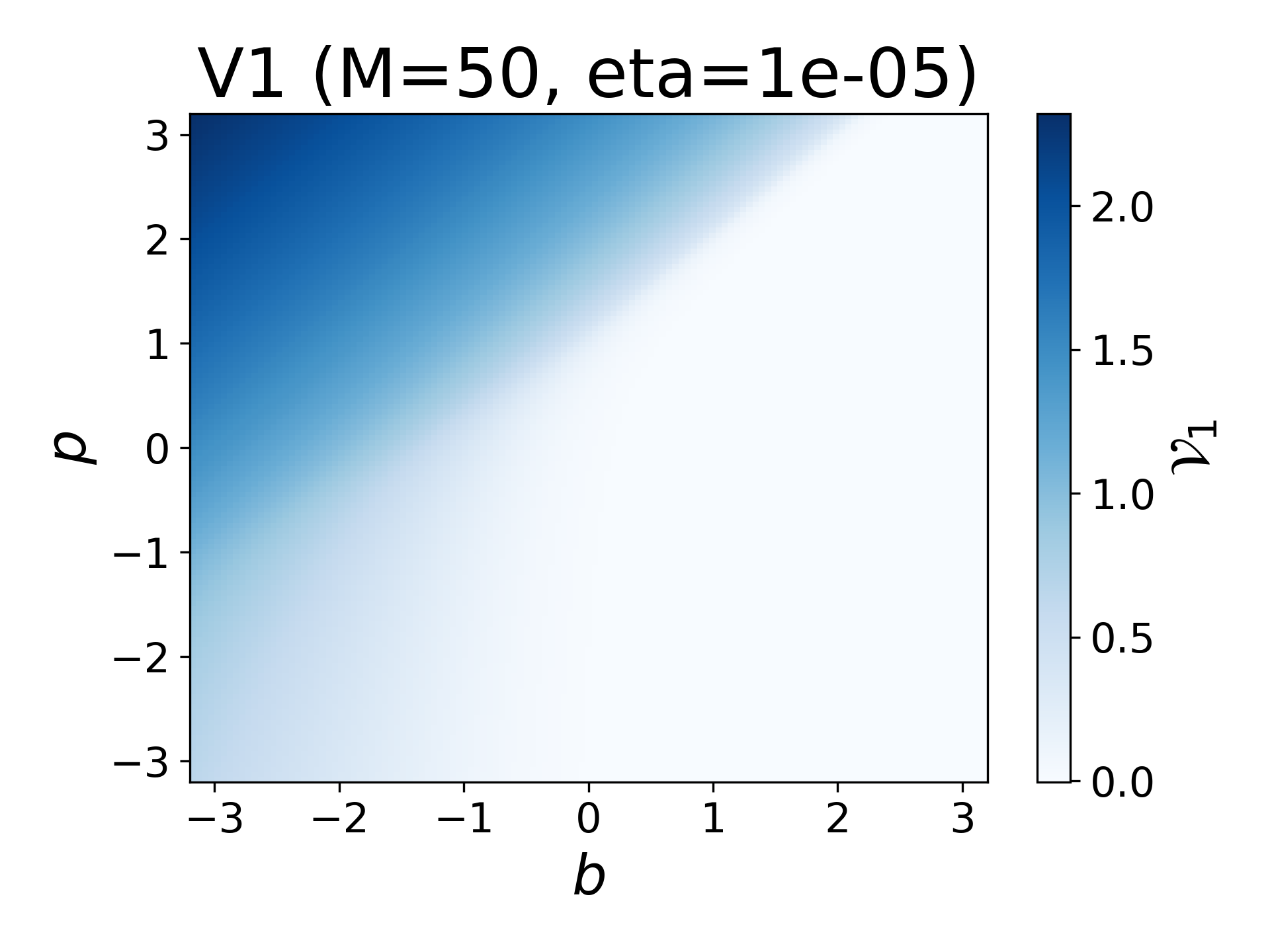}  
    \caption{$M=50$, $\eta=10^{-5}$}
  \end{subfigure}
  
  \caption{Value function $\cV_1(p,b)$ under different $(M,\eta)$}
  \label{fig:HJB-V1-different-M-eta}
\end{figure}

The value functions $\cV_0(p)$ and $\cV_1(p,b)$ under different $(M,\eta)$ are shown in \cref{fig:HJB-V0-different-M-eta,fig:HJB-V1-different-M-eta}, respectively. To avoid the influence of boundary conditions, we focus on the interior region $(p,b) \in [-3.2,3.2] \times [-3.2,3.2]$ far away from the boundaries.
From the plots, we observe two main effects of $M$ and $\eta$:
\begin{itemize}
  \item $\cV_0$ is increasing as $\eta$ increases. This aligns with our theoretical analysis since the entropy is nonpositive.
  \item $\cV_0$ is increasing as $M$ increases. This is because with low $M$, the action space is effectively shrunk which is detrimental to the agent.
\end{itemize}
The monotonicity also matches economic intuitions. $\cV_0(p)$ is monotonically decreasing as $p$ increases. $\cV_1(p,b)$ is monotonically increasing as $p$ increases and $b$ decreases.

\begin{figure}[H]
  \centering
  \begin{subfigure}{0.4\textwidth}
  \includegraphics[width=\textwidth]{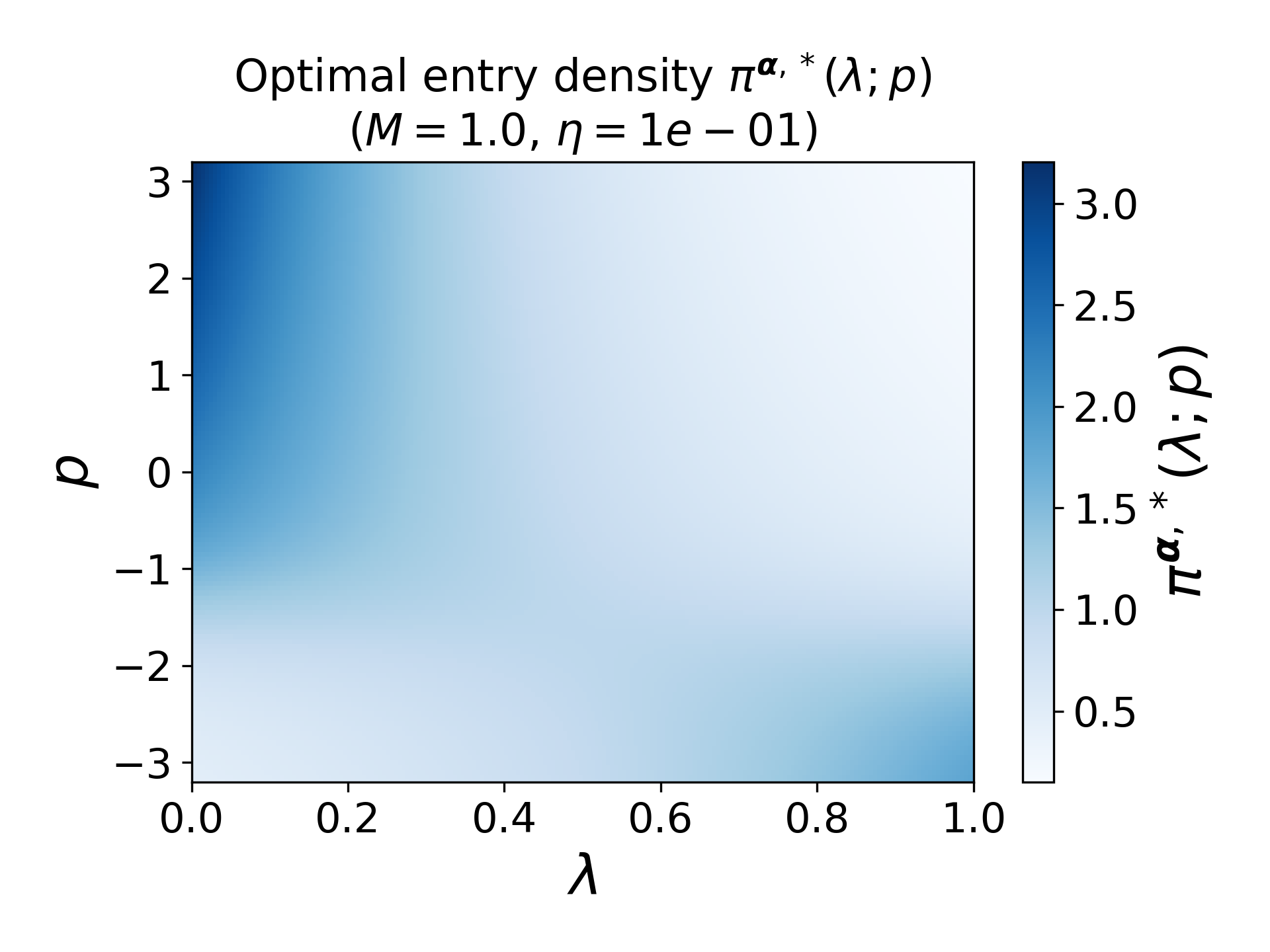}  
  \caption{$M=1$, $\eta=10^{-1}$}
  \end{subfigure}
  \begin{subfigure}{0.4\textwidth}
  \includegraphics[width=\textwidth]{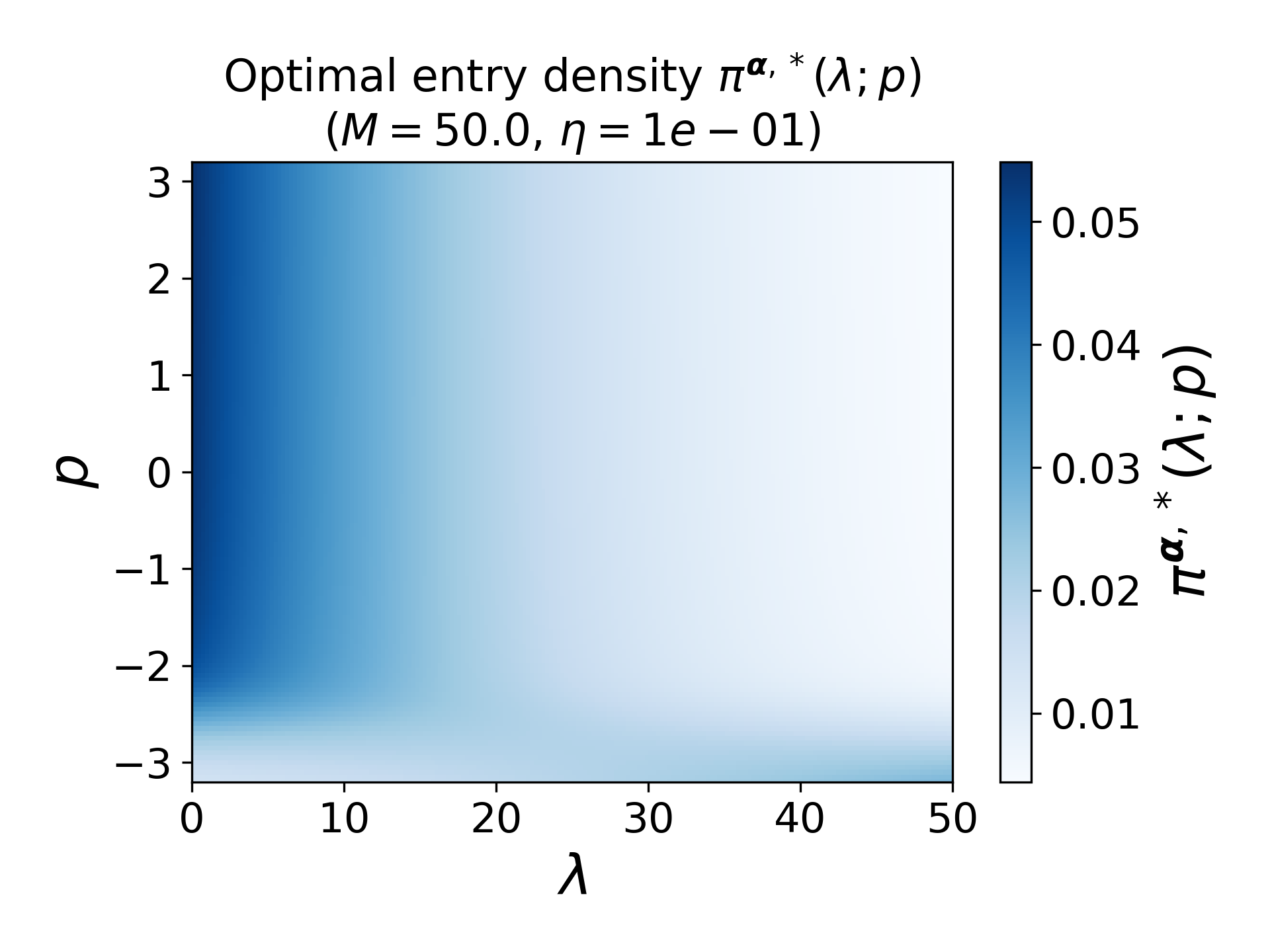}  
  \caption{$M=50$, $\eta=10^{-1}$}
  \end{subfigure}
  \caption{Optimal entry density $\pi^{\bm{\alpha},*}(\lambda;p)$ when $\eta$ is high}
  \label{fig:optimal-density-heatmap-entry-eta-high}
\end{figure}

One advantage of our randomized intensities approach is that the agent has imposed a genuine sampling distribution over the action space (the stopping intensities), rather than choosing a deterministic policy. \cref{fig:optimal-density-heatmap-entry-eta-high} shows the optimal entry density $\pi^{\bm{\alpha},*}(\lambda;p)$ when $\eta$ is high, meaning that the agent is encouraged to explore more. For each fixed $p$, the agent is more likely to adopt a market entry rate indicated by the range of $\lambda$ with a darker region on the heatmap.
Even when $p$ is very low, there are nonzero probabilities of choosing a low entry intensity. Similar observation holds in the case that $p$ is very high. This resonates the key idea of RL that the agent should occasionally consider a less greedy strategy to explore the environment even there are other policies that appear to be more profitable.

\begin{figure}[H]
  \centering
  \begin{subfigure}{0.4\textwidth}
  \includegraphics[width=\textwidth]{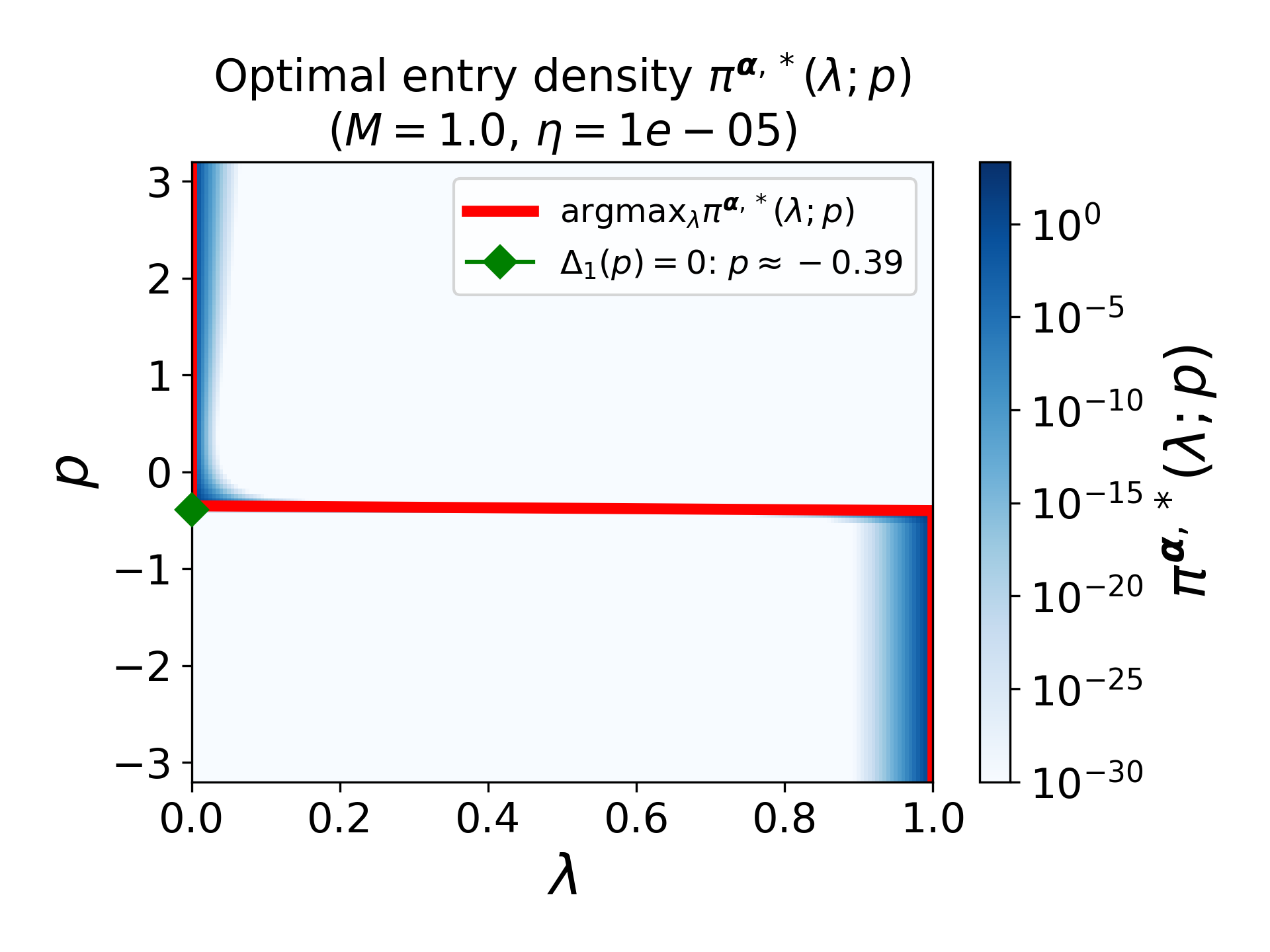}
  \caption{$M=1$, $\eta=10^{-5}$}
  \end{subfigure}
  \begin{subfigure}{0.4\textwidth}
  \includegraphics[width=\textwidth]{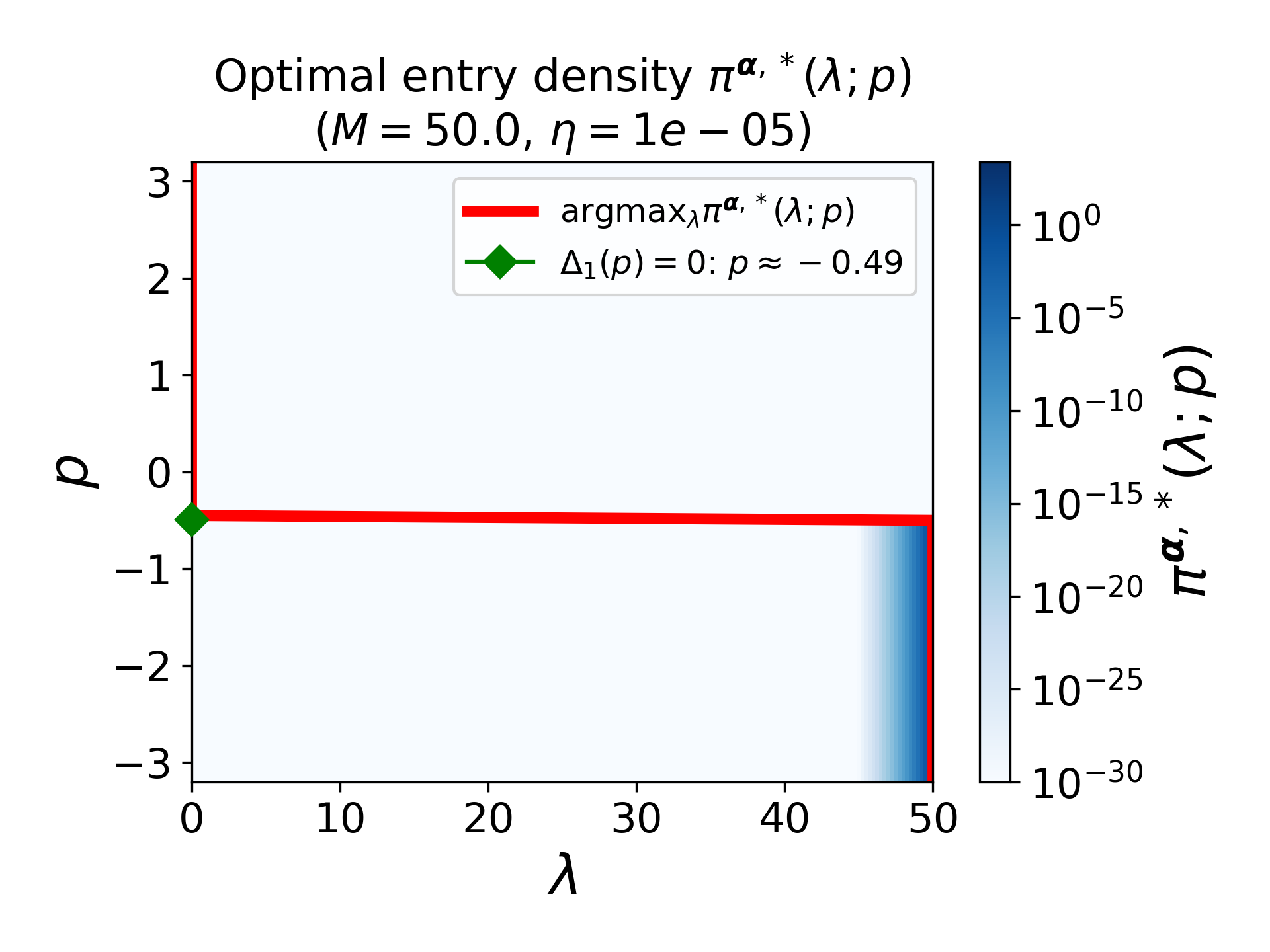}
  \caption{$M=50$, $\eta=10^{-5}$}
  \end{subfigure}
  \caption{Optimal entry density $\pi^{\bm{\alpha},*}(\lambda;p)$ when $\eta$ is small}
  \label{fig:optimal-density-heatmap-entry-eta-small}
\end{figure}

When $\eta$ is small, \cref{fig:optimal-density-heatmap-entry-eta-small} shows that the density becomes more concentrated around $\lambda=0$ for large $p$, and more concentrated around $\lambda=M$ for small $p$. A small $\eta$ suggests that the agent should act greedy and exploration is discouraged. They therefore should consistently enter the market more aggressively for small $p$ (i.e., adopt a high entry rate) and take no action for large $p$ (i.e., adopt an entry rate close to zero). The red line shows the ridge curve $\mathrm{argmax}_\lambda \pi^{\bm{\alpha},*}(\lambda;p)$, which is the intensity $\lambda$ that is most likely to be chosen for each $p$.
By solving $\Delta_1(p)=0$, we find the free boundary for the agent to enter the market, indicated by the green point. When the signal declines across this boundary, the most-likely-chosen intensity jumps from around $\lambda=0$ to around $\lambda=M$.
At the free boundary, the agent is indifferent between entering the market and not entering the market, since the expected utility is the same, and the optimal probability measure becomes uniform on $[0,M]$.

\begin{figure}[H]
  \centering
  \begin{subfigure}{0.4\textwidth}
  \includegraphics[width=\textwidth]{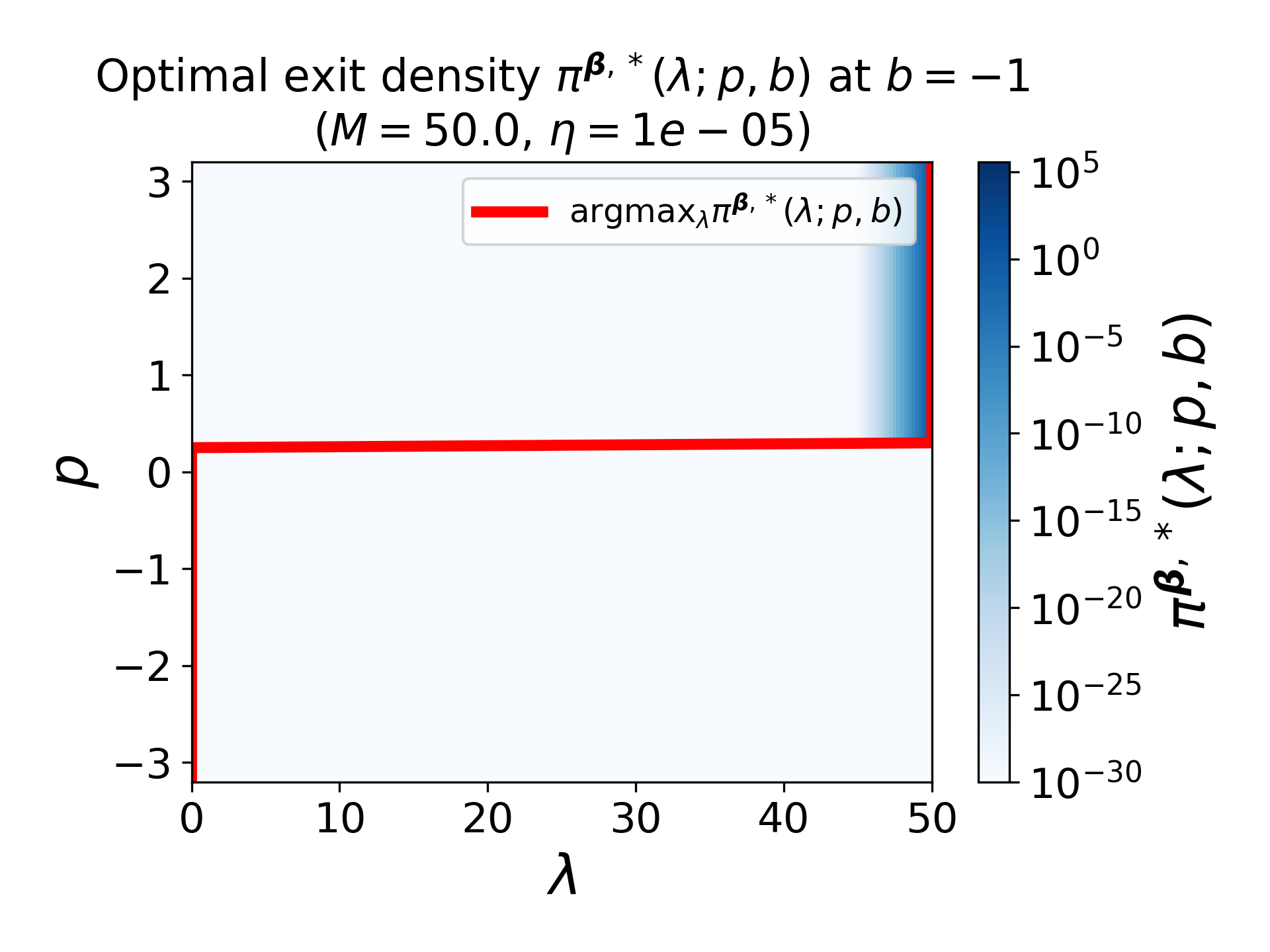}
  \caption{$b=-1$}
  \end{subfigure}
  \begin{subfigure}{0.4\textwidth}
  \includegraphics[width=\textwidth]{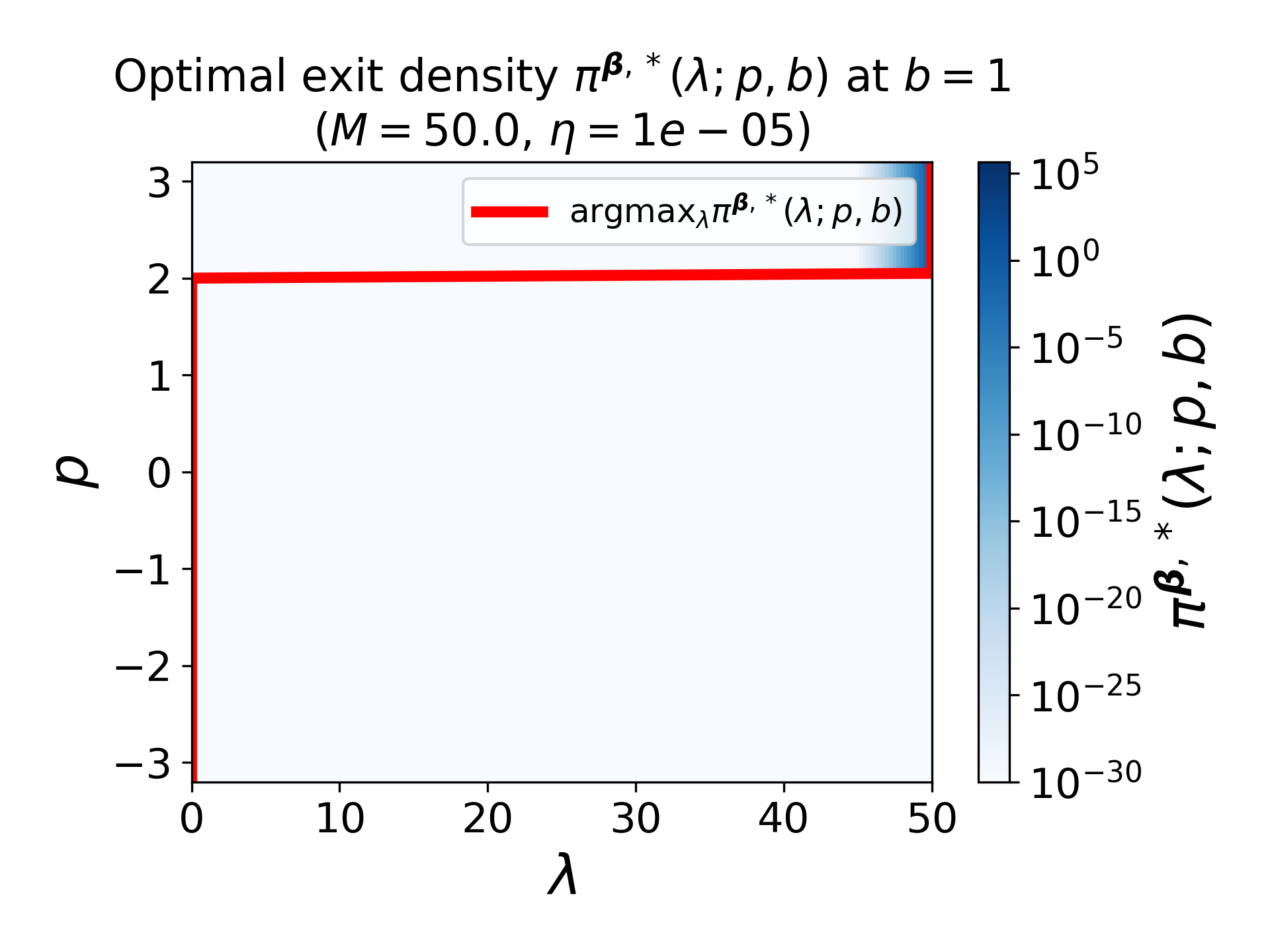}
  \caption{$b=1$}
  \end{subfigure}
  \caption{Optimal exit density $\pi^{\bm{\beta},*}(\lambda;p,b)$ with $M=50$, $\eta=10^{-5}$}
  \label{fig:optimal-density-heatmap-exit}
\end{figure}

To see how the exit free boundary changes with different entry price, we plot the optimal exit density $\pi^{\bm{\beta},*}(\lambda;p,b)$ with $M=50, \eta=10^{-5}$ in \cref{fig:optimal-density-heatmap-exit}. With a fixed pair of $(p,b)$, the agent is more likely to adopt an exit rate $\lambda$ shaded by the darker region on the heatmap. As expected, the exit free boundary increases as $b$ increases, since the agent wants to exit the market at a higher price when the entry price is higher and recall that our agent is loss averse under the Prospect Theory preference.

\begin{figure}[H]
  \centering
  \includegraphics[width=0.4\textwidth]{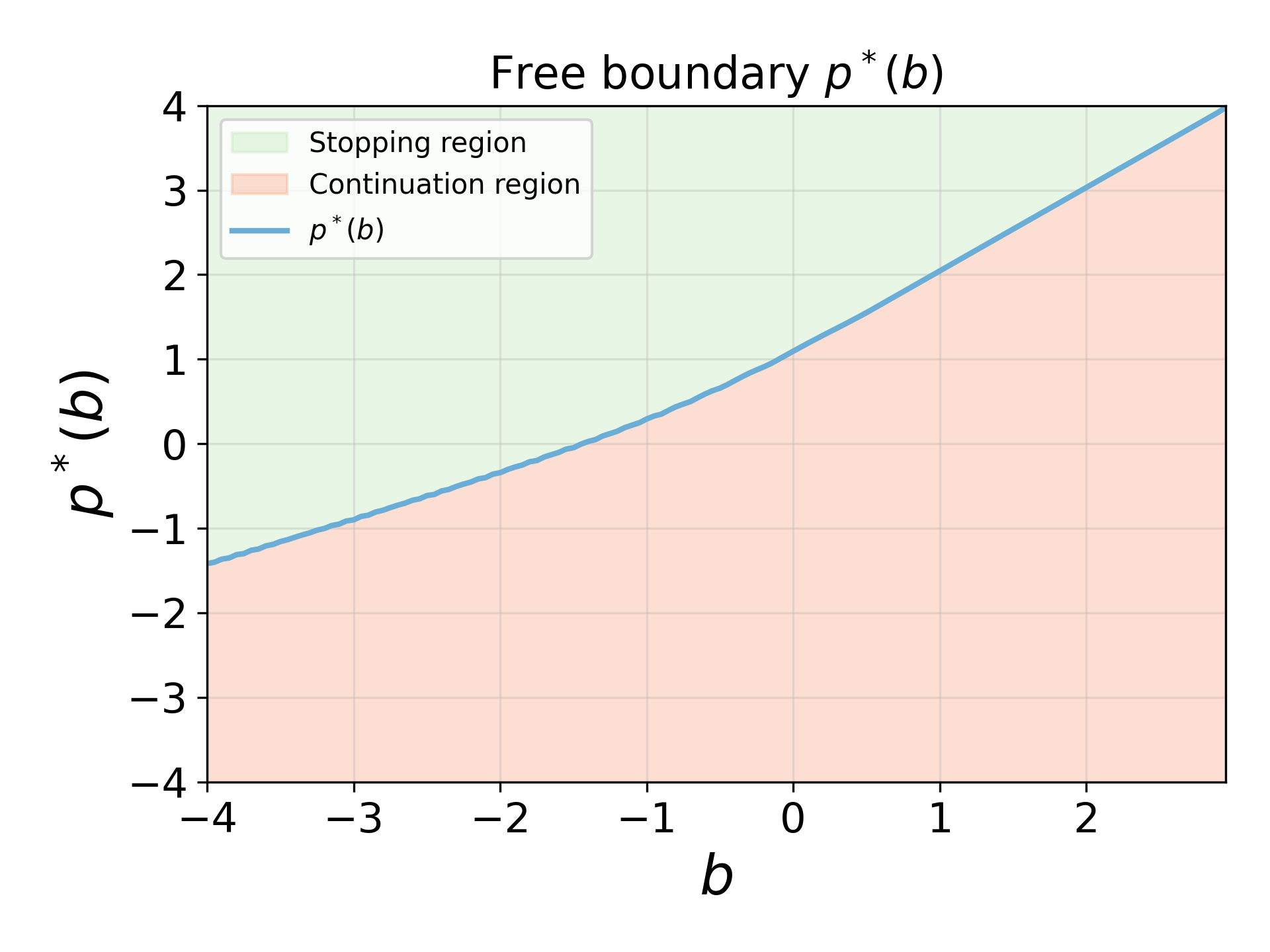}  
  \caption{Free boundary of $p$ as a function of $b$ with $M=50$, $\eta=10^{-5}$}
  \label{fig:free-boundary-p-b}
\end{figure}

To further investigate the exit free boundary, we plot it as a function $p^*(b)$ in \cref{fig:free-boundary-p-b}. It is obtained by finding the root $p^*$ of $\Delta_2(p,b)=0$ for different $b$.
If the agent enters the market at $b$, then they should stay in the market until the signal is equal to or higher than $p^*(b)$.
The region above $p^*(b)$ is the stopping region, and the region below it is the continuation region.
In the continuation region, the signal is not high enough for the agent to close the position and get desired profit. In the stopping region, it is better to close the position and do not undertake more risks.
We see that the free boundary is monotonically increasing as $b$ increases, consistent with the observations from \cref{fig:optimal-density-heatmap-exit}.

A good heuristic rule is to choose $M$ and $\eta$ such that the value functions are insensitive to the values of $M$ and $\eta$. From the above results, we choose $M=50$ and $\eta=10^{-5}$ as a good balance between exploration capacity and entropy regularization penalty such that the original value function can be well approximated.
Then, we vary the values of $\theta$ and $\sigma$ to see how the value functions change with different signal dynamics.

\begin{figure}[H]
  \centering
  \begin{subfigure}{0.4\textwidth}
    \includegraphics[width=\textwidth]{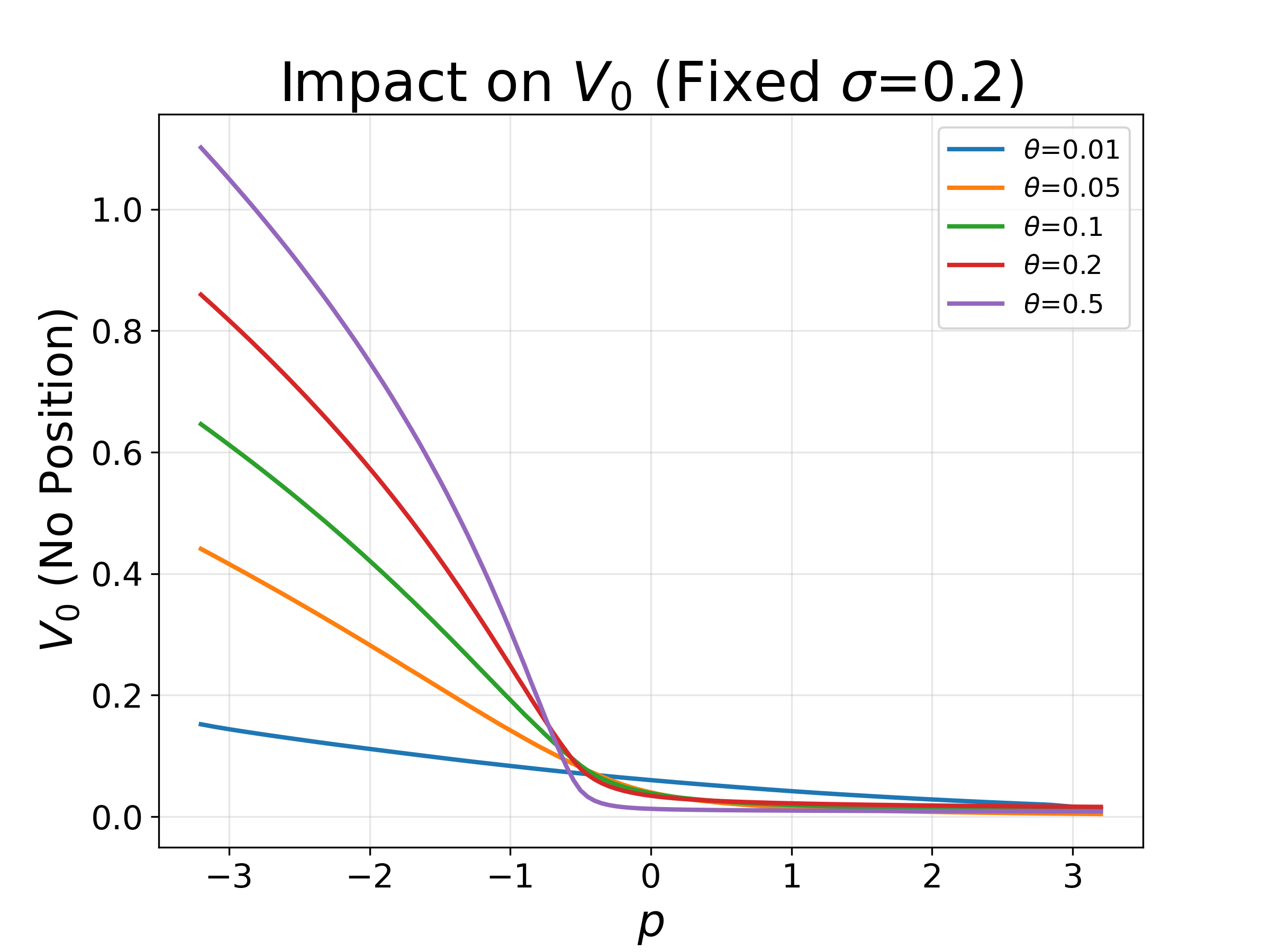}
    \caption{$\sigma=0.2$}
    \label{fig:HJB-V0-different-theta-sigma-sigma0.2}
  \end{subfigure}
  \begin{subfigure}{0.4\textwidth}
    \includegraphics[width=\textwidth]{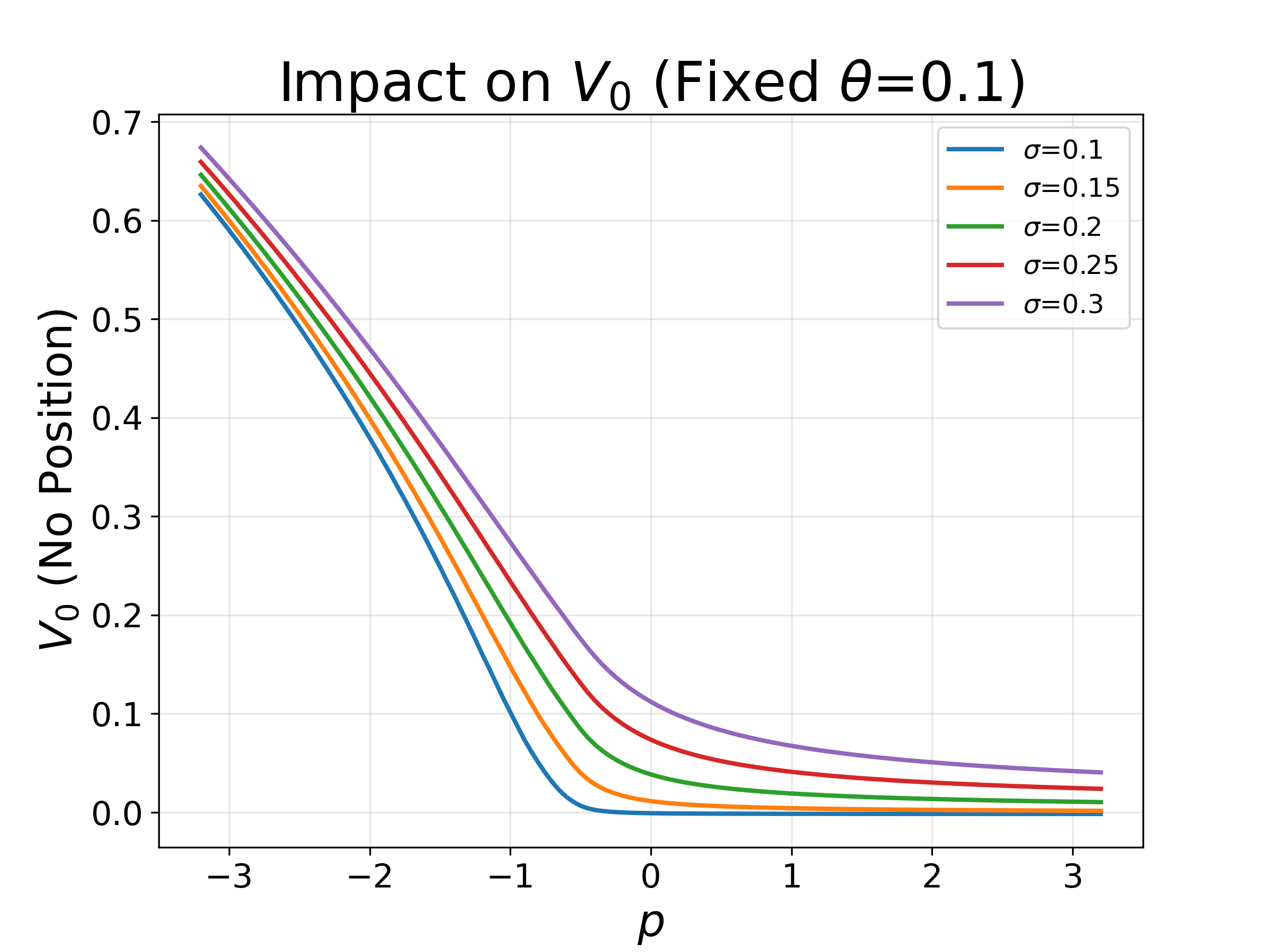}
    \caption{$\theta=0.1$}
    \label{fig:HJB-V0-different-theta-sigma-theta0.1}
  \end{subfigure}

  \caption{Value function $\cV_0(p)$ under different $\theta$ and $\sigma$}
  \label{fig:HJB-V0-different-theta-sigma}
\end{figure}

When $\theta$ increases, the mean-reversion speed of the OU process increases. When the current value of the signal is very low (negative), the strong mean-reversion suggests that the signal is expected to undergo a large increase and hence the trading opportunity becomes more profitable. In contrast, when the current value of the signal is very high, the signal is expected to decline and it is no longer attractive to hold a long position in the spread.\footnote{This conclusion will change if we allow the agent to initiate their trade in either a long or short position of the spread. Then when the signal is very high, the agent could engage statistical arbitrate by taking a short position in the spread. Then we expect $\mathcal{V}_0(p)$ to be a V-shaped function of the signal $p$.} This is validated in \cref{fig:HJB-V0-different-theta-sigma-sigma0.2}.

The effect of mean-reversion on the value function is more complicated around $\bar{p}$. Strong mean-reversion confines the future signal changes to a narrow band around $\bar{p}$: The signal is unlikely to fall far below $\bar{p}$ (limiting the benefit of waiting for a cheaper entry) or to rise far above $\bar{p}$ (limiting the exit profit once a position is opened). Even when the signal is slightly below $\bar{p}$ and the drift is positive, the agent anticipates that the strong mean-reversion will pull down the signal quickly once it exceeds its equilibrium level. Therefore, the reduction in potential round-trip profit decreases $\cV_0$ near $\bar{p}$.

When $\sigma$ increases, the volatility of the OU process increases, leading to a higher chance of the signal moving into the stopping regions (for entry and exit) such that the round-trip profit can be materialized sooner. This should increase the value function $\cV_0$ because the present value of the profit is improved due to the more timely completion of a round-trip trade.
This agrees with the numerical results shown in \cref{fig:HJB-V0-different-theta-sigma-theta0.1}.

\subsubsection{Offline Policy Iteration} \label{sec:offline-policy-iteration-OU-S-shaped-utility}
Since the optimal control has a closed-form expression in terms of the value functions, this can be conveniently estimated via an offline policy iteration algorithm. The idea is to use deep neural networks to parameterize the value functions $\cV_0(p)$ and $\cV_1(p,b)$, compute the optimal exploration policies based on the closed-form feedback control, and then update the value functions based on the TD errors.

We discretize time as $t_l = l \Delta t$ for $l = 0,1,\dots, L$ with some $L, \Delta t>0$, and work on one time step $[t_l,t_{l+1})$. This time discretization is only used for the numerical algorithm, and the value function remains independent of time. On each interval, we freeze the policy and assume the regime can only change at most once at the end point $t_{l+1}$.

Suppose we are given an offline signal path $\{P_{t_l}\}_{l=0}^L$. A key difference of our problem is that we need to augment the given paths to include the regime process $J$ and reference process $B$. To do it, initialize $J_0=0$ and $B_0=0$. For each step $l=0,\dots,L-1$:
\begin{itemize}
  \item If $J_{t_l}=0$ (before entering the market), let the entry policy be $\pi_{t_l}^{\bm{\alpha}}(\cdot)=\pi^{\bm{\alpha},*}(\cdot \ ; P_{t_l})$. Denote the mean intensity as $\bar\lambda_{t_l}^{\bm{\alpha}} = \int_\bM \lambda \pi_{t_l}^{\bm{\alpha}}(\lambda) d\lambda$ and the probability of market entry as $q^{\bm{\alpha}}_{t_l} := 1-\exp{\{-\bar\lambda_{t_l}^{\bm{\alpha}} \Delta t\}}$.
  Draw $Y^{\bm{\alpha}}_l\sim \mathrm{Bernoulli}(q^{\bm{\alpha}}_{t_l})$.
  If $Y^{\bm{\alpha}}_l=1$, set $J_{t_{l+1}}=1$ and $B_{t_{l+1}}=P_{t_{l+1}}$; otherwise set $J_{t_{l+1}}=0$ and $B_{t_{l+1}}=B_{t_l}$.
  \item If $J_{t_l}=1$ (in the market), let the exit policy be $\pi_{t_l}^{\bm{\beta}}(\cdot)=\pi^{\bm{\beta},*}(\cdot \ ; P_{t_l}, B_{t_l})$. Denote the mean intensity as $\bar\lambda_{t_l}^{\bm{\beta}} = \int_\bM \lambda \pi_{t_l}^{\bm{\beta}}(\lambda) d\lambda$ and the probability of market exit as $q^{\bm{\beta}}_{t_l} := 1-\exp{\{-\bar\lambda_{t_l}^{\bm{\beta}} \Delta t\}}$.
  Draw $Y^{\bm{\beta}}_l\sim \mathrm{Bernoulli}(q^{\bm{\beta}}_{t_l})$.
  If $Y^{\bm{\beta}}_l=1$, set $J_{t_{l+1}}=2$ and $B_{t_{l+1}}=B_{t_l}$; otherwise set $J_{t_{l+1}}=1$ and $B_{t_{l+1}}=B_{t_l}$.
  \item If $J_{t_l}=2$ (after exit), keep $J_{t_{l+1}}=2$ and $B_{t_{l+1}}=B_{t_l}$.
\end{itemize}
For each offline signal path, we repeat this simulation $I$ times to obtain multiple $(J,B)$ trajectories, which reduces Monte Carlo variance.
Denote the entropy cost as 
\begin{align*}
  c_{t_l}^{\bm{\alpha}} := \eta \left( \int_\bM \pi_{t_l}^{\bm{\alpha}}(\lambda) \ln \left(M \pi_{t_l}^{\bm{\alpha}}(\lambda)\right) d\lambda \right), \quad c_{t_l}^{\bm{\beta}} := \eta \left( \int_\bM \pi_{t_l}^{\bm{\beta}}(\lambda) \ln \left(M \pi_{t_l}^{\bm{\beta}}(\lambda)\right) d\lambda \right).
\end{align*}
With samples $(P_{t_l}, J_{t_l}, B_{t_l}, P_{t_{l+1}}, J_{t_{l+1}}, B_{t_{l+1}}) = (p_l, j_l, b_l, p_{l+1}, j_{l+1}, b_{l+1})$,
the TD errors in different regimes are given by
\begin{align} 
  \begin{aligned} \label{eq:TD-errors-OU-S-shaped-utility}
  \delta_{0,l} &= \mathds{1}_{\{j_l=0\}} \left(-c_{t_l}^{\bm{\alpha}} \Delta t + e^{-\rho \Delta t} \left( \mathds{1}_{\{j_{l+1}=0\}} \cV_0(p_{l+1}) + \mathds{1}_{\{j_{l+1}=1\}} \cV_1(p_{l+1}, b_{l+1}) \right) - \cV_0(p_l)\right), \\
  \delta_{1,l} &= \mathds{1}_{\{j_l=1\}} \left(-c_{t_l}^{\bm{\beta}} \Delta t + e^{-\rho \Delta t} \left( \mathds{1}_{\{j_{l+1}=1\}} \cV_1(p_{l+1}, b_{l+1}) + \mathds{1}_{\{j_{l+1}=2\}} G(p_{l+1}, b_{l+1}) \right) - \cV_1(p_l, b_l)\right).
 \end{aligned}
\end{align}
We will sum up and take average of both TD errors over all realized paths.
The algorithm is summarized in \cref{alg:offline-policy-iteration-OU-S-shaped-utility}.
The value functions $\cV_0(p)$ and $\cV_1(p,b)$ are parameterized using two-layer neural networks with 32 hidden units and ReLU activation. The time step size is set as $\Delta t = 0.1$, and the total number of time steps is $100$.

\begin{remark}{}{random-initilization}  
In practice, when we simulate $I$ paths of $(J,B)$ for a path of signal,
we do not initialize $(J_0^i,B_0^i) = (0,0)$ for all $i=1,2,\dots,I$. Instead, we initialize half of $J_0$ to be $0$ and half of it to be $1$. When it is initialized to be 1, we randomly initialize $B_0$ with uniform distribution. The simulation approach after $t=0$ does not change. Doing so enables the neural network to observe more possible states with less number of time steps and simulated paths, thus increasing the training efficiency remarkably.
This modification only changes the sampling distribution of training states. It does not change the TD errors or the objective function.
\end{remark}

\begin{figure}[H]
  \centering
  \includegraphics[width=0.8\textwidth]{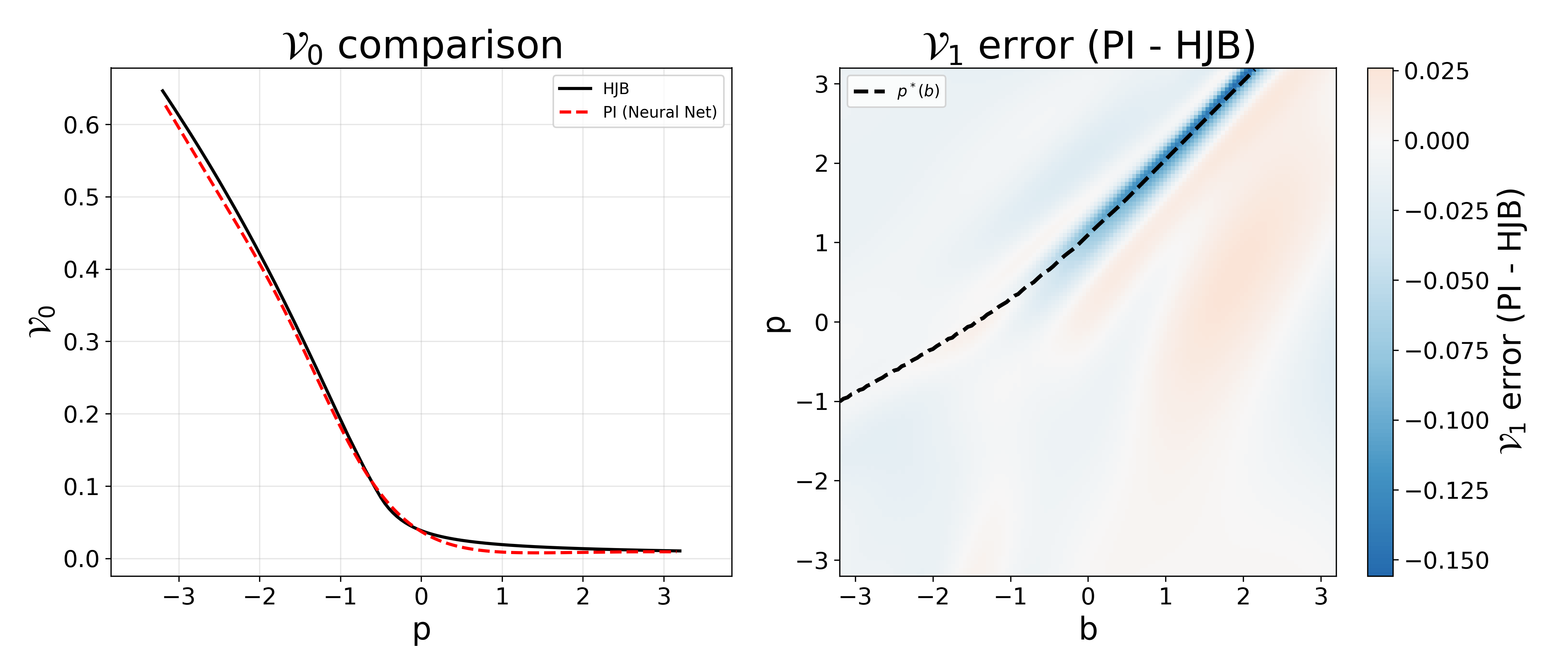}
  \caption{Comparison of value functions $\cV_0(p)$ and $\cV_1(p,b)$ between policy iteration and HJB with $M=50,\eta=10^{-5}$}
  \label{fig:PI-vs-HJB-V0-different-M-eta}
\end{figure}

In \cref{fig:PI-vs-HJB-V0-different-M-eta}, we compare the value functions $\cV_0(p)$ and $\cV_1(p,b)$ obtained from the policy iteration algorithm with the benchmark HJB results under $M=50,\eta=1\times 10^{-5}$.
The results show that the policy iteration algorithm can accurately approximate the value functions.
In the right part of the figure, we see that the error mainly appears around the free boundary $p^*(b)$, which is expected since the region is the most sensitive in most optimal stopping problems.

\section{Conclusion} \label{sec:conclusion}
In this paper, we study a speculative trading problem under a continuous-time exploratory RL framework. Starting from a general diffusion signal process and utility function, we first convert the sequential stopping problem into an optimal control problem over the (deterministic) bounded intensities processes of two Cox processes. To encourage exploration, we further allow randomized policies where the agent chooses probability measure over the intensities and the reward function is regularized by Shannon's entropy. We then derive the HJB system for the value function and the optimal exploratory control associated with the entropy-regularized problem, and we show that the value function converges to that of the original sequential stopping problem when the intensity cap and the exploratory parameters degenerate jointly in a specific manner. Our setup covers many financial applications. We illustrate the numerical implementation and study the comparative statics for a pairs-trading problem featuring OU signal process as well as Prospect Theory preference. Finally, we introduce a model-free RL algorithm which gives very satisfactory performance when benchmarked against the ground-truth solution obtained by the finite difference method.
Future work includes extending the model to multiple round-trip trades where one needs to analyze a more general optimal switching problem.

\appendix

\section{Proofs}

	\subsection{\texorpdfstring{Proof of \cref{lem:wellposedness}}{Proof of Lemma wellposedness}}
	\label{appendix:proof-wellposedness}
	\begin{proof}
		Write $$L(\tau,\nu):=\E{e^{-\rho \nu}U(\gamma P_{\nu}-\iota P_{\tau}-\varPsi-R) \mathds{1}_{\{\nu<\infty\}} \given P_0 = p}.$$
		It is clear that $V_{\text{orig}}(p)\geq 0$ since $(\tau=0,\nu=\infty)$ is admissible. Meanwhile, by the linear growth condition of $U$ (as implied by its H\"older continuity) we have
		\begin{align*}
			L(\tau,\nu)&\leq C\E{e^{-\rho \nu}(1+ |P_{\nu}|+ |P_{\tau}|) \given P_0 = p}\\
			&\leq C\left\{1+\E{e^{-\rho \nu}|P_{\nu}|\given P_0=p}+\E{e^{-\rho \tau}|P_{\tau}|\given P_0=p}\right\}
		\end{align*}
		for any admissible $(\tau,\nu)$. To establish $V_{\text{orig}}(p)< \infty$, it is sufficient to show that $\E{e^{-\rho \nu}|P_{\nu}|\given P_0=p}$ is bounded for all stopping times $\nu$.
		
		An application of Ito's lemma to $(e^{-\rho t} P_t^2)_{t\geq 0}$ gives
		\begin{align}
			e^{-\rho \nu}P^2_{t}= p^2+ \int_0^t e^{-\rho s} \left(2 P_s\mu(P_s)+\sigma^2(P_s)-\rho P_s^2\right)ds+2\int_0^t e^{-\rho s}P_s \sigma(P_s)dW_s
		\label{eq:ito}
		\end{align}
		for all $t\geq 0$. Now define
		\begin{align*}
			\mathcal{A}_2:=\{A_2>0&: \exists C>0 \text{ independent of $p$ and $t$ s.t. }
			\E{P^2_t \given P_0=p}\leq C(1+p^2)e^{A_2 t} \quad \forall t\geq 0\},
		\end{align*}
		where $\mathcal{A}_2$ is non-empty by \cref{lem:moment-estimate-sde}. Following standard localization argument, we define a sequence of stopping times $(\theta_n)_{n\in\mathbb{N}}$ via
		\begin{align*}
			\theta_n:= \inf\left\{t\geq 0: \int_0^t e^{-2\rho s}P_s^2 \sigma^2(P_s)ds \geq n\right\}.
		\end{align*}
		Then for each $n$, the stopped process $t\mapsto \int_0^{t\wedge \theta_n} e^{-\rho s}P_s \sigma(P_s)dW_s$ is a martingale. By applying optional  sampling theorem to the bounded stopping time $\nu\wedge n$, we have $\E{\int_0^{\nu \wedge \theta_n\wedge n}e^{-\rho s}P_s \sigma(P_s)dW_s}=0$. Upon setting $t=\nu\wedge \theta_n\wedge n$ in \eqref{eq:ito} followed by taking expectation on both sides,  
		\begin{align*}
			\E{e^{-\rho (\nu\wedge \theta_n\wedge n)}P^2_{\nu\wedge \theta_n\wedge n} \given P_0=p}&=p^2+\E{\int_0^{\nu\wedge \theta_n\wedge n} e^{-\rho s} \left(2 P_s\mu(P_s)+\sigma^2(P_s)-\rho P_s^2\right)ds\given P_0=p}\\
			&\leq p^2+C\E{\int_0^{\nu\wedge \theta_n\wedge n} e^{-\rho s} (1+P^2_s)ds\given P_0=p},
		\end{align*}
		where we have used the linear growth conditions of $\mu(\cdot)$ and $\sigma(\cdot)$. Since $\nu\wedge \theta_n\wedge n\uparrow \nu$ almost surely as $n\uparrow \infty$,  Fatou's lemma and monotone convergence theorem now give
		\begin{align*}
			\E{e^{-\rho \nu}P^2_{\nu}\given P_0=p}
			&\leq p^2+C\E{\int_0^{\nu} e^{-\rho s} (1+P^2_s)ds\given P_0=p}\\
			&\leq p^2+C\int_0^\infty e^{-\rho s} \E{  (1+ P_s^2) \given P_0=p} ds\\
			&\leq p^2+C\int_0^\infty e^{-\rho s}ds+C(1+p^2)\int_0^\infty e^{-(\rho -A_2)s} ds=p^2+\frac{C}{\rho}+\frac{C(1+p^2)}{\rho-A_2}
		\end{align*}
		for any $A_2\in\mathcal{A}_2$ and any stopping time $\nu$ provided that $\rho>A_2$. Hence, we conclude
		\begin{align*}
			\E{e^{-\rho \nu}|P_{\nu}|\given P_0=p}\leq \left[\E{e^{-2\rho \nu}P^2_{\nu}\given P_0=p}\right]^{\frac{1}{2}}\leq \left[p^2+\frac{C}{2\rho}+\frac{C(1+p^2)}{2\rho-A_2}\right]^{\frac{1}{2}}
		\end{align*}
		provided that $\rho>A_2/2$. Since $A_2\in\mathcal{A}_2$ can be chosen arbitrarily, we deduce $\rho>\hat{A}$ with
		\begin{align}
			\hat{A}:=\frac{\inf\mathcal{A}_2}{2}
			\label{eq:A_hat}
		\end{align} 
		is a sufficient condition under which $V_{\text{orig}}(p)< \infty$.
	\end{proof}

\subsection{\texorpdfstring{Proof of \cref{lem:bound-signal-change-exp-distributed-time}}{Proof of Lemma bound of signal change in exponentially-distributed time}} \label{appendix:proof-bound-signal-change-exp-distributed-time}
\begin{proof}
  While the techniques are standard, we illustrate a full proof completeness.
  From the dynamics of $P_t$,
  \begin{equation*}
    \E{\abs{P_{T} - P_{S}} \given \cF_S} \leq \E{ \int_S^{T} \abs{\mu(P_u)} du \given \cF_S} + \E{ \abs{\int_S^{T} \sigma(P_u) dW_u} \given \cF_S}.
  \end{equation*}
  For the first term, using linear growth of $\mu$, we have
  \begin{align*}
    \E{\int_{S}^{T} \abs{\mu(P_u)} du \given \cF_S} & \leq C \E{\int_{S}^{T} \left(1+\abs{P_u}\right) du \given \cF_S} \\
    & = C \E{\int_0^{T-S} \left(1+\abs{P_{S+t}}\right) dt \given \cF_S} \\
    & = \frac{C}{M} + C \int_0^{\infty} \E{\abs{P_{S+t}} \mathds{1}_{[0,T-S]}(t) \given \cF_S} dt,
  \end{align*}
  where Fubini's theorem is applied. From \cref{lem:moment-estimate-sde}, we get
  \begin{align*}
    \int_0^{\infty} \E{\abs{P_{S+t}} \mathds{1}_{[0,T-S]}(t) \given \cF_S} dt & \leq \int_0^{\infty} \left(\E{ \abs{P_{S+t}}^2 \given \cF_S}\right)^{1/2} \left(\E{ \mathds{1}_{[0,T-S]}(t) \given \cF_S}\right)^{1/2} dt \\
 & \leq \int_0^{\infty} \left(C_2 (1+\abs{P_S}^2)e^{A_2t}\right)^{1/2} e^{-Mt/2} dt \\
    & = \frac{C_2^{1/2} (1+\abs{P_S}^2)^{1/2}}{(M - A_2)/2} \leq \frac{4C_2^{1/2} (1+\abs{P_S}^2)^{1/2}}{M},
  \end{align*}
  provided $M \geq 2A_2$ so that $M-A_2 \geq M/2$. Therefore,
  \begin{align*}
    \E{\int_{S}^{T} \abs{\mu(P_u)} du \given \cF_S} & \leq C \frac{1 + 4C_2^{1/2} (1+\abs{P_S}^2)^{1/2}}{M} \leq C \frac{1 + \abs{P_S}}{M}.
  \end{align*}
  For the second term, using It\^o isometry, we get
  \begin{align*}
    & \E{ \abs{\int_S^{T} \sigma(P_u) dW_u} \given \cF_S} \\
    \leq & \left(\E{ \left(\int_S^{T} \sigma(P_u) dW_u\right)^2 \given \cF_S}\right)^{1/2} = \left(\E{ \int_S^{T} \sigma^2(P_u) du \given \cF_S}\right)^{1/2} \leq \left(\E{ \int_S^{T} C^2 (1+\abs{P_u})^2 du \given \cF_S}\right)^{1/2} \\
    = & C \left(\E{T-S \given \cF_S} + 2 \int_0^{\infty} \E{\abs{P_{S+t}} \mathds{1}_{[0,T-S]}(t) \given \cF_S} dt + \int_0^{\infty} \E{\abs{P_{S+t}}^2 \mathds{1}_{[0,T-S]}(t) \given \cF_S} dt \right)^{1/2}.
  \end{align*}
  We already have the estimates of the first two terms in the bracket. For the last term, from \cref{lem:moment-estimate-sde},
  \begin{align*}
    \int_0^{\infty} \E{\abs{P_{S+t}}^2 \mathds{1}_{[0,T-S]}(t) \given \cF_S} dt & \leq \int_0^{\infty} \left(\E{ \abs{P_{S+t}}^4 \given \cF_S}\right)^{1/2} \left(\E{ \mathds{1}_{[0,T-S]}(t) \given \cF_S}\right)^{1/2} dt \\
    & \leq C_4^{1/2} (1+\abs{P_S}^4)^{1/2} \int_0^{\infty} e^{(A_4/2 - M/2) t} dt \\
    & = \frac{C_4^{1/2} (1+\abs{P_S}^4)^{1/2}}{(M - A_4)/2} \leq \frac{4C_4^{1/2} (1+\abs{P_S}^4)^{1/2}}{M},
  \end{align*}
  provided $M \geq 2A_4$ so that $M-A_4 \geq M/2$. Therefore,
  \begin{align*}
    \E{ \abs{\int_S^{T} \sigma(P_u) dW_u} \given \cF_S} & \leq C \left(\frac{1 + 8C_2^{1/2} (1+\abs{P_S}^2)^{1/2} + 4C_4^{1/2} (1+\abs{P_S}^4)^{1/2}}{M}\right)^{1/2} \\
    & \leq C \frac{(1 + \abs{P_S}^2)^{1/2}}{\sqrt{M}} \leq C \frac{1 + \abs{P_S}}{\sqrt{M}}.
  \end{align*}
  Combining the two terms, we have
  \begin{equation*}
    \E{\abs{P_T - P_S} \given \cF_S} \leq C \left(\frac{1 + \abs{P_S}}{M} + \frac{1+\abs{P_S}}{\sqrt{M}}\right) \leq C \frac{1 + \abs{P_S}}{\sqrt{M}},
  \end{equation*}
  provided $M \geq 1$. Taking expectation on both sides, we obtain the desired result.
\end{proof}

\subsection{\texorpdfstring{Proof of \cref{lem:approximation-errors-modifying-controls}}{Proof of Lemma approximation errors of modifying controls}} \label{appendix:proof-approximation-errors-modifying-controls}
The proof of \cref{lem:approximation-errors-modifying-controls} is lengthy, involving a number of important and complex inequalities. At a high level, we use triangle inequalities to estimate the difference between $\widetilde J(\bm{\pi}^{\varepsilon,\eta})$ and $\widetilde J(\widetilde{\bm{\pi}}^\varepsilon)$, which results in three terms, and we will estimate each term separately in \cref{appendix:proof-estimate-expected-discounted-reward,appendix:proof-estimate-expected-discounted-state-differences,appendix:proof-estimate-I1}.

The key structural property underlying this proof is that both $\bm{\pi}^{\varepsilon,\eta}$ and $\widetilde{\bm{\pi}}^\varepsilon$ live on the same probability space, driven by the same Brownian motion $W$ and exponential variables $(E^a, E^b)$, see discussions in \cref{rem:pathwise-construction,rem:same-poisson-seeds-in-exploratory}. Their state processes differ only through the cumulative effective intensities, enabling a pathwise comparison.

We first state the following integrability results following \cref{assump:assumption-convergence-entropy-regularized-value-function}, where $r$ and $\zeta$ are defined as in \labelcref{eq:assumption-convergence-entropy-regularized-value-function} and $0<\zeta<\rho$.

  \begin{equation} \label{eq:integrability-result-2}
    \begin{aligned}
      D_r := \int_0^\infty e^{-(\rho-\zeta) t} \left(\E{\abs{P_t}^r}\right)^{1/r} dt &\leq \int_0^\infty e^{-(\rho-\zeta) t} \left(1+\E{\abs{P_t}^r}\right) dt \\
      &\leq \int_0^\infty e^{-(\rho-\zeta)t} dt + \int_0^\infty e^{-(\rho-\zeta) t} \E{\abs{P_t}^r} dt < \infty.
    \end{aligned}
  \end{equation}
	
  \begin{equation} \label{eq:integrability-result-1}
    \begin{aligned}
      F_r :=\int_0^\infty e^{-\rho t} \left(\E{\abs{P_t}^r}\right)^{1/r} t^{1-1/r}dt &= \int_0^\infty e^{-(\rho-\zeta/2) t} \left(\E{\abs{P_t}^r}\right)^{1/r} e^{-(\zeta/2) t}t^{1-1/r}dt \\
      &\leq C \int_0^\infty e^{-(\rho-\zeta/2) t} \left(\E{\abs{P_t}^r}\right)^{1/r} dt \\
      &\leq C \int_0^\infty e^{-(\rho-\zeta) t} \left(\E{\abs{P_t}^r}\right)^{1/r} dt = CD_r < \infty.
    \end{aligned}
  \end{equation}

  \begin{equation} \label{eq:integrability-result-3}
    \begin{aligned}
      E_r := \int_0^\infty e^{-(2\rho-\zeta) t} \left(\E{\abs{P_t}^{r}}\right)^{2/r} dt &\leq \int_0^\infty e^{-(2\rho-\zeta) t} \left(1+\E{\abs{P_t}^r}\right) dt \\
      &\leq \int_0^\infty e^{-(\rho-\zeta)t} dt + \int_0^\infty e^{-(\rho-\zeta) t} \E{\abs{P_t}^r} dt < \infty.
    \end{aligned}
  \end{equation}

\begin{proof}

  For $\bm{\pi}^{\varepsilon,\eta} = (\pi^{\varepsilon,\eta,\bm{\alpha}}_t, \pi^{\varepsilon,\eta,\bm{\beta}}_t)_{t \geq 0}$, define their means as $m_t^{\varepsilon,\eta,\bm{\alpha}}$ and $m_t^{\varepsilon,\eta,\bm{\beta}}$ respectively.
  By construction of $\Phi_{m,\delta}$ and \labelcref{eq:mean-error-phi}, for all $t\geq 0$,
  \begin{equation} \label{eq:mean-approximation}
    \abs{m^{\varepsilon,\eta,\bm{\alpha}}_t-\widetilde{m}_t^{\varepsilon,\bm{\alpha}}} \leq \frac{\delta(\eta) M}{2},\qquad
    \abs{m^{\varepsilon,\eta,\bm{\beta}}_t-\widetilde{m}_t^{\varepsilon,\bm{\beta}}} \leq \frac{\delta(\eta) M}{2}.
  \end{equation}
  From the definition, we have
  \begin{align*}
    & \quad \abs{\widetilde J(\bm{\pi}^{\varepsilon,\eta})-\widetilde J(\widetilde{\bm{\pi}}^\varepsilon)} \\
    &= \abs{\E{\int_0^\infty e^{-\rho t} \left(\tilde r_t^{\bm{\pi}^{\varepsilon,\eta}} - \tilde r_t^{\widetilde{\bm{\pi}}^\varepsilon}\right) dt}} \\
    &= \abs{\E{\int_0^\infty e^{-\rho t} \left(\mathds{1}_{\{J_{t-}^{\bm{\pi}^{\varepsilon,\eta}}=1\}} G(P_t, B_t^{\bm{\pi}^{\varepsilon,\eta}}) m_t^{\varepsilon,\eta,\bm{\beta}} - \mathds{1}_{\{J_{t-}^{\widetilde{\bm{\pi}}^\varepsilon}=1\}} G(P_t, B_t^{\widetilde{\bm{\pi}}^\varepsilon}) \widetilde{m}_t^{\varepsilon,\bm{\beta}}\right) dt}} \\
    &\leq \underbrace{\E{\int_0^\infty e^{-\rho t} \abs{\mathds{1}_{\{J_{t-}^{\bm{\pi}^{\varepsilon,\eta}}=1\}} - \mathds{1}_{\{J_{t-}^{\widetilde{\bm{\pi}}^\varepsilon}=1\}}} \abs{G(P_t, B_t^{\bm{\pi}^{\varepsilon,\eta}})} m_t^{\varepsilon,\eta,\bm{\beta}} dt}}_{I_1} \\
    &\quad + \underbrace{\E{\int_0^\infty e^{-\rho t} \mathds{1}_{\{J_{t-}^{\widetilde{\bm{\pi}}^\varepsilon}=1\}} \abs{G(P_t, B_t^{\bm{\pi}^{\varepsilon,\eta}}) - G(P_t, B_t^{\widetilde{\bm{\pi}}^\varepsilon})} m_t^{\varepsilon,\eta,\bm{\beta}} dt}}_{I_2} \\
    &\quad + \underbrace{\E{\int_0^\infty e^{-\rho t} \mathds{1}_{\{J_{t-}^{\widetilde{\bm{\pi}}^\varepsilon}=1\}} \abs{m_t^{\varepsilon,\eta,\bm{\beta}} - \widetilde{m}_t^{\varepsilon,\bm{\beta}}} \abs{G(P_t, B_t^{\widetilde{\bm{\pi}}^\varepsilon})} dt}}_{I_3}.
  \end{align*}
  For $I_3$, from \labelcref{eq:mean-approximation} and the following inequality (see \cref{appendix:proof-estimate-expected-discounted-reward} for the proof):
  \begin{equation} \label{eq:estimate-expected-discounted-reward}
    \E{\int_0^\infty e^{-\rho t} \mathds{1}_{\{J_{t-}^{\widetilde{\bm{\pi}}^\varepsilon}=1\}} \abs{G(P_t, B_t^{\widetilde{\bm{\pi}}^\varepsilon})} dt} \leq CM,
  \end{equation}
  we obtain
  \begin{align*}
    I_3 &\leq \frac{\delta(\eta) M}{2} \E{\int_0^\infty e^{-\rho t} \mathds{1}_{\{J_{t-}^{\widetilde{\bm{\pi}}^\varepsilon}=1\}} \abs{G(P_t, B_t^{\widetilde{\bm{\pi}}^\varepsilon})} dt} \leq C \delta(\eta) M^2.
  \end{align*}
  For $I_2$, from the global H\"older continuity of $G$ and the following inequality (see \cref{appendix:proof-estimate-expected-discounted-state-differences} for the proof):
  \begin{equation} \label{eq:estimate-expected-discounted-state-differences}
    \int_0^\infty e^{-\rho t}\E{\abs{B_t^{\bm{\pi}^{\varepsilon,\eta}} - B_t^{\widetilde{\bm{\pi}}^\varepsilon}}^\kappa} dt \leq C M \delta(\eta)^{\kappa-\kappa/r} + C (M\delta(\eta))^{\kappa/2-\kappa/r},
  \end{equation}
  we have
  \begin{align*}
    I_2 &\leq M \E{\int_0^\infty e^{-\rho t} C \abs{B_t^{\bm{\pi}^{\varepsilon,\eta}} - B_t^{\widetilde{\bm{\pi}}^\varepsilon}}^\kappa dt} \leq C M^2 \delta(\eta)^{\kappa-\kappa/r} + C M^{1+\kappa/2-\kappa/r} \delta(\eta)^{\kappa/2-\kappa/r}.
  \end{align*}
  For $I_1$, we have the following inequality (see \cref{appendix:proof-estimate-I1} for the proof):
  \begin{equation} \label{eq:estimate-I1}
    I_1 \leq CM^2 \delta(\eta)^{1-1/r}.
  \end{equation}
  Combining the estimates of $I_1$, $I_2$, and $I_3$, we conclude that
  \begin{align*}
    \abs{\widetilde J(\bm{\pi}^{\varepsilon,\eta})-\widetilde J(\widetilde{\bm{\pi}}^\varepsilon)}
    &\leq CM^2 \delta(\eta)^{1-1/r} + C M^2 \delta(\eta)^{\kappa-\kappa/r} + CM^{1+\kappa/2-\kappa/r} \delta(\eta)^{\kappa/2-\kappa/r} + CM^2 \delta(\eta).
  \end{align*}
  Since $1>1-1/r \geq \kappa-\kappa/r$,
  when $\delta(\eta) \leq 1$, the bound can be simplified to $C M^2 \delta(\eta)^{\kappa-\kappa/r} + CM^{1+\kappa/2-\kappa/r} \delta(\eta)^{\kappa/2-\kappa/r}$.
\end{proof}

\subsubsection{\texorpdfstring{Proof of \labelcref{eq:estimate-expected-discounted-reward}}{Proof of inequality estimate-expected-discounted-reward}} \label{appendix:proof-estimate-expected-discounted-reward}
\begin{proof}
  By the growth assumptions on $U$ and definition of $G$, there exists $C>0$ such that
  \begin{align*}
    \abs{G(P_t, B_t^{\widetilde{\bm{\pi}}^\varepsilon})} &\leq C \left(1 + \abs{P_t} + \abs{B_t^{\widetilde{\bm{\pi}}^\varepsilon}}\right).
  \end{align*}
  Therefore,
  \begin{align*}
    & \E{\int_0^\infty e^{-\rho t} \mathds{1}_{\{J_{t-}^{\widetilde{\bm{\pi}}^\varepsilon}=1\}} \abs{G(P_t, B_t^{\widetilde{\bm{\pi}}^\varepsilon})} dt} \\
    \leq& \underbrace{C \E{\int_0^\infty e^{-\rho t} \mathds{1}_{\{J_{t-}^{\widetilde{\bm{\pi}}^\varepsilon}=1\}} dt}}_{I_{3.1}} + \underbrace{C \E{\int_0^\infty e^{-\rho t} \mathds{1}_{\{J_{t-}^{\widetilde{\bm{\pi}}^\varepsilon}=1\}} \abs{P_t} dt}}_{I_{3.2}} + \underbrace{C \E{\int_0^\infty e^{-\rho t} \mathds{1}_{\{J_{t-}^{\widetilde{\bm{\pi}}^\varepsilon}=1\}} \abs{B_t^{\widetilde{\bm{\pi}}^\varepsilon}} dt}}_{I_{3.3}}.
  \end{align*}
  For the first term, since $\rho>0$, $I_{3.1} \leq C\int_0^\infty e^{-\rho t} dt = C/\rho < \infty$.
  For the second term, we have
  \begin{align*}
    I_{3.2} &\leq C \E{\int_0^\infty e^{-\rho t} \abs{P_t} dt} 
    \leq C \int_0^\infty e^{-\rho t} \left(\E{\abs{P_t}^r}\right)^{1/r} dt 
    \leq C D_r < \infty,
  \end{align*}
  where we used the integrability result \labelcref{eq:integrability-result-2}.
  Let $\tilde\tau^\varepsilon$ and $\tilde\nu^\varepsilon$ be the entry and exit times driven by average intensities $(\widetilde{m}_t^{\varepsilon,\bm{\alpha}})_{t\ge0}$ and $(\widetilde{m}_t^{\varepsilon,\bm{\beta}})_{t\ge0}$.
  For the third term,
  \begin{align*}
    I_{3.3} &= C\E{\abs{P_{\tilde\tau^\varepsilon}} \int_{\tilde\tau^\varepsilon}^{\tilde\nu^\varepsilon}e^{-\rho t}dt} \leq C\frac{1}{\rho}\E{\abs{P_{\tilde\tau^\varepsilon}} e^{-\rho\tilde\tau^\varepsilon}} \\
    &= C\frac{1}{\rho}\E{\int_0^\infty e^{-\rho t}\abs{P_t}\widetilde{m}_t^{\varepsilon,\bm{\alpha}}\exp{\left(-\int_0^t \widetilde{m}_s^{\varepsilon,\bm{\alpha}} ds\right)} dt} \\
    &\leq C \frac{1}{\rho}M\E{\int_0^\infty e^{-\rho t}\abs{P_t} dt} \leq C \frac{1}{\rho}M < \infty,
  \end{align*}
  where we used the bound for $I_{3.2}$ and
  $ \Prob{\tilde\tau^\varepsilon \in dt \given \cF_t} = \widetilde{m}_t^{\varepsilon,\bm{\alpha}} \exp{\left(-\int_0^t \widetilde{m}_s^{\varepsilon,\bm{\alpha}} ds\right)} dt$.
  Combining the three estimates, we conclude that
  \begin{align*}
    \E{\int_0^\infty e^{-\rho t} \mathds{1}_{\{J_{t-}^{\widetilde{\bm{\pi}}^\varepsilon}=1\}} \abs{G(P_t, B_t^{\widetilde{\bm{\pi}}^\varepsilon})} dt} &\leq I_{3.1} + I_{3.2} + I_{3.3} \leq C+CM,
  \end{align*}
  for some finite constant $C$ depending only on $(\rho,\mu,\sigma,G)$ and the initial condition. When $M\geq1$, the bound simplifies to $CM$.
\end{proof}

\subsubsection{\texorpdfstring{Proof of \labelcref{eq:estimate-expected-discounted-state-differences}}{Proof of inequality estimate-expected-discounted-state-differences}} \label{appendix:proof-estimate-expected-discounted-state-differences}
\begin{proof}
  Let $\tilde\tau^\varepsilon$ and $\tau^{\varepsilon,\eta}$ be the entry times driven by average intensities $(\widetilde{m}_t^{\varepsilon,\bm{\alpha}})_{t\ge0}$ and $(m_t^{\eta,\varepsilon,\bm{\alpha}})_{t\ge0}$, respectively.
  Since both $\tilde\tau^\varepsilon$ and $\tau^{\varepsilon,\eta}$ are constructed from the same exponential variables $E^a$ via different cumulative effective intensities, and $E^a$ is independent of $\cF_t$, we have,
  for any $t\geq 0$, 
  \begin{align*}
    \Prob{\tau^{\varepsilon,\eta}\leq t < \tilde\tau^\varepsilon \given \cF_t} & = \Prob{\int_0^t \widetilde{m}_s^{\varepsilon,\bm{\alpha}} ds < E^a \leq \int_0^t m_s^{\varepsilon,\eta,\bm{\alpha}} ds \given \cF_t} \\
  & = \left(\exp(-\int_0^t \widetilde{m}_s^{\varepsilon,\bm{\alpha}} ds) - \exp(-\int_0^t m_s^{\varepsilon,\eta,\bm{\alpha}} ds)\right)^+ \\
  & \leq \abs{\int_0^t \widetilde{m}_s^{\varepsilon,\bm{\alpha}} - m_s^{\varepsilon,\eta,\bm{\alpha}} ds} \leq \frac{M}{2}\delta(\eta) t.
  \end{align*}
  The same argument applies to $\Prob{\tilde\tau^\varepsilon\leq t < \tau^{\varepsilon,\eta} \given \cF_t}$.
  Since
  \begin{align*}
    B_t^{\bm{\pi}^{\varepsilon,\eta}} - B_t^{\widetilde{\bm{\pi}}^\varepsilon}
     = & P_{\tau^{\varepsilon,\eta}} \mathds{1}_{\{\tau^{\varepsilon,\eta}\leq t\}} - P_{\tilde\tau^\varepsilon} \mathds{1}_{\{\tilde\tau^\varepsilon\leq t\}} \\
    = & P_{\tau^{\varepsilon,\eta}} \mathds{1}_{\{\tau^{\varepsilon,\eta} \leq t < \tilde\tau^\varepsilon\}} - P_{\tilde\tau^\varepsilon} \mathds{1}_{\{\tilde\tau^\varepsilon \leq t < \tau^{\varepsilon,\eta}\}} + (P_{\tau^{\varepsilon,\eta}} - P_{\tilde\tau^\varepsilon}) \mathds{1}_{\{\tau^{\varepsilon,\eta} \vee \tilde\tau^\varepsilon \leq t\}},
  \end{align*}
  we have
  \begin{align*}
  &\int_0^\infty e^{-\rho t}\E{\abs{B_t^{\bm{\pi}^{\varepsilon,\eta}} - B_t^{\widetilde{\bm{\pi}}^\varepsilon}}} dt \\
  \leq & \underbrace{\int_0^\infty e^{-\rho t} \E{\abs{P_{\tau^{\varepsilon,\eta}} \mathds{1}_{\{\tau^{\varepsilon,\eta} \leq t < \tilde\tau^\varepsilon\}}}} dt}_{I_{2.1}} + \underbrace{\int_0^\infty e^{-\rho t} \E{\abs{P_{\tilde\tau^\varepsilon} \mathds{1}_{\{\tilde\tau^\varepsilon \leq t < \tau^{\varepsilon,\eta}\}}}} dt}_{I_{2.2}} \\
  & + \underbrace{\int_0^\infty e^{-\rho t} \E{\abs{P_{\tau^{\varepsilon,\eta}} - P_{\tilde\tau^\varepsilon}} \mathds{1}_{\{\tau^{\varepsilon,\eta} \vee \tilde\tau^\varepsilon \leq t\}}} dt}_{I_{2.3}}.
  \end{align*}
  We can bound the first two terms using the earlier probability estimates. Take the first term as an example,

  \begin{align*}
    I_{2.1} &= \int_0^\infty e^{-\rho t} \E{\abs{P_{\tau^{\varepsilon,\eta}} \mathds{1}_{\{\tau^{\varepsilon,\eta} \leq t \}} \mathds{1}_{\{\tau^{\varepsilon,\eta} \leq t < \tilde\tau^\varepsilon\}}}} dt \\
    &\leq \int_0^\infty e^{-\rho t} \left(\E{\abs{P_{\tau^{\varepsilon,\eta}}}^r \mathds{1}_{\{\tau^{\varepsilon,\eta} \leq t \}}}\right)^{1/r} \left(\Prob{\tau^{\varepsilon,\eta} \leq t < \tilde\tau^\varepsilon}\right)^{1-1/r} dt \\
    &\leq C \int_0^\infty e^{-\rho t} \left(\int_0^t \E{\abs{P_s}^r m_s^{\varepsilon,\eta,\bm{\alpha}} \exp{\left(-\int_0^s m_u^{\varepsilon,\eta,\bm{\alpha}} du\right)}} ds \right)^{1/r} \left(M\delta(\eta) t\right)^{1-1/r} dt \\
    &\leq C \int_0^\infty e^{-\rho t} \left( e^{(\rho-\xi) t}\int_0^t e^{-(\rho-\xi) s} \E{\abs{P_s}^r} ds\right)^{1/r} M^{1/r} \left(M\delta(\eta) t\right)^{1-1/r} dt \\
    &\leq CK_r^{1/r} M \delta(\eta)^{1-1/r} \int_0^\infty e^{-\rho t} e^{(\rho-\xi)t/r} t^{1-1/r} dt \leq C M \delta(\eta)^{1-1/r},
  \end{align*}
  where we used $\Prob{\tau^{\varepsilon,\eta} \in dt \given \cF_t} = m_t^{\varepsilon,\eta,\bm{\alpha}} \exp{\left(-\int_0^t m_s^{\varepsilon,\eta,\bm{\alpha}} ds\right)} dt$. Similarly, $I_{2.2} \leq C M \delta(\eta)^{1-1/r}$.
  For $I_{2.3}$,
  \begin{align*}
    I_{2.3} &= \E{\int_{\tau^{\varepsilon,\eta} \vee \tilde\tau^\varepsilon}^\infty e^{-\rho t} \abs{P_{\tau^{\varepsilon,\eta}} - P_{\tilde\tau^\varepsilon}} dt} \\
    &= \frac{1}{\rho} \E{e^{-\rho (\tau^{\varepsilon,\eta} \vee \tilde\tau^\varepsilon)} \abs{P_{\tau^{\varepsilon,\eta}} - P_{\tilde\tau^\varepsilon}}} \\
    &\leq \underbrace{\frac{1}{\rho} \E{e^{-\rho (\tau^{\varepsilon,\eta} \vee \tilde\tau^\varepsilon)} \abs{\int_{\tau^{\varepsilon,\eta} \wedge \tilde\tau^\varepsilon}^{\tau^{\varepsilon,\eta} \vee \tilde\tau^\varepsilon} \mu(P_t)dt}}}_{I_{2.3.1}} + \underbrace{\frac{1}{\rho} \E{e^{-\rho (\tau^{\varepsilon,\eta} \vee \tilde\tau^\varepsilon)} \abs{\int_{\tau^{\varepsilon,\eta} \wedge \tilde\tau^\varepsilon}^{\tau^{\varepsilon,\eta} \vee \tilde\tau^\varepsilon} \sigma(P_t)dW_t}}}_{I_{2.3.2}}.
  \end{align*}
  Denote $A_t:=\{\tau^{\varepsilon,\eta}\leq t<\tilde\tau^\varepsilon\} \cup \{\tilde\tau^\varepsilon\leq t<\tau^{\varepsilon,\eta}\}$. We have $\Prob{A_t}\leq M\delta(\eta)t$.
  For $I_{2.3.1}$,
  \begin{align*}
    I_{2.3.1} &\leq \frac{1}{\rho} \E{\int_{\tau^{\varepsilon,\eta} \wedge \tilde\tau^\varepsilon}^{\tau^{\varepsilon,\eta} \vee \tilde\tau^\varepsilon} e^{-\rho t}\abs{\mu(P_t)}dt} \\
    & = \frac{1}{\rho} \E{\int_0^\infty e^{-\rho t} \abs{\mu(P_t)} \mathds{1}_{A_t} dt} \\
    &\leq \frac{1}{\rho}\int_0^\infty e^{-\rho t} \left(\E{\abs{\mu(P_t)}^r}\right)^{1/r} \left(\Prob{A_t}\right)^{1-1/r} dt \\
    &\leq C \frac{1}{\rho} \int_0^\infty e^{-\rho t} \left(1+\E{\abs{P_t}^r}^{1/r}\right) \left(M\delta(\eta)t\right)^{1-1/r} dt \\
    & \leq C\frac{1}{\rho}\left(M\delta(\eta)\right)^{1-1/r}\left(\int_0^\infty e^{-\rho t} t^{1-1/r} dt + D_r \sup_{t\geq0}(t^{1-1/r}e^{-\zeta t})\right) \\
    &\leq C \left(M\delta(\eta)\right)^{1-1/r},
  \end{align*}
  where we used the integrability result \labelcref{eq:integrability-result-2}.
  For $I_{2.3.2}$,
  consider $Y_t :=
  \int_0^t \sigma(P_s) \mathds{1}_{A_s} dW_s$. We have $Y_{\tau^{\varepsilon,\eta} \wedge \tilde\tau^\varepsilon} = 0$ and $Y_{\tau^{\varepsilon,\eta} \vee \tilde\tau^\varepsilon} = \int_{\tau^{\varepsilon,\eta} \wedge \tilde\tau^\varepsilon}^{\tau^{\varepsilon,\eta} \vee \tilde\tau^\varepsilon} \sigma(P_s) dW_s$.
  Applying It\^o's lemma on $e^{-2\rho t} Y_t^2$,
  \[
  d(e^{-2\rho t} Y_t^2) = -2\rho e^{-2\rho t}Y_t^2 dt + e^{-2\rho t} \sigma(P_t)^2 \mathds{1}_{A_t} dt + 2e^{-2\rho t} Y_t \sigma(P_t)\mathds{1}_{A_t} dW_t.
  \]
  Using standard localization argument, Fatou's Lemma and monotone convergence as in the proof of \cref{lem:wellposedness}, while integrating from $0$ to $\tau^{\varepsilon,\eta} \vee \tilde\tau^\varepsilon$ and taking expectation,
  \[
  \E{e^{-2\rho (\tau^{\varepsilon,\eta} \vee \tilde\tau^\varepsilon)} Y_{\tau^{\varepsilon,\eta} \vee \tilde\tau^\varepsilon}^2} = \E{e^{- 2\rho (\tau^{\varepsilon,\eta} \vee \tilde\tau^\varepsilon)} \left(\int_{\tau^{\varepsilon,\eta} \wedge \tilde\tau^\varepsilon}^{\tau^{\varepsilon,\eta} \vee \tilde\tau^\varepsilon} \sigma(P_t)dW_t\right)^2} \leq \E{\int_{\tau^{\varepsilon,\eta} \wedge \tilde\tau^\varepsilon}^{\tau^{\varepsilon,\eta} \vee \tilde\tau^\varepsilon} e^{-2\rho t}\sigma(P_t)^2 dt}.
  \]
  Therefore,
  \begin{align*}
    I_{2.3.2} &\leq \frac{1}{\rho} \left(\E{e^{- 2\rho (\tau^{\varepsilon,\eta} \vee \tilde\tau^\varepsilon)} \left(\int_{\tau^{\varepsilon,\eta} \wedge \tilde\tau^\varepsilon}^{\tau^{\varepsilon,\eta} \vee \tilde\tau^\varepsilon} \sigma(P_t)dW_t\right)^2}\right)^{1/2} \\
    &\leq \frac{1}{\rho} \left(\E{ \int_{\tau^{\varepsilon,\eta} \wedge \tilde\tau^\varepsilon}^{\tau^{\varepsilon,\eta} \vee \tilde\tau^\varepsilon} e^{-2\rho t} \sigma(P_t)^2 dt}\right)^{1/2} \\
    &= \frac{1}{\rho} \left(\E{ \int_0^\infty e^{-2\rho t} \sigma(P_t)^2 \mathds{1}_{A_t} dt}\right)^{1/2} \\
    & \leq \frac{1}{\rho} \left(\int_0^\infty e^{-2\rho t} \left(\E{\abs{\sigma(P_t)}^{r}}\right)^{2/r} \left(\Prob{A_t}\right)^{1-2/r} dt\right)^{1/2} \\
    & \leq C \frac{1}{\rho} \left(\int_0^\infty e^{-2\rho t} \left(1+\left(\E{\abs{P_t}^{r}}\right)^{2/r}\right) \left(M\delta(\eta)t\right)^{1-2/r} dt\right)^{1/2} \\
    & \leq C \frac{1}{\rho}\left(M\delta(\eta)\right)^{1/2-1/r} \left(\int_0^\infty e^{-2\rho t} t^{1-2/r} dt + E_{r}\sup_{t\geq0}(t^{1-2/r}e^{-\zeta t})\right)^{1/2} \\
    & \leq C \left(M\delta(\eta)\right)^{1/2-1/r},
  \end{align*}
  where we used the integrability result \labelcref{eq:integrability-result-3}.
  Combining the estimates for $I_{2.3.1}$ and $I_{2.3.2}$, we have $I_{2.3} \leq C (M\delta(\eta))^{1/2-1/r} + C (M\delta(\eta))^{1-1/r}$.
  Combining the estimates for the three terms, we obtain
  \begin{align*}
    \int_0^\infty e^{-\rho t}\E{\abs{B_t^{\bm{\pi}^{\varepsilon,\eta}} - B_t^{\widetilde{\bm{\pi}}^\varepsilon}}} dt \leq C M \delta(\eta)^{1-1/r} + C (M\delta(\eta))^{1/2-1/r} + C (M\delta(\eta))^{1-1/r}.
  \end{align*}
  Apply Jensen's inequality on $(\Omega,\bP)$ and $([0,\infty),\rho e^{-\rho t} dt)$,
  \begin{align*}
    \int_0^\infty e^{-\rho t}\E{\abs{B_t^{\bm{\pi}^{\varepsilon,\eta}} - B_t^{\widetilde{\bm{\pi}}^\varepsilon}}^\kappa} dt &\leq \rho^{\kappa-1}\left(\int_0^\infty e^{-\rho t}\E{\abs{B_t^{\bm{\pi}^{\varepsilon,\eta}} - B_t^{\widetilde{\bm{\pi}}^\varepsilon}}} dt\right)^\kappa \\
    & \leq C M \delta(\eta)^{\kappa-\kappa/r} + C (M\delta(\eta))^{\kappa/2-\kappa/r} + C (M\delta(\eta))^{\kappa-\kappa/r}.
  \end{align*}
  When $M\delta(\eta) \leq 1$, the bound simplifies to $C M \delta(\eta)^{\kappa-\kappa/r} + C (M\delta(\eta))^{\kappa/2-\kappa/r}$.

\end{proof}

\subsubsection{\texorpdfstring{Proof of \labelcref{eq:estimate-I1}}{Proof of inequality estimate-I1}} \label{appendix:proof-estimate-I1}
\begin{proof}
  $I_1$ can be estimated as
  \begin{align*}
    I_1 &\leq M\int_0^\infty e^{-\rho t}\E{\abs{G(P_t,B_t^{\bm{\pi}^{\varepsilon,\eta}})}\mathds{1}_{\{J_{t-}^{\bm{\pi}^{\varepsilon,\eta}} \neq J_{t-}^{\widetilde{\bm{\pi}}^\varepsilon}\}}} dt \\
    &\leq M\int_0^\infty e^{-\rho t} \left(\E{\abs{G(P_t,B_t^{\bm{\pi}^{\varepsilon,\eta}})}^{r}}\right)^{1/r} \left(\Prob{J_{t-}^{\bm{\pi}^{\varepsilon,\eta}} \neq J_{t-}^{\widetilde{\bm{\pi}}^\varepsilon}}\right)^{1-1/r} dt,
  \end{align*}
  where we used the fact that $\abs{\mathds{1}_{\{J_{t-}^{\bm{\pi}^{\varepsilon,\eta}}=1\}} - \mathds{1}_{\{J_{t-}^{\widetilde{\bm{\pi}}^\varepsilon}=1\}}} \leq \mathds{1}_{\{J_{t-}^{\bm{\pi}^{\varepsilon,\eta}}\neq J_{t-}^{\widetilde{\bm{\pi}}^\varepsilon}\}}$.
  We have the estimate that
  \begin{align} \label{eq:estimate-G-r}
    \left(\E{\abs{G(P_t,B_t^{\bm{\pi}^{\varepsilon,\eta}})}^{r}}\right)^{1/r} &\leq C \left(1 + \left(\E{\abs{P_t}^{r}}\right)^{1/r} + \left(\E{\abs{B_t^{\bm{\pi}^{\varepsilon,\eta}}}^{r}}\right)^{1/r}\right).
  \end{align}
  The following mismatch probability quantifies how the common-seed setup discussed in \cref{rem:pathwise-construction,rem:same-poisson-seeds-in-exploratory} limits the probability that their regime indicators disagree (see \cref{appendix:proof-mismatch-probability} for the proof):
  \begin{align} \label{eq:mismatch-probability}
    \Prob{J_{t-}^{\bm{\pi}^{\varepsilon,\eta}} \neq J_{t-}^{\widetilde{\bm{\pi}}^\varepsilon}} \leq M\delta(\eta)t.
  \end{align}
  Using \labelcref{eq:estimate-G-r} and \labelcref{eq:mismatch-probability}, we have
  \begin{align*}
    I_1 
    &\leq \underbrace{C M^{2-1/r} \delta(\eta)^{1-1/r} \int_0^\infty e^{-\rho t} t^{1-1/r} dt}_{I_{1.1}} \\
    &\quad + \underbrace{C M^{2-1/r} \delta(\eta)^{1-1/r} \int_0^\infty e^{-\rho t} \left(\E{\abs{P_t}^{r}}\right)^{1/r} t^{1-1/r} dt}_{I_{1.2}} \\
    &\quad + \underbrace{C M^{2-1/r} \delta(\eta)^{1-1/r} \int_0^\infty e^{-\rho t} \left(\E{\abs{B_t^{\bm{\pi}^{\varepsilon,\eta}}}^{r}}\right)^{1/r} t^{1-1/r} dt}_{I_{1.3}}.
  \end{align*}
  Since $\rho>0$, we have $I_{1.1} \leq CM^{2-1/r} \delta(\eta)^{1-1/r}$.
  Using \labelcref{eq:integrability-result-1}, $I_{1.2} \leq C M^{2-1/r} \delta(\eta)^{1-1/r}$.
  For $I_{1.3}$,
  \begin{align*}
    I_{1.3} &\leq C M^{2-1/r} \delta(\eta)^{1-1/r} \left(\int_0^\infty e^{-\rho t}\E{\abs{B_t^{\bm{\pi}^{\varepsilon,\eta}}}^{r}} dt\right)^{1/r} \left(\int_0^\infty e^{-\rho t}t dt\right)^{1-1/r} \\
    &\leq C M^{2-1/r} \delta(\eta)^{1-1/r} \left(\int_0^\infty e^{-\rho t}\E{\abs{P_{\tau^{\varepsilon,\eta}}}^{r}\mathds{1}_{\{\tau^{\varepsilon,\eta}\leq t\}}} dt\right)^{1/r} \\
    &= C M^{2-1/r} \delta(\eta)^{1-1/r} \left(\frac{1}{\rho}\E{\abs{P_{\tau^{\varepsilon,\eta}}}^{r} e^{-\rho \tau^{\varepsilon,\eta}}}\right)^{1/r} \\
    &\leq C M^{2-1/r} \delta(\eta)^{1-1/r} \left(\frac{M}{\rho}\int_0^\infty e^{-\rho t}\E{\abs{P_t}^{r}} dt\right)^{1/r} \\
    &\leq C M^{2-1/r} \delta(\eta)^{1-1/r} \left(\frac{M}{\rho}K_r \right)^{1/r} \leq C M^2 \delta(\eta)^{1-1/r},
  \end{align*}
  where we used $K_r$ in \labelcref{eq:assumption-convergence-entropy-regularized-value-function}.
  Combining the estimates for $I_{1.1}$, $I_{1.2}$, and $I_{1.3}$, we have $I_1 \leq CM^{2-1/r}\delta(\eta)^{1-1/r}+CM^2\delta(\eta)^{1-1/r}$. When $M \geq 1$, the bound simplifies to $I_1 \leq CM^2 \delta(\eta)^{1-1/r}$.
\end{proof}

\subsubsection{\texorpdfstring{Proof of \labelcref{eq:mismatch-probability}}{Proof of inequality mismatch-probability}} \label{appendix:proof-mismatch-probability}
\begin{proof}
  Since both controls operate on the same probability space, the regime processes $J_t^{\bm{\pi}^{\varepsilon,\eta}}$ and $J_t^{\widetilde{\bm{\pi}}^\varepsilon}$ are driven by the same exponential variables $E^a$ and $E^b$, see discussions in \cref{rem:pathwise-construction,rem:same-poisson-seeds-in-exploratory}.
  The entry (resp. exit) events of each process are determined by whether $E^a$ (resp. $E^b$) has been exceeded by the respective cumulative effective intensity. A regime mismatch can only happen when $E^a$ (resp. $E^b$) falls in the gap between two cumulative effective intensities.
  Following \labelcref{eq:cumulative-hazard-randomized}, denote their cumulative effective intensities as
  \begin{align*}
    \bar{\Lambda}_t^{\bm{\alpha},1} &:= \int_0^t \int_\bM \lambda \pi_s^{\varepsilon,\eta,\bm{\alpha}}(\lambda) d\lambda \mathds{1}_{\{J_{s-}^{\bm{\pi}^{\varepsilon,\eta}}=0\}} ds, \quad \bar{\Lambda}_t^{\bm{\beta},1} := \int_0^t \int_\bM \lambda \pi_s^{\varepsilon,\eta,\bm{\beta}}(\lambda) d\lambda \mathds{1}_{\{J_{s-}^{\bm{\pi}^{\varepsilon,\eta}}=1\}} ds, \\
    \bar{\Lambda}_t^{\bm{\alpha},2} &:= \int_0^t \int_\bM \lambda \widetilde{\pi}^{\varepsilon,\bm{\alpha}}_s(\lambda) d\lambda \mathds{1}_{\{J_{s-}^{\widetilde{\bm{\pi}}^\varepsilon}=0\}} ds, \quad \bar{\Lambda}_t^{\bm{\beta},2} := \int_0^t \int_\bM \lambda \widetilde{\pi}^{\varepsilon,\bm{\beta}}_s(\lambda) d\lambda \mathds{1}_{\{J_{s-}^{\widetilde{\bm{\pi}}^\varepsilon}=1\}} ds.
  \end{align*}
  Denote the maximum and minimum
  between $J_t^{\bm{\pi}^{\varepsilon,\eta}}$ and $J_t^{\widetilde{\bm{\pi}}^\varepsilon}$ as
  \begin{align*}
    J_t^{\max} := \max\{J_t^{\bm{\pi}^{\varepsilon,\eta}}, J_t^{\widetilde{\bm{\pi}}^\varepsilon}\}, \quad J_t^{\min} := \min\{J_t^{\bm{\pi}^{\varepsilon,\eta}}, J_t^{\widetilde{\bm{\pi}}^\varepsilon}\}.
  \end{align*}
  Similarly, for the cumulative effective intensities, denote
  \begin{align*}
    \bar{\Lambda}^{\bm{\alpha},\max}_t := \max\{\bar{\Lambda}_t^{\bm{\alpha},1}, \bar{\Lambda}_t^{\bm{\alpha},2}\}, \quad \bar{\Lambda}^{\bm{\alpha},\min}_t := \min\{\bar{\Lambda}_t^{\bm{\alpha},1}, \bar{\Lambda}_t^{\bm{\alpha},2}\}, \\
    \bar{\Lambda}^{\bm{\beta},\max}_t := \max\{\bar{\Lambda}_t^{\bm{\beta},1}, \bar{\Lambda}_t^{\bm{\beta},2}\}, \quad \bar{\Lambda}^{\bm{\beta},\min}_t := \min\{\bar{\Lambda}_t^{\bm{\beta},1}, \bar{\Lambda}_t^{\bm{\beta},2}\}.
  \end{align*}
  One can get
$   \{J_{t-}^{\bm{\pi}^{\varepsilon,\eta}} \neq J_{t-}^{\widetilde{\bm{\pi}}^\varepsilon}\} = \{J_{t-}^{\min}=0, J_{t-}^{\max}\geq 1\} \cup \{J_{t-}^{\min}=1, J_{t-}^{\max}=2\}$.
  We have the following bounds:
  \begin{align*}
    \Prob{J_{t-}^{\min}=0, J_{t-}^{\max}\geq 1} &= \Prob{\bar{\Lambda}^{\bm{\alpha},\min}_t < E^a \leq \bar{\Lambda}^{\bm{\alpha},\max}_t} \\
    &= \E{\Prob{\bar{\Lambda}^{\bm{\alpha},\min}_t < E^a \leq \bar{\Lambda}^{\bm{\alpha},\max}_t \given \cF_t}} \\
    &= \E{\exp(-\bar{\Lambda}^{\bm{\alpha},\min}_t) - \exp(-\bar{\Lambda}^{\bm{\alpha},\max}_t)} \\
    &\leq \E{\abs{\bar{\Lambda}^{\bm{\alpha},\max}_t - \bar{\Lambda}^{\bm{\alpha},\min}_t}} = \E{\abs{\bar{\Lambda}_t^{\bm{\alpha},1} - \bar{\Lambda}_t^{\bm{\alpha},2}}} \\
    &\leq \E{\int_0^t \abs{m_s^{\varepsilon,\eta,\bm{\alpha}} - \widetilde{m}_s^{\varepsilon,\bm{\alpha}}} ds} \leq \frac{M \delta(\eta)}{2} t,
  \end{align*}
  where $\E{\abs{\bar{\Lambda}_t^{\bm{\alpha},1} - \bar{\Lambda}_t^{\bm{\alpha},2}}} \leq \E{\int_0^t \abs{m_s^{\varepsilon,\eta,\bm{\alpha}} - \widetilde{m}_s^{\varepsilon,\bm{\alpha}}} ds}$ follows from the four cases:
  \begin{enumerate}
    \item If $J_{t-}^{\bm{\pi}^{\varepsilon,\eta}}=0$ and $J_{t-}^{\widetilde{\bm{\pi}}^\varepsilon}=0$, it is obvious.
    \item If $J_{t-}^{\bm{\pi}^{\varepsilon,\eta}}=1$ and $J_{t-}^{\widetilde{\bm{\pi}}^\varepsilon}=1$, both $\bar{\Lambda}_t^{\bm{\alpha},1}$ and $\bar{\Lambda}_t^{\bm{\alpha},2}$ are equal to the exponential variable $E^a$.
    \item If $J_{t-}^{\bm{\pi}^{\varepsilon,\eta}}=0$ and $J_{t-}^{\widetilde{\bm{\pi}}^\varepsilon}=1$, then $\bar{\Lambda}_t^{\bm{\alpha},1} < E^a \leq \bar{\Lambda}_t^{\bm{\alpha},2}$. Thus, $\abs{\bar{\Lambda}_t^{\bm{\alpha},1} - \bar{\Lambda}_t^{\bm{\alpha},2}} = \bar{\Lambda}_t^{\bm{\alpha},2} - \bar{\Lambda}_t^{\bm{\alpha},1} = \bar{\Lambda}_t^{\bm{\alpha},2} - \int_0^t m_s^{\varepsilon,\eta,\bm{\alpha}} ds \leq \int_0^t \widetilde{m}_s^{\varepsilon,\bm{\alpha}} ds - \int_0^t m_s^{\varepsilon,\eta,\bm{\alpha}} ds \leq \int_0^t \abs{m_s^{\varepsilon,\eta,\bm{\alpha}} - \widetilde{m}_s^{\varepsilon,\bm{\alpha}}} ds$.
    \item If $J_{t-}^{\bm{\pi}^{\varepsilon,\eta}}=1$ and $J_{t-}^{\widetilde{\bm{\pi}}^\varepsilon}=0$, then $\bar{\Lambda}_t^{\bm{\alpha},2} < E^a \leq \bar{\Lambda}_t^{\bm{\alpha},1}$. Thus, $\abs{\bar{\Lambda}_t^{\bm{\alpha},1} - \bar{\Lambda}_t^{\bm{\alpha},2}} = \bar{\Lambda}_t^{\bm{\alpha},1} - \bar{\Lambda}_t^{\bm{\alpha},2} = \bar{\Lambda}_t^{\bm{\alpha},1} - \int_0^t \widetilde{m}_s^{\varepsilon,\bm{\alpha}} ds \leq \int_0^t m_s^{\varepsilon,\eta,\bm{\alpha}} ds - \int_0^t \widetilde{m}_s^{\varepsilon,\bm{\alpha}} ds \leq \int_0^t \abs{m_s^{\varepsilon,\eta,\bm{\alpha}} - \widetilde{m}_s^{\varepsilon,\bm{\alpha}}} ds$.
  \end{enumerate}

  Similarly, $\Prob{J_{t-}^{\min}=1, J_{t-}^{\max}=2} \leq \Prob{\bar{\Lambda}^{\bm{\beta},\min}_t < E^b \leq \bar{\Lambda}^{\bm{\beta},\max}_t} \leq \frac{M \delta(\eta)}{2} t.$
  Combining the two bounds, we have $\Prob{J_{t-}^{\bm{\pi}^{\varepsilon,\eta}} \neq J_{t-}^{\widetilde{\bm{\pi}}^\varepsilon}} \leq M \delta(\eta) t$.
\end{proof}

\section{Algorithm} \label{appendix:algorithms}

\begin{algorithm}[H]
  \caption{Offline Policy Iteration}
  \label{alg:offline-policy-iteration-OU-S-shaped-utility}
  \textbf{Require:} Offline data $\{P_{t_l}^n\}_{l=0,\dots,L}^{n=1,\dots,N}$, parameters $(\rho,\theta,\bar p,\sigma,M,\eta)$, initial value function parameters $\Theta^{(0)}_0$, $\Theta^{(0)}_1$, learning rate $\texttt{lr}$, batch size $M_{\text{batch}}$, time step size $\Delta t$, max iterations $K_{\max}$, simulations per path $I$.\\
  \textbf{Initialize:} value function approximators $\cV_0(p;\Theta^{(0)}_0)$ and $\cV_1(p,b;\Theta^{(0)}_1)$.\\
  \textbf{For} $k=0,1,\dots,K_{\max}-1$ \textbf{do}:
  \begin{enumerate}
    \item Simulate $\{J_{t_l}^{n,i,k}\}_{l=0,\dots,L}^{n=1,\dots,N; i=1,\dots,I}$ and $\{B_{t_l}^{n,i,k}\}_{l=0,\dots,L}^{n=1,\dots,N; i=1,\dots,I}$ from Bernoulli random variables with success probabilities given by the optimal policies derived from $\cV_0(p;\Theta^{(k)}_0)$ and $\cV_1(p,b;\Theta^{(k)}_1)$.
    \item Compute TD errors $\{\delta_{0,l}^{n,i,k}, \delta_{1,l}^{n,i,k}\}_{l=0,\dots,L-1}^{n=1,\dots,N; i=1,\dots,I}$ using \labelcref{eq:TD-errors-OU-S-shaped-utility} with value functions $\cV_0(p;\Theta^{(k)}_0)$ and $\cV_1(p,b;\Theta^{(k)}_1)$.
    \item Sum all TD errors to get the average loss
    \begin{align*}
    \texttt{loss}^{(k)} = \frac{1}{\sum_{n=1}^N \sum_{i=1}^I \sum_{l=0}^{L-1} \mathds{1}_{\{J_{t_l}^{n,i,k} \in \{0,1\}\}}} \sum_{n=1}^N \sum_{i=1}^I \sum_{l=0}^{L-1} \left( (\delta_{0,l}^{n,i,k})^2 + (\delta_{1,l}^{n,i,k})^2 \right).
    \end{align*}
    \item Update value function parameters using gradient descent and get $\Theta^{(k+1)}_0$ and $\Theta^{(k+1)}_1$.
  \end{enumerate}
  \textbf{Output:} learned value functions $\cV_0(p;\Theta^{(K_{\max})}_0)$ and $\cV_1(p,b;\Theta^{(K_{\max})}_1)$.
\end{algorithm}

\end{document}